\documentclass[12pt]{article}

\usepackage{amssymb,amsmath,mathrsfs,tikz,graphicx}
\def\Journal#1#2#3#4{{#1} {\bf #2}, #3 (#4)}

\def\CQG{\em Class. Quantum Grav.}

\def\PRD{\em Phys. Rev. D }
\def\GRG{\em Gen. Rel. Grav.}

\def\JMP{\em J. Math. Phys.}

\def\PRL{\em Phys. Rev. Lett.}

\def\espaitemps{({\cal V},g)}
\def\varietat{{\cal V}}

\def\xiv{\vec \xi }
\def\etav{\vec \eta}
\def\lie{{\pounds}}

\def\AH{\mbox{AH}}

\def\S{\Sigma}

\def\scri{\mathscr{J}}
\def\B{\mathscr{B}}
\def\R{\mathscr{R}}

\def\rr{\R_0}

\def\be{\begin{equation}}
\def\ee{\end{equation}}
\def\bea{\begin{eqnarray}}
\def\eea{\end{eqnarray}}
\def\bean{\begin{eqnarray*}}
\def\eean{\end{eqnarray*}}

\newtheorem{prop}{Proposition}[section]
\newtheorem{theorem}{Theorem}[section]
\newtheorem{result}{Result}[section]

\newtheorem{lemma}{Lemma}[section]
\newtheorem{coro}{Corollary}[section]
\newtheorem{defi}{Definition}
\newtheorem{pr}{Property}

\def\proof{\noindent{\em Proof.\/}\hspace{3mm}}
\def\fin{\hfill \rule{2.5mm}{2.5mm}\\ \vspace{0mm}}

\begin{document}

\title{The region with trapped surfaces in spherical symmetry, its core, and their boundaries}
\author{Ingemar Bengtsson$^1$ and Jos\'e M. M. Senovilla$^2$ \\ \\
$^1$ Stockholms Universitet, AlbaNova, Fysikum,\\
S-106 91 Stockholm, Sweden\\
ingemar@physto.se\\
$^2$ F\'{\i}sica Te\'orica, Universidad del Pa\'{\i}s Vasco, \\
Apartado 644, 48080 Bilbao, Spain \\ 
josemm.senovilla@ehu.es}
\date{}
\maketitle
\begin{abstract} 
We consider the region $\mathscr{T}$ in spacetime containing future-trapped closed surfaces and its boundary $\B$, and derive some of their general properties. We then concentrate on the case of spherical symmetry, but the methods we use are general and applicable to other situations. We argue that closed trapped surfaces have a non-local property, ``clairvoyance", which is inherited by $\B$. We prove that $\B$ is not a marginally trapped tube in general, and that it can have portions in regions whose whole past is flat. For asymptotically flat black holes, we identify a general past barrier, well inside the event horizon, to the location of $\B$ under physically reasonable conditions. We also define the core $\mathscr{Z}$ of the trapped region as that part of $\mathscr{T}$ which is indispensable to sustain closed trapped surfaces. We prove that the unique spherically symmetric dynamical horizon is the boundary of such a core, and we argue that this may serve to single it out. To illustrate the results, some explicit examples are discussed, namely Robertson-Walker geometries and the imploding Vaidya spacetime.
\end{abstract} 

PACS: 04.70.BW, 04.20.Cv

\section{Introduction}

If a black hole is a thing, does it have a boundary? If so, where is it? For 
stationary black holes the event horizon seems to be the obvious answer \cite{Haw}, 
but for evolving black holes the situation is less clear. Obviously, all black holes must be formed (a dynamical process) and then undergo accretion and other evolutionary processes. Furthermore, as a matter of principle 
isolated black holes always evolve through Hawking radiation, and it has been 
argued that---strictly speaking---the event horizon may not even exist, due to 
quantum gravity effects close to the singularity \cite{Haji}. In any case it 
has been argued that the event horizon is `unreasonably global' \cite{AG}. 
Thus there are exact solutions---such as the imploding Vaidya 
spacetime \cite{V}---where an 
observer can cross the event horizon even though her entire causal past is a 
piece of flat Minkowski space \cite{AK1}. In numerical 
relativity this problem is posed sharply because one wants to identify the 
boundary of a black hole in an initial data set, not by inspection of the full 
solution. 

It is then natural to turn to closed trapped surfaces, the hallmark of 
gravitational collapse \cite{P2}. Efficient algorithms to identify the region 
where they occur within a given initial data set do exist \cite{T}.
The boundary of such a spatial region is called 
the apparent horizon \cite{HE}, and it is in itself a marginally outer trapped 
surface \cite{KH,AM}. It `evolves' to form hypersurfaces foliated by such surfaces 
\cite{AMS}, which we refer to as apparent 3-horizons. In general hypersurfaces 
foliated by marginally trapped surfaces are called marginally trapped tubes \cite{AG,B,BBGV}. There 
are still problems, because the location of the horizons obtained in this way 
depends strongly on an arbitrary slicing into space and time. In general 
there will be many marginally trapped tubes in a given black hole spacetime, 
and they weave through each other in a complicated way \cite{AG}.

Another possibility is to locate the boundary of the region through which 
closed trapped surfaces pass, or the boundary of the region through which 
outer trapped surfaces pass. For closed trapped surfaces both null expansions 
are negative, while for outer trapped surfaces no condition is imposed on the
inner expansion. It is important to make this distinction, because the two 
regions do not coincide in general. It was conjectured by Eardley \cite{E} 
that the region defined by outer trapping coincides with the event horizon. 
Ben-Dov \cite{BD} proved this for the Vaidya spacetime, but he also proved that 
the region containing trapped surfaces with both expansions negative lies 
strictly inside it. Nevertheless 
it can still happen that an observer can cross a closed trapped surface even 
though her entire causal past is a piece of flat Minkowski spacetime, as we 
showed in an earlier paper \cite{BS}. This proves that closed trapped surfaces have highly non-local properties too, they are {\em clairvoyant}: they are ``aware" of things that happen elsewhere, very far away with no causal connection.

We focus on trapped surfaces with both expansions 
negative, partly because marginally trapped surfaces form marginally 
trapped tubes---such as future outer trapping horizons \cite{Hay,Hay1} and 
dynamical horizons \cite{AK1}---having particularly interesting 
properties with regard to energy fluxes and the like, and partly 
simply for definiteness. We concentrate on spherically symmetric imploding 
spacetimes. We originally thought that this restriction would enable us to 
fully characterise the boundary $\B$ of the trapped region---the region 
through which closed trapped surfaces pass---but in fact we did not 
succeed in this. Still we are able to give what we think is a quite coherent 
picture of the trapped regions that occur in spherically symmetric spacetimes, and in particular we identify a past barrier for the location of closed trapped surfaces, and of marginally trapped tubes. This boundary turns out to have some quite non-local properties, and may penetrate flat regions of the spacetime. We will further prove that in spherically symmetric spacetimes the boundary $\B$ can never in itself contain any marginally trapped surfaces, so that it cannot be a marginally trapped tube. 

Once we have learnt that the boundary $\B$ of the trapped region suffers from non-local properties which are related to those of the event horizon, and that it is not a marginally trapped tube, we put forward a novel idea that may allow us to determine a preferred marginally trapped tube. We define the core $\mathscr{Z}$ of the trapped region as the region which is indispensable to maintain closed trapped surfaces. This core turns out to be generically smaller than the trapped region, and its boundary may thus be used as a definition of the black hole. We will actually identify a particular core in spherically symmetric spacetimes, and we will prove that its boundary is the unique spherically symmetric marginally trapped tube. It remains as an interesting question to know if this is the unique marginally trapped tube which is the boundary of a core.

In outline, our paper is organised as follows: Section \ref{sec:preli} contains 
reminders about trapped surfaces, the fauna of different cases that 
can occur, and how they can be characterised in a way convenient for 
our purposes by the mean curvature vector. Section \ref{sec:TandB} defines the trapped 
region and its boundary $\B$, and gives their basic properties. Throughout 
the paper we provide proofs of all statements that we make. Section \ref{sec:fun} 
gives some basic results 
on which we build the rest of the paper. The arguments are purely geometrical 
and based on the interplay between (causal) vector fields and surfaces with 
special properties of their mean curvature vectors. Many of these 
results are already known \cite{MS,S2,S3} but we do offer some 
sharpenings, such as Theorem \ref{th:no-min} and Corollary \ref{cor:noposK}. Section \ref{sec:RW} is an interlude dealing 
with Robertson-Walker spacetimes. In this case the general results of 
section 4 are sufficient to pin down the boundary $\B$ exactly. Section \ref{sec:general} 
introduces spherically symmetric spacetimes and their unique spherically 
symmetric apparent 3-horizon. In section 
\ref{sec:notAH} we discuss perturbations of the resulting round marginally trapped 
tubes. We use the stability operator that describes how the outer expansion 
varies when a marginally outer trapped surface is deformed \cite{CC,AMS1}, 
and give a proof that one can always find trapped surfaces that extend 
to both sides of the spherically symmetric apparent 3-horizon. We also show that the region of the perturbed trapped sphere that is inside the apparent 3-horizon can be made arbitrarily small. Section \ref{sec:imploding} 
discusses imploding and asymptotically flat spherically symmetric spacetimes 
in general. In section \ref{sec:Kodama} we identify and discuss a past barrier 
through which future trapped surfaces cannot pass. It is based 
on the presence of the hypersurface forming Kodama vector field \cite{Ko}, 
and our restriction is a definite improvement on previous results 
\cite{BD, AG}. We also prove that any trapped surface must lie at least 
partly inside the spherically symmetric apparent 3-horizon. In section \ref{sec:converse}
we discuss the precise 
location and the properties of the boundary $\B$ in a generic spherically 
symmetric and imploding spacetime. We demonstrate that the boundary $\B$ of the trapped region cannot be a marginally trapped tube. Section \ref{sec:new} 
raises a new issue: Given that trapped surfaces 
must be confined at least partly within the spherically symmetric apparent 
horizon, what is the minimal region that must be removed from spacetime 
in order for it to have no closed trapped surfaces at all? We call this 
the core of the trapped region. We show that the spherically symmetric apparent 3-horizon is the boundary of one such core. Finally, the important example provided 
by the Vaidya spacetime is treated in an Appendix.

\section{Preliminaries: the trapped-surface family}\label{sec:preli}
Trapped surfaces are the basic objects to be studied in this paper. Thus, we start by providing their definition and their types, and by fixing our notation. Standard references  are \cite{HE,HP,Kr,P2,Wald}.

Let $\espaitemps$ be a 4-dimensional causally orientable 
spacetime with metric $g_{\mu\nu}$ of signature $-,+,+,+$. 
Let $S$ denote a connected 2-dimensional
surface with local intrinsic coordinates $\{\lambda^A\}$ ($A,B,\dots =2,3$) imbedded in
$\varietat$ by the $C^3$ parametric equations
$$
x^{\alpha} =\Phi ^{\alpha}(\lambda^A) \hspace{1cm}
(\alpha,\beta, \dots =0,1,2,3)
$$
where $\{x^{\alpha}\}$ are local coordinates for $\varietat$.
The tangent vectors $\vec{e}_A$ of $S$ are locally given by
\bean
\vec{e}_A \equiv e^{\mu}_A \left.\frac{\partial}{\partial x^{\mu}}
\right\vert _S \equiv
\frac{\partial \Phi^{\mu}}{\partial \lambda^A}
\left.\frac{\partial}{\partial x^{\mu}}\right\vert_S
\eean
so that the first fundamental form of $S$ in $\varietat$ is
$$
\gamma _{AB}\equiv \left.g_{\mu \nu}\right\vert_S\frac{\partial \Phi^{\mu}}
{\partial \lambda^A}\frac{\partial \Phi^{\nu}}{\partial \lambda^B} 
$$
which collects the scalar products $g\left(\vec{e}_A,\vec{e}_B \right)$.
From now on, we shall assume that $\gamma _{AB}$ is positive definite so that $S$ is
a {\it spacelike} surface. Then, the two linearly independent normal one-forms
$k_{\mu}^{\pm}$ to $S$ can be chosen to be null and future directed
everywhere on $S$, so they satisfy
$$
k_{\mu}^{\pm}e^{\mu}_A=0,\ \ k^{+}_{\mu} k^{+ \mu}=0 ,\ \ 
 k^{-}_{\mu} k^{- \mu}=0 ,\hspace{3mm}  k_{\mu}^+ k^{-\mu}=-1 , 
$$
where the last equality incorporates a condition of normalization.
Obviously, there still remains the freedom
\begin{equation}
k^+_{\mu} \longrightarrow k'^+_{\mu}=\sigma^2 k^+_{\mu}, \hspace{1cm}
k^-_{\mu} \longrightarrow k'^-_{\mu}=\sigma^{-2} k^-_{\mu} \label{free}
\end{equation}
where $\sigma^2$ is a positive function defined only on $S$.

The standard splitting into tangential and normal 
directions to $S$ leads to a formula relating the covariant 
derivatives on $\espaitemps$ and on $(S,\gamma)$ \cite{Kr,O}:
$$
\nabla_{\vec{e}_{A}}\vec{e}_{B}=\overline{\Gamma}^{C}_{AB}\vec{e}_{C}-\vec{K}_{AB}
$$
where $\overline{\Gamma}^{C}_{AB}$ are the coefficients of the Levi-Civita 
connection $\overline\nabla$ of 
$\gamma$ (i.e. $\overline\nabla\gamma =0$),
and $\vec{K}_{AB}$ is the shape tensor (also called second fundamental form vector, and extrinsic curvature vector) of $S$ in $\espaitemps$. Note that $\vec{K}_{AB}=\vec{K}_{BA}$ is symmetric, and
orthogonal to $S$, from where we deduce
$$
\vec{K}_{AB}=-K^-_{AB}\vec k^+ -K^+_{AB}\vec k^- \, .
$$
Here, $K^{\pm}_{AB}$ are two symmetric covariant tensor fields defined on $S$ and called the two null (future) second fundamental forms of $S$ in $\espaitemps$. They are explicitly defined by
$$
K^{\pm}_{AB} \equiv -k^{\pm}_{\mu}e^{\nu}_A\nabla_{\nu}e^{\mu}_B = e^{\mu}_Be^{\nu}_A\nabla_{\nu}k^{\pm}_{\mu} .
$$
The shape tensor gives the difference between the projection to $S$ of the covariant derivative and the intrinsic derivative on $S$ by means of the fundamental relation
\be
e^{\mu}_{A}e^{\nu}_{B}\nabla_{\mu}v_{\nu}|_{S}=\overline\nabla_{A} 
\overline{v}_{B}+v_{\mu}|_{S} K^{\mu}_{AB}
\label{nablas2}
\ee
where, for all $v_{\mu}$ we denote by
$$
\overline{v}_{B}\equiv v_{\mu}|_{S}\,\, e^{\mu}_{B}
$$
its projection to $S$.

The mean curvature vector of $S$ in $\espaitemps$ \cite{O,Kr} is the trace of the shape tensor
$$
\vec{H}\equiv \gamma^{AB}\vec{K}_{AB}
$$
where $\gamma^{AB}$ is the contravariant metric on $S$: $\gamma^{AC}\gamma_{CB}=\delta^A_B$. Observe that $\vec H$ is orthogonal to $S$ and that
$$
\vec{H}\equiv -\theta^-\vec{k}^+ - \theta^+\vec{k}^-
$$
where
\begin{equation}
\theta^{\pm} \equiv \gamma^{AB}K^{\pm}_{AB} \label{tr}
\end{equation}
are the traces of the null second fundamental forms, usually called the (future) null expansions.
Clearly, $\vec H$ is invariant under transformations (\ref{free}). 

The class of {\em weakly future-trapped} (f-trapped from now on) surfaces are characterized by having
$\vec H$ pointing to the future everywhere on $S$, and similarly for weakly past 
trapped. There are three important subcases that deserve their own name: (i) the traditional f-trapped surfaces have $\vec H\neq \vec 0$ timelike everywhere on $S$; (ii) marginally f-trapped surfaces have $\vec H\not\equiv \vec 0$ null everywhere on $S$; and (iii) minimal surfaces have $\vec H \equiv \vec 0$ on $S$. 

These conditions can be equivalently expressed in terms of the signs of the expansions as follows:

\begin{center}
\begin{tabular}{c|c|c|l}
$\vec H$ & Null expansions & Type of surface \\
\hline
causal future & $\theta^+\leq 0, \theta^-\leq 0$ & weakly f-trapped \\
{\bf Subcases:} & & \\
zero & $\theta^+=\theta^-=0$ & minimal \\
null and future $\not\equiv \vec 0$ & $\theta^+= 0, \theta^-\leq 0$ & marginally f-trapped \\
null and future $\not\equiv \vec 0$ & $\theta^+\leq 0, \theta^-= 0$ & marginally f-trapped \\
timelike future & $\theta^+<0, \theta^-<0$ & f-trapped\\
\end{tabular}
\end{center}
This is to be compared with \cite{Wald,AG,HE,S4}, as sometimes different names are given to the same objects, and vice versa. In particular, weakly f-trapped surfaces were called nearly f-trapped in \cite{MS,S4}. Here we will follow the previous nomenclature which pretends to respect standard names as much as possible. See \cite{S4} for further details. In the case that $\vec H$ is proportional to one of the null normals but realizing both causal orientations the surface is said to be null dual, or marginally $(\pm)$-trapped, where the $\pm$ refers to the direction with vanishing expansion ---as the definition is equivalent to having either $\theta^+$ or $\theta^-$ vanishing. In the literature they are usually referred as MOTS (``marginally outer trapped surfaces"), by declaring the direction with vanishing expansion to be ``outer".

For completeness and future reference, we also mention that a surface is called {\em untrapped} if the mean curvature vector is spacelike everywhere, or equivalently, if both expansions have opposite signs.

\section{Definition and basic properties of $\mathscr{T}$}\label{sec:TandB}
In this paper, we will be concerned with the following sets in $\espaitemps$, see also \cite{Hay}.
\begin{defi}
The future-trapped region $\mathscr{T}$ is defined as the set of points $x\in \varietat$ such that $x$ lies on a closed future-trapped surface.
\end{defi}
Since the characterization of f-trapped surfaces is the negativity of both null expansions $\theta^\pm$ as defined in (\ref{tr}), the following general property follows easily.
\begin{pr}\label{pr:Topen}
The future-trapped region $\mathscr{T}$ is an open set.
\end{pr}
\proof Take any $x\in \mathscr{T}$ and let $S\ni x$ be a closed f-trapped surface.
We can perturb any such $S$ along an arbitrary direction $\vec q$ defined on $S$, by moving any point $y\in S$ along $\vec q$ a distance $\epsilon >0$. The deformed surface $S_\epsilon$ has future null expansions $\theta^\pm_\epsilon= \theta^\pm +\epsilon \delta_{\vec q}\, \theta^\pm +O(\epsilon^2)$, where $\theta^\pm$ are the null expansions on $S$ and the variations $\delta_{\vec q}\, \theta^\pm $ are given by precise  formulas involving $\vec q$, the future null normals to $S$ and the geometric properties of $S$, see e.g. \cite{AMS,AMS1,AM,Hay,Gal,CC}. As $\theta^\pm <0$ and $\delta_{\vec q}\, \theta^\pm $ are continuous on $S$, we can always choose $\epsilon (\vec q)$ small enough such that $\theta^\pm_\epsilon <0$ for {\em any} given direction $\vec q$. This implies that there exists a small neighborhood $U(x)$ of $x\in S$ such that $U(x)\subset \mathscr{T}$.\fin

In general, $\mathscr{T}$ does not have to be connected. Any connected component of $\mathscr{T}$ will thus be termed as a ``connected f-trapped region''.

It is clear that $\mathscr{T}$ can be empty, or it can be the whole spacetime. An example of the former case $\mathscr{T}=\emptyset$ is provided by any globally static spacetime \cite{MS}. Examples of the latter case $\mathscr{T}=\varietat$ are de Sitter spacetime or some contracting Robertson-Walker geometries, see section \ref{sec:RW}. 
However, in asymptotically flat black-hole type spacetimes \cite{Haw,HE,Wald} neither of these cases will happen, because there are no trapped surfaces near spatial infinity, and there will appear future-trapped surfaces in the black hole region. In these cases $\mathscr{T}$ will have a boundary on $\espaitemps$. 
\begin{defi}\label{def:B}
We denote by $\B$ the boundary of the future-trapped region $\mathscr{T}$:
$$
\B \equiv \partial \mathscr{T} \, .
$$
\end{defi}
\begin{pr}\label{pr:Bclosed}
$\B$ being the boundary of an open set, it is itself a {\em closed} set without boundary. 
Moreover $\B \cap \mathscr{T}=\emptyset$.
\end{pr}
\begin{pr}\label{pr:2sides}
$\B$ is also the boundary of the untrapped region defined by the set of points $x\notin \mathscr{T}$, that is, such that $x$ does not lie on any closed f-trapped surface. 
\end{pr}
We remark that $\B$ is not necessarily connected.

$\mathscr{T}$ and $\B$ are genuine spacetime objects, independent of any foliations or initial Cauchy data sets. Therefore, $\mathscr{T}$ and $\B$ are different in nature from the trapped regions and their boundaries contained in given slices, as recently studied in \cite{AM}.

The symmetries of the spacetimes respect $\mathscr{T}$ and $\B$.  More precisely:
\begin{result}\label{res:Bandsym}
If $G$ is the group of isometries of the spacetime $\espaitemps$, then $\mathscr{T}$ is invariant under the action of $G$, and the transitivity surfaces of $G$, relative to points of $\B$, remain in $\B$.
\end{result}
\proof Take any point $x\in \mathscr{T}$. Then there is a closed f-trapped surface $S$ passing through $x$. By moving $S$ via the motion group $G$, and as f-trapped surfaces are moved to f-trapped surfaces by isometries, it follows that the whole transitivity surface of the group  passing through $x$ lies in $\mathscr{T}$. Similarly, let $y\in \B$, so that any small neighborhood of $y$ intersects $\mathscr{T}$. By moving one such small neighborhood using $G$ one similarly deduces that the transitivity surface of $G$ relative to $y$ is part of $\B$. \fin

An implication of this result is that no globally defined Killing vector can be transversal to $\B=\partial\mathscr{T}$. Actually, this also holds for homothetic Killing vectors (vector fields $\xiv$ satisfying $(\lie_{\xiv}g)_{\mu\nu}=2 c g_{\mu\nu}$) as they also respect f-trapped surfaces (if $c>0$) \cite{CM}.

\begin{coro}\label{cor:sym}
Let $G_n$ be the (global) continuous group of isometries of $\espaitemps$ where $n$ is its dimension (i.e., the number of linearly independent Killing vectors) and let $m$ be the dimension of its surfaces of transitivity.
\begin{enumerate}
\item If $n\geq 4$ and $m=4$, that is, if the spacetime is homogeneous, then $\B=\emptyset$ and either $\mathscr{T}=\emptyset$ or $\mathscr{T}=\varietat$.
\item If $n\geq 3$ and $m=3$, then either $\B=\emptyset$ or each connected component of $\B$ is one of the 3-dimensional surfaces of transitivity.
\item If $n\geq 2$ and $m=2$, then either $\B=\emptyset$ or each connected component of $\B$ is a hypersurface without boundary foliated by the 2-dimensional surfaces of transitivity.
In particular, in arbitrary spherically symmetric spacetimes, $\B$ (if not empty) is a spherically symmetric hypersurface without boundary.
\end{enumerate}
\end{coro}
\proof
Point 1 is immediate. Points 2 and 3 follow because any connected component of $\B$ cannot have a boundary and cannot be given by isolated 2-dimensional surfaces of transitivity, as these would contradict its basic Properties \ref{pr:Bclosed} and \ref{pr:2sides}.
\fin

What is the possible relevance of $\B$? Apart from answering natural questions such as ``where can there be closed f-trapped surfaces?'' or ``is this event part of a closed f-trapped surface?'', the location of $\B$ provides important physical information due to the fundamental relevance of closed trapped surfaces in the development of black holes and singularities \cite{HP,P2,P5,S}. More importantly, it provides a precise limit as to where dynamical horizons or marginally trapped tubes can develop. One could also hope that $\B$ is related to the surface of a dynamical black hole. We are going to see that this suffers from the same problems as other candidates.

\section{Fundamental results}\label{sec:fun}
In this section we present the main results that will allow us to put restrictions on the location of the region $\mathscr{T}$ containing the closed future-trapped surfaces of a spacetime and its boundary $\B$. These results are fully general, and can be obtained within the framework of the interplay between generalized symmetries and submanifolds with special properties of their mean curvature vector. The underlying ideas come from \cite{MS,S,S2,S3}.

We start with the main formula to be used in what follows. Let $\xiv$ be an
arbitrary $C^1$ vector field on $\varietat$ defined on a neighbourhood 
of $S$. Recalling the identity $(\lie_{\xiv} g)_{\mu\nu}=
\nabla_{\mu}\xi_{\nu}+\nabla_{\nu}\xi_{\mu}$ where 
$\lie_{\xiv}$ denotes the Lie derivative with respect to $\xiv$, one gets on using (\ref{nablas2})
$$
(\lie_{\xiv} g|_{S})_{\mu\nu}\, e^{\mu}_Ae^{\nu}_B=\overline\nabla_{A}\overline\xi_{B}+
\overline\nabla_{B}\overline\xi_{A}+2 \xi_{\mu}|_{S} K^{\mu}_{AB}\, .
$$
Contracting now with $\gamma^{AB}$ we arrive at
\be
\fbox{$\displaystyle{\frac{1}{2}
P^{\mu\nu}(\lie_{\xiv} g|_{S})_{\mu\nu} =
\overline\nabla_{C}\overline\xi^{C}+ \xi_\rho H^\rho}$}
\label{main}
\ee
where 
$$
P^{\mu\nu}\equiv \gamma^{AB} e^{\mu}_Ae^{\nu}_B
$$
is the orthogonal projector that projects any object to its part tangent to $S$. 

The elementary formula (\ref{main}) is very useful and permits one to obtain many interesting results about the existence of weakly trapped closed surfaces in the presence of generalized symmetries, see \cite{MS,S2,S3}. For instance: 
\begin{enumerate}
\item If $S$ is minimal ($\vec H =\vec 0$), integrating (\ref{main}) and using Gauss' theorem one finds that the divergence term does not contribute {\em whenever} $S$ is closed (that is, compact without boundary), ergo
$$
\oint_{S}P^{\mu\nu}(\lie_{\xiv} g|_{S})_{\mu\nu}=0 \, .
$$
Observe that this relation must be satisfied for {\em all} possible vector fields $\xiv$. Therefore, closed minimal surfaces are very rare.
\item If $\xiv$ is a Killing vector, then the left hand side of 
(\ref{main}) vanishes. Integrating the righthand side on $S$ we get for closed $S$
$$
\oint_{S}\xi_\rho H^\rho =0 \, .
$$
Therefore, if the Killing vector $\xiv$ is timelike on $S$, then $S$ cannot be weakly
trapped (neither future nor past), unless it is minimal.
\end{enumerate}

The following is an important consequence of formula (\ref{main}).
\begin{lemma}\label{lem:basic}
Let $\xiv$ be a vector field which is future-pointing on a region $\R\subset \varietat$, and let $S$ be a surface contained in $\R$ such that $P^{\mu\nu}(\lie_{\xiv} g|_{S})_{\mu\nu} \geq 0$. Then, $S$ cannot be closed and weakly f-trapped unless $\xi_\mu H^\mu=0$ and $P^{\mu\nu}(\lie_{\xiv} g|_{S})_{\mu\nu} = 0$.
\end{lemma}
\proof 
Integrating (\ref{main}) on the closed $S$, the divergence term integrates to zero and we get
$$
\oint_{S}\xi_\rho H^\rho =\frac{1}{2} \oint_S P^{\mu\nu}(\lie_{\xiv} g|_{S})_{\mu\nu} \geq 0
$$
which implies that $\vec H$ cannot be future pointing all over $S$, unless $\xi_\mu H^\mu=P^{\mu\nu}(\lie_{\xiv} g|_{S})_{\mu\nu} =0$. \fin

{\bf Remarks:}
\begin{itemize}
\item
For non-minimal $S$, the exceptional case $\xi_\mu H^\mu=0$ implies that $S$ is marginally f-trapped and that $\xiv|_S$ is null and proportional to $\vec H$.
\item 
Notice that only the averaged condition $\oint_S P^{\mu\nu}(\lie_{\xiv} g|_{S})_{\mu\nu} \geq 0$ is needed here, so that $P^{\mu\nu}(\lie_{\xiv} g|_{S})_{\mu\nu}$ can be negative somewhere on $S$ as long as the averaged formula holds.
\end{itemize}

Important instances where Lemma \ref{lem:basic} can be applied are given by the {\em conformal Killing vectors} (see e.g. \cite{Exact}) and the {\em Kerr-Schild vector fields} \cite{CHS}. The former satisfy
\be
(\lie_{\vec\xi}g)_{\mu\nu}=2\psi g_{\mu\nu} \label{CKV}
\ee
for some function $\psi$, so that $P^{\mu\nu}(\lie_{\xiv} g|_{S})_{\mu\nu} =4\psi|_S$. Thus, the condition on Lemma \ref{lem:basic} requires simply that $\psi|_S\geq 0$.
On the other hand, Kerr-Schild vector fields are characterized by
\be
(\lie_{\vec\xi}g)_{\mu\nu}=2h\ell_\mu \ell_\nu, \hspace{1cm} (\lie_{\vec\xi}\ell)_{\mu}=b\ell_\mu
\label{KSVF}
\ee
for some functions $h$ and $b$, where $\ell_\mu$ is a fixed {\em null} one-form field ($\ell_\mu\ell^\mu=0$). Therefore $P^{\mu\nu}(\lie_{\xiv} g|_{S})_{\mu\nu} =2h\, \bar\ell_A \bar\ell^A$ and the condition on the Lemma \ref{lem:basic} requires now $h|_S\geq 0$.

Stronger results can be found for the case where $\xiv$ is a hypersurface-orthogonal vector field, that is to say,
$$
\xi_{[\mu}\nabla_{\nu}\xi_{\rho]}=0 \hspace{2mm} \Longleftrightarrow \hspace{2mm} \xi_{\mu}=-F \partial_{\mu} \tau
$$
for some local functions $F>0$ and $\tau$. Then,  $\xiv$ is orthogonal to the hypersurfaces $\tau =$const., which are called the level hypersurfaces. In this case we have the following fundamental result.
\begin{theorem}\label{th:no-min}
Let $\xiv$ be a vector field which is future-pointing and hypersurface-orthogonal on a region $\R\subset \varietat$ with level hypersurfaces $\tau =$const.\ and let $S$ be a f-trapped surface. Then, $S$ cannot have a local minimum of $\tau$ at any point $q\in \R$ where $P^{\mu\nu}(\lie_{\xiv} g)_{\mu\nu}|_q\geq 0$.

In the case that $S$ is a weakly f-trapped surface the conclusion still holds unless
\be
\left.\frac{\partial^2\bar\tau}{\partial\lambda^A\partial\lambda^B}\right|_q=0 \hspace{3mm} \mbox{and} \hspace{3mm}
P^{\mu\nu}(\lie_{\xiv} g)_{\mu\nu}|_q=0 \hspace{3mm} \mbox{and} \hspace{3mm} \left.\xi_\rho H^\rho\right|_q=0 .\label{nueva}
\ee

\end{theorem}
\proof Let $q\in S\cap \R$ be a point where $S$ has a local extreme of $\tau$. Then, $\xiv$ is orthogonal to $S$ at $q$, that is to say, $\bar\xi_A|_q=0$. Another way of stating the same is that
$$
\left.\frac{\partial\bar\tau}{\partial \lambda^A}\right|_q=0
$$
where $\tau =\bar\tau(\lambda^A)$ is the local parametric expression of $\tau$ in terms of the local coordinates $\{\lambda^A\}$ of $S$ (note that $\bar\xi_A=-\bar F\partial\bar\tau/\partial \lambda^A$ with $\bar F \equiv F|_S$). Using the previous expression, we can now compute the divergence $\overline\nabla_{C}\overline\xi^{C}$ at $q$:
$$
\left.\overline\nabla_{A}\overline\xi^{A}\right|_q=\left.\gamma^{AB}\overline\nabla_A\left(-\bar F\frac{\partial\bar\tau}{\partial\lambda^B}\right)\right|_q=\left.-\bar F\gamma^{AB}\frac{\partial^2\bar\tau}{\partial\lambda^A\partial\lambda^B}\right|_q\, .
$$
Introducing this in formula (\ref{main}) we get at $q$
$$
\left.\bar F\gamma^{AB}\frac{\partial^2\bar\tau}{\partial\lambda^A\partial\lambda^B}\right|_q=\left.-\frac{1}{2}
P^{\mu\nu}(\lie_{\xiv} g)_{\mu\nu} \right|_q+
\left.\xi_\rho H^\rho\right|_q\leq \left.\xi_\rho H^\rho\right|_q
$$
so that, if $S$ has a future pointing (possibly vanishing) $\vec H|_q$ we deduce
$$
\left.\gamma^{AB}\frac{\partial^2\bar\tau}{\partial\lambda^A\partial\lambda^B}\right|_q\leq 0
$$
which, given that $\gamma^{AB}$ is positive definite, implies that $\partial^2\bar\tau/\partial\lambda^A\partial\lambda^B|_q$ cannot be positive (semi)-definite. Therefore, $\bar\tau$ cannot have a local minimum at $q$.\fin

{\bf Remarks:} 
\begin{enumerate}
\item Observe that $S$ does not need to be compact, nor contained in $\R$.
\item Notice that it is enough to assume $P^{\mu\nu}(\lie_{\xiv} g)_{\mu\nu}|_q\geq 0$ only at the points $q$ that are local extrema of $\tau$ on $S$. The sign of $P^{\mu\nu}(\lie_{\xiv} g|_{S})_{\mu\nu}$ can thus be left arbitrary everywhere else on $S$ where $\bar\xi_A\neq 0$.
\item Let us stress that the possibility with a positive semi-definite $\partial^2\bar\tau/\partial\lambda^A\partial\lambda^B|_q$ is also excluded, so that $\bar\tau$ cannot even be constant along one direction at $q$. The only exceptional possibility is given by the case identified in the theorem satisfying (\ref{nueva}). If $\xiv|_q$ is timelike, the last in (\ref{nueva}) implies that $\vec H|_q=\vec 0$. 

Therefore, letting aside this exceptional possibility, $\tau$ will always decrease at least along one tangent direction in $T_qS$. It follows that, under the conditions of the theorem, starting from any point $x\in S\cap\R$ one can always follow a connected path along $S\cap\R$ with decreasing $\tau$.
\item Theorem \ref{th:no-min} applies in particular (but not only!) to (i) {\em static} Killing vectors of course, (ii) hypersurface-orthogonal conformal Killing vectors (\ref{CKV}) with $\psi\geq 0$, and (iii) hypersurface-orthogonal Kerr-Schild vector fields (\ref{KSVF}) with $h\geq 0$.
\item The results in section IV of \cite{BD} ---that any f-trapped surface penetrating a flat portion of the spacetime cannot have a minimum of ``inertial time" there, so that they have to ``bend down in time"--- are simple consequences of the more general Theorem \ref{th:no-min}, which applies not only to flat spacetimes but in general to any static region, and to much more general cases as remarked above.
\end{enumerate}

Another important result for hypersurface-orthogonal vector fields is (see also \cite{MS,S2,S3}):
\begin{theorem}\label{th:untrapped}
Let $\xiv$ be a vector field which is future-pointing and hypersurface-orthogonal on a region $\R\subset \varietat$. Then, all spacelike surfaces $S$ (compact or not) contained in one of the level hypersurfaces $\tau=$const. within $\R$ have
\be
2\xi_\rho H^\rho =P^{\mu\nu}(\lie_{\xiv} g|_{S})_{\mu\nu}  .\label{tau=c}
\ee
In particular, at any point $x\in S$ such that $P^{\mu\nu}(\lie_{\xiv} g)_{\mu\nu} |_x \geq 0$, $S$ has a mean curvature vector which is not timelike future-pointing, and it can be future-pointing null or zero only if $P^{\mu\nu}(\lie_{\xiv} g)_{\mu\nu}|_x=\xi_\mu H^\mu|_{x}=0$.
\end{theorem}
\proof Let $S\cap\R$ be a (portion of a) surface contained in one of the  hypersurfaces $\tau=$const. Then $\bar\xi_A=0$ all over $S\cap\R$ so that from (\ref{main}) we deduce (\ref{tau=c}). This immediately implies the rest of results.\fin

From Theorems \ref{th:untrapped} and \ref{th:no-min} we deduce the following general property.
\begin{coro}\label{cor:noposK}
No f-trapped surface (closed or not) can touch a spacelike hypersurface to its past at a single point, or have a 2-dimensional portion contained in the hypersurface, if the latter has a positive semi-definite second fundamental form.
\end{coro}
\proof As the hypersurface is spacelike its normal vector, say $\xiv$, is timelike, can be extended to be hypersurface-orthogonal with level function $\tau$ and can be chosen to be future-pointing. Then, the projection of $(\lie_{\xiv}g)_{\mu\nu}$ to the hypersurface is proportional to its second fundamental form. Thus, if this were positive semi-definite it would follow that $P^{\mu\nu}(\lie_{\xiv} g|_{S})_{\mu\nu} \geq 0$ where $P^{\mu\nu}$ is the projector to any such surface $S$, and the result follows.\fin

\begin{figure}[!ht]
\includegraphics[height=5cm]{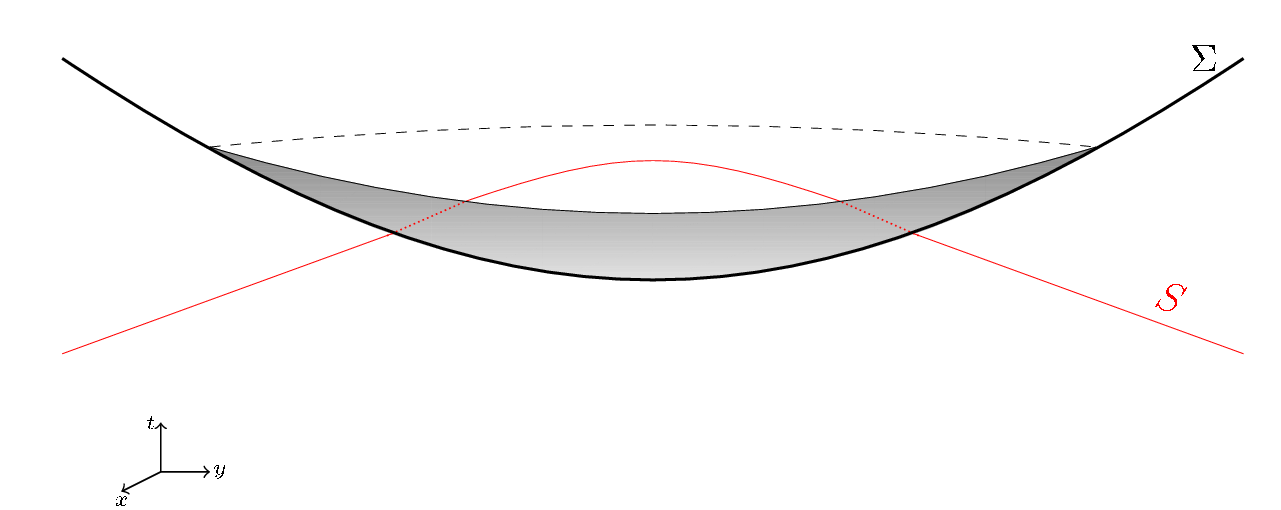}
\caption{\footnotesize{Schematic diagram of the situation relevant to Corollary \ref{cor:noposK}. For simplicity the picture is drawn in flat spacetime, one dimension suppressed. The trapped surface $S$, represented here by a red line, has to bend down in time $t$. $\S$ represents a spacelike hypersurface with positive definite second fundamental form.}}
\label{fig:intuition}
\end{figure}

An intuitive explanation of this result is presented in figure \ref{fig:intuition}.

\section{Application to Robertson-Walker geometries}\label{sec:RW}
Take the Robertson-Walker spacetimes, with line-element given by \cite{Exact,Wald}
\be
ds^2=-dt^2+a^2(t)\,\,  d\S^2_{k}\label{RW}
\ee 
where $d\S^2_{k}$ is the standard metric on a maximally symmetric 3-dimensional space with normalized sectional curvature $k=\pm 1,0$, and $a(t)$ is a function of the preferred time $t$ called the scale factor. The future is defined by increasing values of $t$. The above line-element possesses a timelike conformal Killing vector  given by $\xiv = a(t)\partial_{t}$ such that $\psi$ in (\ref{CKV}) is simply $\dot a \equiv da/dt$. Furthermore, $\xiv$ is obviously hypersurface-orthogonal
$$
\xi_{\mu}dx^\mu = -a(t) dt \, ,
$$
with level surfaces given by the preferred hypersurfaces $t=$const. $\xiv$ has been used to derive results on closed weakly f-trapped surfaces in \cite{MS,S2} and more recently on MOTS in \cite{CM}.
\begin{result}\label{adotnotpos}
No closed weakly f-trapped surface $S$ can be fully contained in the region $\dot a\geq 0$ of a Robertson-Walker spacetime, unless $k=1$ and $S$ is a minimal surface imbedded in a 3-sphere $t=t_{0}$ with $\dot a(t_{0})=0$.
\end{result}
\proof 
Direct application of Lemma \ref{lem:basic} implies \cite{MS,S2} that the only open possibility is given by $\vec H =\vec 0$ and $\dot a|_{S}=0$. Given that the hypersurface $t=t_{0}$ has vanishing second fundamental form, any such minimal surface must be minimal {\em within} the hypersurface $t=t_{0}$. This immediately rules out the cases $k=0,-1$, as there are no compact minimal surfaces imbedded in Euclidean or hyperbolic spaces \cite{My,Os}.\fin

Similarly, from Theorem \ref{th:no-min} follows that
\begin{result}\label{nominadotpos}
No closed weakly f-trapped surface can have a local minimum of $t$ at the region with $\dot a\geq 0$. Thus, the minimum of $t$ 
must be attained at a hypersurface $t=\tilde t$ with $\dot a|_{\tilde t}<0$, if they exist.
\end{result}

As a consequence, f-trapped closed surfaces are absent in generic {\em expanding} Robertson-Walker models. They can only be present in models with contracting phases. Actually, it is well-known ---e.g. \cite{HE,S}--- that there are closed trapped round spheres at any $t=$constant slice with $\dot a \neq 0$ and $\dot a^2 +k >0$. This last condition is simply the positivity of $\varrho +\Lambda$, where $\varrho$ is the energy density relative to the preferred observer $\partial_{t}$ and $\Lambda$ the cosmological constant. Given the homogeneity of the maximally symmetric slices, this implies that there are closed trapped round spheres through {\em any} point of the the slices with $\dot a \neq 0$ and $\dot a^2 +k >0$. These spheres are, of course, past-trapped if $\dot a >0$ and f-trapped if $\dot a <0$. The only remaining possibility ---keeping $\dot a^2 +k >0$--- is that of 3-sphere slices ($k=1$) with $\dot a =0$, in which case all round spheres in the slice are untrapped except for the equatorial one which is a minimal surface.

From Corollary \ref{cor:sym} we already know that the boundary $\B$ must be constituted by $t=$constant slices. Thus, $\B$ splits into two different disconnected sets, $\B^-$ and $\B^+$, according to whether the closed f-trapped surfaces lie locally to the future or past of $\B$, respectively. Combining this with the arguments in the previous paragraph we can deduce the following.
\begin{result} 
On a general Robertson-Walker spacetime with line-element (\ref{RW}) and $\dot a^2 +k \geq 0$
\begin{itemize}
\item
If $\dot a \geq 0$ everywhere, then $\mathscr{T} =\emptyset$ and thus $\B=\emptyset$.
\item
If $\dot a \leq 0$ everywhere, then $\mathscr{T} =\varietat$ and thus $\B=\emptyset$.
\item In the case that $\dot a$ changes sign,
the past boundary $\B^-$, if non-empty, is necessarily contained in the region with $\dot a= 0$ and $\ddot a <0$: 
$$\B^- \subset \{\dot a=0\} \cap \{\ddot a <0\} .$$
\end{itemize}
\end{result}
\proof
The first two points are direct consequences of Results \ref{adotnotpos} and \ref{nominadotpos}. To prove the third point, suppose that a f-trapped surface $S$ enters into a region with $\dot a>0$. Then, $S$ cannot have a minimum of $t$ there due to Result \ref{nominadotpos}. If $S$ is closed, $S$ must attain the minimum of $t$, hence $S$ has to extend to the past for all values of $t$ until it crosses the first slice with $\dot a=0$ and $\ddot a >0$, entering into a region with $\dot a <0$. But we know that there are closed f-trapped round spheres through every point of such a region, so the boundary $\B^-$ cannot be there. Proceeding towards the past, we either encounter another slice with $\dot a =0$ and $\ddot a<0$ or not. In the second possiblility, there is no boundary $\B^-$ for the connected component of $\mathscr{T}\supset S$. In the first case, either that slice is such a boundary $\B^-$, or else there are closed f-trapped surfaces crossing the slice towards the past, that is, entering into another region with $\dot a >0$. We can then repeat the argument from the beginning until one of the slices with $\dot a =0$ and $\ddot a<0$ is the past boundary of the mentioned connected f-trapped region or there is no such a boundary.\fin

The last point in the previous result can be re-phrased as saying that the past boundary $\B^-$ is constituted by preferred slices with $\dot a=0$, which have a vanishing second fundamental form and therefore are maximal and totally geodesic, and such that the spacetime starts to re-collapse there. 
\begin{coro}
If the Robertson-Walker geometry (with $\dot a^2 +k \geq 0$) has an initial expanding epoch, then either $\B=\emptyset$ ---if it is expanding forever---, or the first re-collapsing time is always part of $\B^-$.
\end{coro}

There is the open question whether the future boundary $\B^+$ can be non-empty. If so, one easily obtains that it is necessarily contained in the region with $\dot a > 0$. (Actually, one can further prove that $\B^+ \subset \{\dot a >0\} \cap \{\ddot a \geq 0\} $, but this will not be necessary in what follows.) The question of whether
there can be weakly f-trapped surfaces intersecting both expanding and contracting regions is answered in the positive, as follows from de Sitter spacetime or in more general cases from the results in \cite{FHO}.

An important consequence of all the above is the following
\begin{result}\label{res:noWTSinB}
On a general Robertson-Walker spacetime with line-element (\ref{RW}) and $\dot a^2 +k \geq 0$, the boundary $\B$ of the future-trapped region $\mathscr{T}$ does not contain any non-minimal weakly f-trapped surface (closed or not).
\end{result} 
\proof Take any slice $t=t_{0}$ which is part of $\B$, so that $\dot a (t_{0})\geq 0$, and choose any surface $S$, closed or not, imbedded into $t=t_{0}$. Then $\xiv$ is normal to $S$ so that $\bar\xi_{A}=0$ and Theorem \ref{th:untrapped} implies that $\xi_{\rho}|_{S}H^\rho =\dot a(t_{0})\geq 0$. Thus, $\vec H$ has to be spacelike, possibly zero, or past-pointing everywhere on $S$ (the past-pointing possibility is forbidden if $\dot a(t_{0})=0$, i.e., within $\B^-$.)\fin

We will see that this turns out to be a rather general property in spherical symmetry, so that $\B$ will never contain closed weakly f-trapped surfaces.

Observe that, in the case $k=0$ (say; a similar reasoning works for the case $k={-1}$), if $\B^-$ is non-empty, there are f-trapped round spheres as close to $\B^-$ as we like. What happens if we try to approach $\B^-$ following a sequence of such f-trapped round spheres? The answer is simply that they get larger and larger, and if we go to the limit approaching $\B^-$, they actually break and become non-compact minimal planes. 

\begin{figure}[!ht]
\includegraphics[height=10cm]{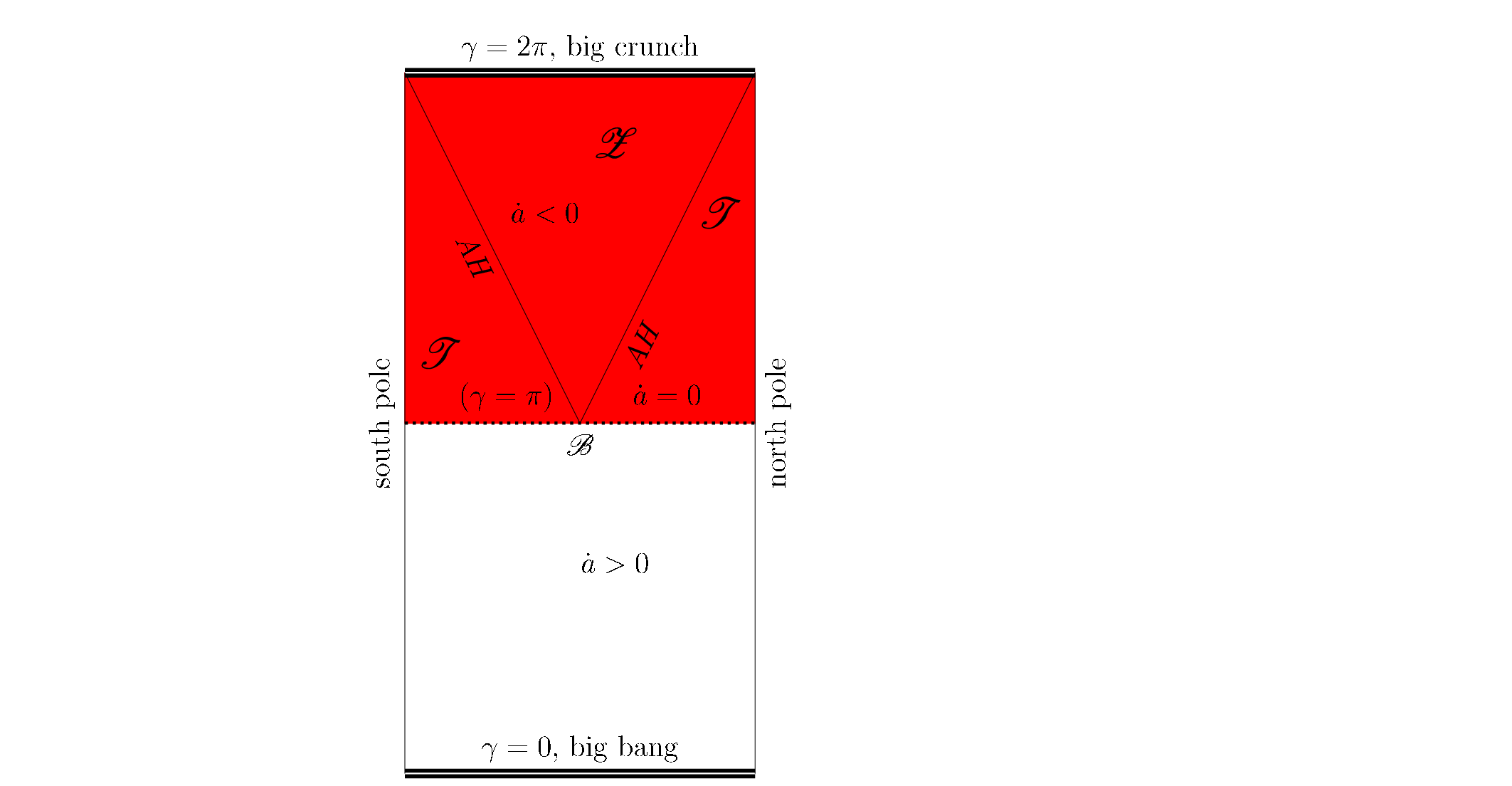}
\caption{\footnotesize{Conformal diagram of the Robertson-Walker spacetime (\ref{RW}) for $k=1$ and dust. As usual, null radial lines are drawn at 45$^o$ and the future direction is upwards. Closed f-trapped surfaces exist in the colored region (in red, this is $\mathscr{T}$), defined by $\dot a <0$. The boundary $\B$ is the maximal hypersurface with $\dot a =0$, the re-collapsing time, shown here as a dotted horizontal line. A marginally trapped tube ---correspoding to the apparent 3-horizon AH to be defined in section \ref{sec:general}--- foliated by marginally f-trapped round spheres in also shown. The region $\mathscr{Z}$ to the future of the AH is a core of the trapped region, as defined in section \ref{sec:new}. If $\mathscr{Z}$ is removed from the spacetime, then there are no remaining closed f-trapped surfaces.}}
\label{fig:RW}
\end{figure}

As we have seen, not all the $\dot a=0$ hypersurfaces will be part of the boundary in general. An illustrative example is given by de Sitter spacetime, which has $k=1$ and $a(t)=\cosh t$, so that $\dot a$ is negative, zero, or positive for $t<0$ , $t=0$ or $t>0$, respectively. Thus, de Sitter spacetime has f-trapped round spheres in any slice with $t<0$. However, de Sitter spacetime is maximally symmetric, and therefore all of its points are fully equivalent, so that there are f-trapped round spheres through every point. Hence, $\B=\emptyset$ and $\mathscr{T} =\varietat$ in de Sitter spacetime. 

A different, more standard example, is provided by the closed Friedman model for dust and $\Lambda =0$, defined by $k=1$ with the following parametric form for $a(t)$:
$$
a(\gamma ) =C(1-\cos \gamma ), \,\,\, t(\gamma) =C (\gamma -\sin \gamma ),\,\,\,
\gamma\in (0,2\pi)
$$
where $C$ is a constant. Obviously $\dot a = \sin \gamma /(1-\cos \gamma)$. Now there are closed f-trapped round spheres through all points of any slice with $t>C\pi$ ($\gamma > \pi$), as then $\dot a <0$. However, no closed f-trapped surface can enter the region with $t<C\pi$ ($\gamma < \pi$), as otherwise they would reach their minimum value of $t$ at a slice with $\dot a>0$. Thus, in this case $\B =\B^- = \{t=C\pi\}$. 
The Penrose diagram of this situation is depicted in figure \ref{fig:RW}.

Obviously, the same holds for arbitrary models, with any value of $k$ (but keeping $\dot a^2 +k\geq 0$), such that $a(t)$ has a unique maximum value, that is, for models with one expanding phase, a unique re-collapsing time, and one contracting phase.

Observe that, in accordance with our general results, the slice $t=0$ in de Sitter spacetime has $\dot a=0$ but $\ddot a >0$, while in the previous models with a unique re-collapsing time, such as the closed Robertson-Walker dust model of Figure \ref{fig:RW}, $\ddot a <0$ everywhere.  

\section{Spherically symmetric spacetimes and the apparent 3-horizon AH}\label{sec:general}

A spacetime is spherically symmetric if it admits an SO(3) group of isometries. This group acts transitively on round spheres embedded in space-time. Let $d\Omega^2$ denote the standard metric on the unit round spheres, then the line-element can always be written as
$$
ds^2 = dL^2 + r^2d\Omega^2
$$
for some Lorentzian 2-dimensional metric $dL^2$ and where $r$ is constant on each round sphere. We assume from now on that $dr \neq 0$, hence $r$ can be chosen as one of the coordinates for the metric $dL^2$. It is called the area coordinate because the preferred round spheres have an area equal to $4\pi r^2$. We shall use the invariantly defined {\em mass function} 
$$
m\equiv r(1- \nabla_\mu r \nabla^{\mu} r)/2 .
$$
By choosing $v$ as a coordinate labeling the incoming radial null geodesics, the spherically symmetric line-element can then be written as
\be
ds^2=-e^{2\beta}\left(1-\frac{2m(v,r)}{r}\right)dv^2+2e^\beta dvdr+r^2d\Omega^2 \label{gds2}
\ee
with $\beta(v,r)$. The future is defined by increasing values of $v$. These are called the advanced Eddington-Finkelstein coordinates. They may not be globally defined in some spacetimes, but they are well adapted to our purposes: describing the cases with incoming matter and radiation. 

The future-directed radial null geodesic vector fields are given by
\be
\vec\ell = -e^{-\beta}\partial_r , \hspace{1cm} 
\vec k=\partial_v +\frac{1}{2}\left(1-\frac{2m}{r}\right)e^{\beta}\partial_r\label{ks}
\ee
so that
$$
\ell_{\mu}dx^\mu =-dv , \hspace{1cm} k_{\mu}dx^\mu =e^{\beta}dr - \frac{1}{2}\left(1-\frac{2m}{r}\right)e^{2\beta}dv \, .
$$
They satisfy $k_{\mu}\ell^\mu =-1$.

The mean curvature vector for each round sphere $\vec H_{sph}$ can be easily computed
\be
\vec H_{sph}=\frac{2}{r}\left(e^{-\beta}\partial_{v}+\left(1-\frac{2m}{r}\right)\partial_{r}\right).
\label{Hspheres}
\ee
Taking into account that for each round sphere the two future-pointing null normals are $\vec k_{sph}^+ =\vec k$ and $\vec k_{sph}^- =\vec \ell$, the null expansions can be read off from (\ref{Hspheres}):
\be
\theta_{sph}^+ =\frac{e^{\beta}}{r}\left(1-\frac{2m}{r}\right), \hspace{1cm}
\theta_{sph}^-=-\frac{2e^{-\beta}}{r} .\label{thetaspheres}
\ee
Thus, these round spheres are f-trapped if and only if $r<2m$, and they are marginally f-trapped if $r=2m$. The set defined by
$$
\mbox{AH:} \hspace{2mm} r-2m(v,r)=0
$$
is formed by hypersurfaces foliated by marginally f-trapped round spheres. Thus, each of these hypersurfaces is a ``marginally trapped tube" \cite{AG}. They are called the spherically symmetric  ``apparent 3-horizons", as each of its marginally f-trapped round spheres is an apparent horizon ---in the sense of \cite{HE}, see also \cite{KH}--- for a time slice that respects the symmetry. And it is unique with these properties:
\begin{result}\label{res:AHunique}
AH are the unique spherically symmetric hypersurfaces ---with respect to the given SO(3) group--- foliated by marginally f-trapped surfaces.
\end{result}
{\bf Remarks:} 
\begin{itemize}
\item There are Lorentzian manifolds with several SO(3) groups of isometries, such as flat, de Sitter, or Robertson-Walker spacetimes (simply change the origin of coordinates). This is why we have to fix the group defining the spherical symmetry. However, in generic situations the isometry group SO(3) is unique, and then the AH is unique in absolute sense.
\item The {\em foliation} hypothesis is crucial here, as there can be marginally trapped surfaces different from round spheres, and actually of any genus, imbedded in spherically symmetric hypersurfaces. Explicit examples are given in \cite{FHO} for Robertson-Walker spacetimes with $k=1$, where marginally trapped surfaces of any genus are imbedded in $t=$ const. slices.
\item The assumption of spherical symmetry for the hypersurface is essential here, as there are non-spherically-symmetric marginally trapped tubes in spherically symmetric spacetimes. This is a general property, as we will show in subsection \ref{subsec:pertnotiso}. 
Explicit examples for closed ($k=1$) Robertson-Walker spacetimes are given in \cite{FHO}. 
\end{itemize}

\proof
That AH is the unique spherically symmetric set constituted by marginally f-trapped round spheres is obvious by construction. Suppose then that there is a spherically symmetric hypersurface $\Sigma$ foliated by marginally f-trapped closed surfaces $\{S_{\lambda}\}$. Pick up any such surface, say $S_0$. 
By using the SO(3) group of isometries, move $S_0$ to obtain a new, necessarily marginally f-trapped, closed surface $S'_0$ in $\Sigma$. This surface must be tangent at some point $x$ to one of the foliating surfaces, say $S_{\lambda_0}$. But then, using the maximum principle as in the proof of Proposition 3.1 of \cite{AG} one can deduce that $S'_0 =S_{\lambda_0}$, and a fortiori, that $S_0=S'_0 =S_{\lambda_0}$, so that $S_0$ must be tangent to the SO(3) Killing vectors. The proof in \cite{AG} assumed that $\Sigma$ was spacelike, however this is not necessary and the reasoning works equally well for timelike $\Sigma$. Actually, it can be seen that the result holds for null $\Sigma$ as long as its null generator does not point along the null direction with vanishing expansion of $S_\lambda$. In the remaining case in which the null generator of $\Sigma$ points along the direction of vanishing null expansion of the $S_\lambda$, all possible cross-sections of $\Sigma$ have the corresponding null expansion vanishing---in particular, the round spheres in $\Sigma$ are marginally f-trapped---, and therefore the result holds true too. \fin

The exceptional situation that has arisen in the previous proof is only possible in cases where AH happens to have portions of isolated-horizon type \cite{AK1}, that is, null portions such that their null generators point along $\vec k_{sph}^+ =\vec k$. To understand when this occurs, note that
the normal one-form to AH is 
$$
n_\mu dx^\mu=\left(1-2\frac{\partial m}{\partial r}\right)dr-2\frac{\partial m}{\partial v} dv
$$
whose norm is
$$
n_\mu n^\mu |_{AH}=-\left.4e^{-\beta} \frac{\partial m}{\partial v}\left( 1-2\frac{\partial m}{\partial r}\right)\right|_{AH}
$$
so that AH is null at any point $x\in \AH$ with $\partial m/\partial v|_x=0$. Moreover, AH can be null at points where $\partial m/\partial r =1/2$, and it can also be timelike or spacelike. Therefore, AH is a dynamical horizon \cite{AK1,BI} (a spacelike marginally trapped tube) on the region where it is spacelike, and an isolated horizon \cite{AK1} on any open region where $m=m(r)$. This isolated horizon portion of AH, if non-empty, is characterized also by:
$$
G_{\mu\nu}k^\mu k^\nu |_{AH}=0 \Longleftrightarrow \mbox{isolated-horizon portion of AH} \equiv \AH^{iso}
$$
where $G_{\mu\nu}$ is the Einstein tensor of the spacetime. This will be relevant for the perturbations of AH to be studied in the next section.
On the other hand, the dynamical horizon portion of AH is {\em generic} in the sense of \cite{AG}, and therefore it is actually an outer trapping horizon in the sense of Hayward \cite{Hay}. In an open region with $m=r/2+f(v)$ the AH is null, but it is {\em not} an isolated horizon as the null normal to AH is not the null normal with vanishing expansion.

A simple illustrative example of these different possibilities is provided by the Robertson-Walker spacetimes: fixing an origin of coordinates arbitrarily, the corresponding spherically symmetric apparent 3-horizon is a marginally trapped tube \cite{AG} with the property of being spacelike, null or timelike if $(\varrho+p)(\varrho -3p+4\Lambda)$ is less, equal or greater than zero, see pp.779-780 in \cite{S} ---see also \cite{BBGV}---, where $\varrho$ and $p$ are the energy density and pressure, respectively ($\varrho +\Lambda =3(\dot a^2+k)/a^2$, $p-\Lambda=-2\ddot a/a -(\dot a^2+k)/a^2$.) A timelike AH is explicitly shown in figure \ref{fig:RW}. An example of a null AH which is not an isolated horizon is provided by the Robertson-Walker spacetime with $p=\varrho/3$ and $\Lambda =0$ \cite{FST}. These results are obviously related to the instability of MOTS in Robertson-Walker geometries proven in \cite{CM} if $\varrho +\Lambda >0$, $\varrho +p \geq 0$ and $\varrho -3p+4\Lambda \geq 0$.

Observe that AH can be empty (e.g. in flat spacetime), but this will only happen if there are no marginally f-trapped round spheres on the entire spacetime. As our aim is to study the region with closed f-trapped surfaces and its boundary, we will assume that they certainly exist. In this situation, AH cannot be empty. Under reasonable hypotheses, and for general asymptotically flat initial data sets (so that the cosmological constant $\Lambda=0$), one then knows \cite{D} that there is a regular complete future null infinity $\scri^+$ (so that close to infinity the round spheres are untrapped). The event horizon (EH) is defined as the boundary of the causal past of future null infinity: $\partial J^-(\scri^+)$ \cite{Haw,HE,Wald}, hence it is a null hypersurface by definition. In our case, it is also spherically symmetric. 

The apparent 3-horizon AH does not need to be connected, and it can have as many connected components as desired, even if $m(v,r)$ is bounded everywhere by a finite positive mass $M$. As an elementary example, take the case with $m=m(r)$, $m(r)\leq M<\infty$, so that AH is null everywhere. The connected components of AH are given by the null hypersurfaces $r=r_i$ where $r_i$ are the positive roots of the equation $r-2m(r)=0$. In general, we will only be interested in the particular connected component of AH which is related to the event horizon (EH) of an asymptotically flat end (in the example above, the one with the largest $r_i$). This particular connected component of $\AH$ will be denoted by $\AH_1$:
$$
\AH_1 \equiv \{\mbox{Portion of AH that merges with or is asymptotic to the EH}\}
$$

The region where the round spheres are untrapped will be denoted by
$$
\rr: \hspace{1cm} r-2m(v,r)>0 
$$
and we also use the notation
$$
\R=\rr\cup \AH \,\,\, (\Longleftrightarrow \R : r-2m(v,r)\geq 0).
$$
Note that $\rr$ can actually be empty. For instance, in cases where all the round spheres are f-trapped (this happens for example in the Kantowski-Sachs models \cite{Exact}). However, $\rr$ can never be empty in the asymptotically flat cases considered herein, because then the round spheres close to spacelike infinity must be untrapped. Notice, similarly, that the whole spacetime may sometimes coincide with $\R$, so that no round sphere is f-trapped, but they still can be marginally f-trapped. An example is provided by the extreme Reissner-Nordstr\"om solution which has a degenerate horizon \cite{HE,Exact}. We avoid this situation due to the assumption of the existence of f-trapped spheres.

\section{Perturbations of round spheres on $\AH$}\label{sec:notAH}
We are going to use the stability operator \cite{AMS,AMS1} to probe the possible perturbations of marginally f-trapped round spheres with the aim, in particular, of ascertaining if there can be closed f-trapped surfaces traversing AH \cite{GW}. We will also find a characterization of the AH in terms of the Einstein tensor, as well as other interesting results. Related conclusions were derived in full generality, by the same method of perturbation, in \cite{BF}. We will not restrict the causal character of AH, though, and we will also prove that the deformed surfaces can be made genuinely f-trapped while extending to both sides of $\AH\backslash \AH^{iso}$.

Choose any connected component of AH, so that this is a spherically symmetric marginally f-trapped tube: a hypersurface foliated by marginally f-trapped round spheres (defined by constant values of $r$ and $v$). As explained in the previous section, the future null normals are given by (\ref{ks}) with $\vec k_{sph}^-=\vec\ell$ and $\vec k_{sph}^+=\vec k$, so that the corresponding expansions were presented in (\ref{thetaspheres}), which restricted to AH are:
$$
\theta_{sph}^-=-e^{-\beta}\frac{2}{r} <0 , \hspace{1cm} \theta_{sph}^+=0\, .
$$
We can now perturb any such marginally f-trapped round sphere, say  $\varsigma$, along a direction $f\vec n$ defined on $\varsigma$ and orthogonal to it. Following \cite{AMS1}, $f$ is any function defined on $\varsigma$ and we describe $\vec n$ by means of the outward vector field
\be
\vec n =-\left. \vec\ell +\frac{n_{\mu}n^{\mu}}{2}\vec k  \right|_{\varsigma} \label{n}
\ee
which is normalized with respect to the fixed null directions such that
$$
k_\mu n^\mu=1.
$$
Observe that the causal character of $\vec n$ is unrestricted.

Deform the round sphere $\varsigma$ by going orthogonally to $\varsigma$ along $f\vec n$ a distance $\epsilon$. The formula for the variation of the expansion $\theta^+$ can be found in many references, e.g. in \cite{AMS,AMS1,AM,BF,Hay,Gal,CC}, but the version better adapted to our purposes is that appearing in \cite{AMS1} because it keeps the norm of $\vec n$ free. Given the marginal character of $\varsigma$ and its spherical symmetry most of the terms in the variation formula vanish and the outer expansion of the perturbed surface $\varsigma_\epsilon$ is given by
\be
\theta^+_\epsilon = \epsilon \delta_{f\vec n} \theta^+ +O(\epsilon^2),\hspace{1cm}
\delta_{f\vec n} \theta^+=-\Delta_{\varsigma}f+f\left(\frac{1}{r_{\varsigma}^2}-\left.G_{\mu\nu}k^\mu u^{\nu}\right|_\varsigma\right)
\label{deltatheta}
\ee
where $r_{\varsigma}$ is the constant value of $r$ on $\varsigma$, 
$\Delta_{\varsigma}$ is the Laplacian on $\varsigma$, and $\vec u$ is the following vector field orthogonal to $\varsigma$ and $\vec n$:
$$
\vec u =\vec\ell +\frac{n_{\mu}n^{\mu}}{2}\vec k, \hspace{5mm} u_{\mu}n^\mu=0, 
\hspace{5mm} u_{\mu}u^\mu=-n^\mu n_\mu .
$$

\begin{figure}[!ht]
\includegraphics[width=12cm]{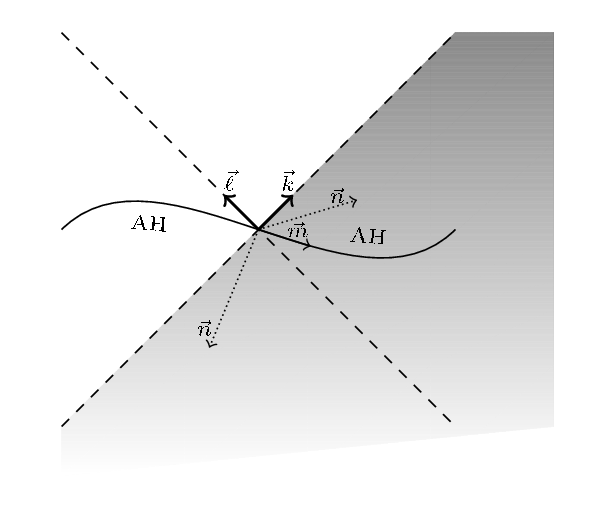}
\caption{\footnotesize{This is the scheme for the vector fields involved in the perturbation analysis. The spherically symmetric null directions are represented at $45^o$, defined by the radial null vector fields $\vec \ell$ and $ \vec k$ given in (\ref{ks}). The vertex at the centre represents a marginally f-trapped round sphere in the apparent 3-horizon AH, which is also drawn nearby. The vector $\vec m$, as defined in the main text, is tangent to AH. Thus, perturbing the initial round sphere in the direction of $c\vec m$ with constant $c$ leads to another marginally f-trapped round sphere in AH. By letting $n_{\mu}n^{\mu}\in (-\infty,\infty)$, the general vector $\vec n$ defined in (\ref{n}) always points towards the shadowed region, though its direction may depend on the point of the sphere, and it can be spacelike, null or timelike. The AH can also have any causal character.}}
\label{fig:Pert}
\end{figure}

Observe, by the way, that selecting $f=$constant (\ref{deltatheta}) informs us that the vector $\vec u$ such that $1/r_{\varsigma}^2-\left.G_{\mu\nu}k^\mu u^{\nu}\right|_\varsigma=0$ produces no variation on $\theta^+$, so that the corresponding $\vec n$ is tangent to the AH simply leading to other marginally f-trapped round spheres on AH. Let us call such a vector field $\vec m$, so that
\be
\frac{1}{r_{\varsigma}^2}-\left.G_{\mu\nu}k^\mu \ell^{\nu}\right|_\varsigma -
 \left.\frac{m_\rho m^\rho}{2}G_{\mu\nu}k^\mu k^{\nu}\right|_\varsigma =0 \label{m}
\ee
together with 
$$
 \vec m =-\left. \vec\ell +\frac{m_{\mu}m^{\mu}}{2}\vec k \right|_{\varsigma} 
$$
characterizes $AH\backslash \AH^{iso}$, since $\vec m$ is the unique spherically symmetric direction tangent to it. The exceptional isolated-horizon portion $\AH^{iso}$ has the null $\vec k$ as the tangent vector field. Observe that the only forbidden direction ---due to the normalization used--- for $\vec n$ (and $\vec m$) is that defined by $\vec k$ (which would correspond to $n_{\mu}n^{\mu}\rightarrow \pm \infty$). The situation is depicted in figure \ref{fig:Pert}. 

Whether $\vec m$ is spacelike, null or timelike (and accordingly for AH) depends on the magnitude of $G_{\mu\nu}k^\mu \ell^{\nu}|_\varsigma$ and the sign of $G_{\mu\nu}k^\mu k^{\nu}|_\varsigma$. This sign is non-negative if the null energy condition is assumed. In the next section we will actually assume the stronger dominant energy condition, so that for most of our purposes the positive sign must be kept in mind. Recall, though, that the condition $G_{\mu\nu}k^\mu k^{\nu}=0$ defines the portions of AH which are isolated horizons. We have to consider both cases separately.

\subsection{Perturbations on $\AH^{iso}$}
Assume that $G_{\mu\nu}k^\mu k^{\nu}=0$ holds on a region so that we are dealing with $\AH^{iso}$. This can be seen equivalent to the condition $\partial m/\partial v =0$. From the variation formula (\ref{deltatheta}) we deduce that
$$
\delta_{f\vec n} \theta^+=-\Delta_{\varsigma}f+f\left(\frac{1}{r_{\varsigma}^2}-\left.G_{\mu\nu}k^\mu \ell^{\nu}\right|_\varsigma\right)
$$
so that the perturbed expansion is {\em independent of the direction} of deformation $\vec n$. One can check that  $G_{\mu\nu}k^\mu \ell^{\nu}|_\varsigma =(2/r_\varsigma^2)\partial m/\partial r|_\varsigma$ and the previous relation can be rewritten as
\be
\Delta_{\varsigma}f-f\frac{1}{r_{\varsigma}^2}\left(1-2\left.\frac{\partial m}{\partial r}\right|_\varsigma\right)=- \delta_{f\vec n} \theta^+ \, .
\label{deltathetaiso}
\ee
Notice that the term between round brackets will generally be positive ---for instance if $m=$ const.---, and it will certainly be so for any part of $\AH^{iso}$ related to an asymptotically flat end, because then $r-2m(r)$ changes from negative to positive values. Eq.(\ref{deltathetaiso}) can be seen as an equation $L(f)=- \delta_{f\vec n} \theta^+$ where $L=\Delta_{\varsigma} -(1/r_\varsigma^2)(1-2\partial m/\partial r|_\varsigma)$ is an elliptic operator on $\varsigma$, and thus it is adapted for direct application of the maximum principle, see e.g. \cite{Sp} or Appendix 3 in \cite{BF}. In particular, if $\delta_{f\vec n} \theta^+$ is non-positive everywhere it follows that $f$ must be negative everywhere on $\varsigma$. 

Combining this with the known fact that arbitrary perturbations along the null generator $\vec k$ of the isolated horizon $\AH^{iso}$ produce marginally f-trapped surfaces \cite{BF} we obtain the following theorem.
\begin{theorem}\label{th:pertiso}
On any isolated-horizon portion $\AH^{iso}$ of $\AH$ arbitrary deformations of its round spheres within $\AH^{iso}$ lead to marginally f-trapped surfaces. Moreover, if $\AH^{iso}$ is such that $1\geq 2\partial m/\partial r$, any other possible perturbation leading to weakly f-trapped surfaces has $f< 0$, so that the deformed surfaces lie strictly outside the region $\R$. 
\end{theorem}


Observe that from (\ref{deltathetaiso})
$$
\left(1- 2\left.\frac{\partial m}{\partial r}\right)\right|_\varsigma \oint_{\varsigma} f =-\oint_{\varsigma}\delta_{f\vec n} \theta^+
$$
and, given that the righthand side can be chosen as small as desired, the minimum of the non-positive function $f$ can be made as small in magnitude as needed. In concrete situations, one can even use the freedom on choosing the variation vector $\vec n$ if this helps. 

\subsection{Perturbations on $\AH\backslash \AH^{iso}$}\label{subsec:pertnotiso}
Let us now consider the parts of $\AH$ with $G_{\mu\nu}k^\mu k^{\nu}> 0$. 
From figure \ref{fig:Pert} we deduce that the perturbation along $f\vec n$ will enter into the region with f-trapped round spheres (which is trivially part of $\mathscr{T}$ and can be identified because $\vec k$ always points into it) at points with
$$
f(n_{\mu}n^{\mu} - m_{\mu}m^{\mu}) >0 .
$$
For easy control of these signs we note that, according to (\ref{deltatheta}-\ref{m}), 
\be
(G_{\mu\nu}k^\mu k^{\nu}|_{\varsigma}) f(n_{\mu}n^{\mu} - m_{\mu}m^{\mu}) =-2(\Delta_{\varsigma}f+\delta_{f\vec n}\theta^+).\label{deltatheta1}
\ee
An interesting conclusion arises by integrating this equality on $\varsigma$
$$
(G_{\mu\nu}k^\mu k^{\nu}|_{\varsigma})\oint_{\varsigma} f (n_{\mu}n^{\mu} - m_{\mu}m^{\mu}) =-2\oint_{\varsigma} \delta_{f\vec n}\theta^+
$$
from where we deduce the following facts:
\begin{itemize}
\item spherically symmetric deformations, defined by having constant $f$ and $n_\mu n^\mu$, are uninteresting because they only lead to untrapped round spheres in the region $\rr$ if $f(n_{\mu}n^{\mu} - m_{\mu}m^{\mu}) <0$ and to f-trapped round spheres outside $\R$ if $f(n_{\mu}n^{\mu} - m_{\mu}m^{\mu}) >0$. 
\item the deformed surface can be f-trapped --with a negative definite sign of the variation--- only if $f(n_{\mu}n^{\mu} - m_{\mu}m^{\mu})$ is positive somewhere. Hence, a f-trapped surface (obtained in this way) must lie at least partially in the region where the round spheres are f-trapped. This will turn out to be a fully general result (Corollary \ref{cor:NOr>2m}). 
\item the deformed surface can be untrapped only if $f(n_{\mu}n^{\mu} - m_{\mu}m^{\mu})$ is somewhere negative. 
\item if the deformed surface lies entirely within $\rr$ ---so that $f(n_{\mu}n^{\mu} - m_{\mu}m^{\mu})< 0$ everywhere ---, then $\delta_{f\vec n}\theta^+$ must be positive  somewhere.
\item if the deformed surface lies entirely outside $\R$, then $\delta_{f\vec n}\theta^+$ must be negative somewhere.
\end{itemize}
Note that (\ref{deltatheta}) is also adapted for direct application of the maximum principle, as it takes the form $L(f)=-\delta_{f\vec n}\theta^+$ where now the elliptic operator $L=\Delta_{\varsigma}+(G_{\mu\nu}k^\mu k^{\nu}|_{\varsigma})(n_{\mu}n^{\mu} - m_{\mu}m^{\mu})/2$. Thus we also have
\begin{itemize}
\item All possible perturbations with $n_{\mu}n^{\mu} - m_{\mu}m^{\mu}\leq 0$ and leading to $\delta_{f\vec n}\theta^+\leq 0$ everywhere are such that $f$ is negative everywhere. Thus, all perturbed weakly f-trapped surfaces with $n_{\mu}n^{\mu} \leq m_{\mu}m^{\mu}$ are strictly outside $\rr$.
\end{itemize}

In order to construct examples of f-trapped deformed surfaces which lie partly in $\rr$ we choose to consider perturbations such that 
$$
n_{\mu}n^{\mu} - m_{\mu}m^{\mu} >0 .
$$
For this choice the deformed surface enters the region with f-trapped 
round spheres if $f>0$, and it enters $\rr$ if $f<0$. We introduce 
a constant $a_0$. We will aim for f-trapped surfaces for 
which 
$$
(G_{\mu\nu}k^\mu k^{\nu}|_{\varsigma})\, a_{0}(n_{\mu}n^{\mu}- m_{\mu}m^{\mu}) +2 \delta_{f\vec n}\theta^+ = 0 . 
$$
By our assumptions this implies that $\delta_{f\vec n}\theta^+ < 0$ if $a_{0}>0$, 
so that the deformed surface is f-trapped. 
Next we set
$$
f\equiv a_{0}+\tilde f 
$$
for some as yet undetermined function $\tilde f$. Equation (\ref{deltatheta1}) becomes 
\be
(G_{\mu\nu}k^\mu k^{\nu}|_{\varsigma})(n_{\mu}n^{\mu} - m_{\mu}m^{\mu})\tilde f + 2\Delta_{\varsigma}\tilde f = 0 . \label{deltatheta2}
\ee
We conclude that our assumptions require that  
\be
\frac{1}{2}(G_{\mu\nu}k^\mu k^{\nu}|_{\varsigma})(n_{\mu}n^{\mu} - m_{\mu}m^{\mu}) = -\frac{\Delta_{\varsigma}\tilde f}{\tilde f} >  0 . \label{finequal}
\ee
This is a (mild) restriction on the function $\tilde f$. A simple 
solution is to choose $\tilde f$ to be an eigenfunction of the Laplacian $\Delta_{\varsigma}$, say $\tilde f =\sum_{m=-l}^l c_{lm}Y_{l}^m$ for a fixed $l\in \mathbb{N}$ and constants $c_{lm}$, where 
$Y_{l}^m$ are the spherical harmonics. Then, on using $\Delta_\varsigma Y_l^m =- \frac{ l (l +1)}{r_{\varsigma}^2}Y_l^m$ the deformation direction $\vec n$ is determined by
$$
n_{\mu}n^{\mu} - m_{\mu}m^{\mu}=\frac{2}{G_{\mu\nu}k^\mu k^{\nu}|_{\varsigma}}\frac{l(l+1)}{r^2_{\varsigma} } >0
$$
and the variation of the expansion then reads
$$
\delta_{f\vec n} \theta^+=-a_0 \frac{ l (l+1)}{r_{\varsigma}^2} 
$$
which is negative (resp.\ positive) for all $a_0 >0$ (resp.\ $a_{0}<0$.) 
As the other expansion was initally negative, by choosing $\epsilon$ very small we can always achieve that 
$\theta^-_\epsilon=\theta_{sph}^- +\epsilon \delta_{f\vec n}\theta^- +O(\epsilon^2)$ is also negative and therefore the deformed $\varsigma_\epsilon$ is f-trapped (resp.\ untrapped). Throughout we 
assume that the deformation is small enough so that the we can rely 
on the first order perturbation.  

It only remains to check that $f$ realizes all signs, so that the deformed surface criss-crosses AH. Given that
$$
f=a_0 +\sum_{m=-l}^l c_{lm}Y_{l}^m
$$
it is enough to adjust the constants $c_{lm}$ to achieve this goal. For instance, the choice $c_{lm}=0$ for $m\neq 0$ and $c_{l0}<-a_0$ if $a_{0}>0$ (or $c_{l0}>-a_{0}$ if $a_{0}<0$) will do, so that $f$ has the sign of $a_{0}$ at the region where $Y_{l}^0=P_{l} \leq 0$, and the opposite sign around the north pole of $\varsigma$ where $P_l>0$ ($P_l$ are the Legendre polynomials).\footnote{The idea behind the argument 
just given was  communicated to us by R.M. Wald, who informed us that it arose in conversations with G. Galloway \cite{GW}.}

Thus, we have proven the following theorem.
\begin{theorem}\label{th:AHnotB}
In arbitrary spherically symmetric spacetimes there are closed f-trapped, as well as untrapped, surfaces (topological spheres) penetrating both sides of the apparent 3-horizon AH at any region where $G_{\mu\nu}k^\mu k^{\nu}|_{AH} >0$.

Therefore, any part of AH which is not an isolated horizon belongs to the f-trapped region $\mathscr{T}$, so that these parts of AH never belong to the boundary $\B$.
\end{theorem}

We remark that the previous reasoning is independent of the causal character of AH, which can be spacelike, null or timelike. The only restriction is that $G_{\mu\nu}k^\mu k^\nu |_\varsigma >0$. Observe also that the original round sphere has a positive Gaussian curvature, and thus the deformed f-trapped surfaces penetrating both sides of AH will also have, for sufficiently small $\epsilon$, positive Gaussian curvature. This disproves a conjecture by Hayward \cite{Hay3}. Actually, explicit examples of the same kind but going far away from AH were presented in \cite{ABS}. 

We can now address the non-uniqueness of dynamical horizons. The perturbation argument tells us that there are f-trapped surfaces penetrating into both sides of $\AH\backslash \AH^{iso}$. We also know that there are untrapped round spheres lying just outside it. 
If $\AH$ is spacelike this means that we can find a spacelike hypersurface having such 
an outer trapped sphere as its inner boundary and an untrapped round sphere as its outer 
boundary, and such that it contains a path connecting the boundaries and lying entirely outside $\AH$ (that is, inside $\rr$). There is a theorem that ensures that such a spacelike hypersurface 
necessarily contains a marginally (outer) trapped surface \cite{AM}. By construction 
it has a part lying inside $\rr$, and we know that it must penetrate outside $\R$. 
Moreover, generically such a surface `evolves' into a marginally outer trapped tube 
\cite{AMS}. As long as we stay sufficiently close to $\AH$ all the marginally 
outer trapped surfaces in the argument will be inner trapped as well. Thus we 
have obtained:

\begin{coro}\label{cor:lars}
In arbitrary spherically symmetric spacetimes there are marginally trapped 
tubes penetrating both sides of the apparent 3-horizon $AH$ at any region where $G_{\mu\nu}k^\mu k^{\nu}|_{AH} >0$.
\end{coro}

Explicit examples in Robertson-Walker spacetimes can be found in \cite{FHO}, and in the Vaidya spacetime in \cite{NJKS}.

As a final question, we wonder how small the fraction of the closed f-trapped surface that extends outside $\R$ can be made. This will be relevant in section \ref{sec:new}, when we will ask the question of whether or not the complement of $\R$ is the optimal set to be removed from spacetime in order to get rid of all closed f-trapped surfaces. With the assumptions used in the proof of Theorem \ref{th:AHnotB} we see that 
this means that we must produce a $C^2$ 
function $\tilde f$ defined on the sphere and obeying the inequality 
(\ref{finequal}), and which is positive only in a region that we can 
make arbitrarily small. If we choose a sufficiently small constant 
$a_0$ the last requirement implies that the region where 
the surface extends outside $\R$ can be made arbitrarily small. To find 
such a function it is convenient to introduce stereographic coordinates 
$\{\rho, \phi\}$ on the sphere, so that the Laplacian takes the form
$$
\Delta_{\varsigma} = \Omega^{-1} \left( \partial_\rho^2 + 
\frac{1}{\rho}\partial_\rho + \frac{1}{\rho^2}\partial_\phi^2 \right) \ , 
\hspace{8mm} \Omega = \frac{4r_{\varsigma}^2}
{(1+\rho^2)^2} .
$$
A solution to the problem as stated is 
\be
\tilde f (\rho ) = \left\{ \begin{array}{lll} c_1
\left( e^{\frac{1}{2a}(2a-\rho^2)} - 1\right) & & \rho^2 < 4a \\ 
\\ \frac{8c_1a}{e}\frac{1}{\rho^2} -c_1(1+e^{-1}) & & \rho^2 > 4a \ . 
\end{array} \right. 
\ee   
This function is $C^2$ (and can be further smoothed if necessary), and it 
is positive only if $\rho^2 < 2a$, that is on a disk surrounding the origin 
whose size can be chosen at will. The function obeys 
$$
- \frac{\Delta_{\varsigma} \tilde f}{\tilde f} = \left\{ \begin{array}{lll} 
\frac{\Omega^{-1}}{a^2}\frac{2a-\rho^2}
{1-e^{-\frac{1}{2a}(2a-\rho^2)}} & & \rho^2 < 4a \\ 
\\ \frac{32a\Omega^{-1}}{\rho^4}\frac{\rho^2}{(e+1)\rho^2-8a} \ , & & \rho^2 > 4a \ . 
\end{array} \right. 
$$ 
This is always larger than zero.  
Thus we have proven the following important result.
\begin{theorem}\label{th:tiny}
In spherically symmetric spacetimes, there are closed f-trapped surfaces (topological spheres) penetrating both sides of the apparent 3-horizon $\AH\backslash\AH^{iso}$ with arbitrarily small portions outside the region $\R$.
\end{theorem}

\section{General imploding, asymptotically flat, spherically symmetric spacetime}\label{sec:imploding}
In this section we present the restrictions on the mass function in order to describe the case of inflow of matter and radiation, satisfying the dominant energy condition, entering into an initially flat spacetime and leading to the formation of a black hole.

If the Einstein field equations hold (with vanishing cosmological constant), the dominant energy condition \cite{HE} requires, among other restrictions, that the following inequalities hold, e.g. \cite{FST2,D}:
\bea
\frac{\partial \beta}{\partial r}\geq 0, \label{dec1}\\
 \frac{\partial m}{\partial r} \geq \frac{r}{2}\left(1-\frac{2m}{r}\right) \frac{\partial \beta}{\partial r}, \label{dec2}\\ 
\frac{\partial m}{\partial v}\geq -e^{\beta}\frac{\partial \beta}{\partial r}\frac{r}{4}\left(1-\frac{2m}{r}\right)^2\label{dec3} .
\eea
From (\ref{dec1}-\ref{dec2}) one can deduce (see, e.g., \cite{FST2})
$$
\frac{\partial (me^{\beta})}{\partial r}\geq \frac{r}{2}\frac{\partial e^\beta}{\partial r}\geq 0
$$
which implies that, at any null hypersurface $v=v_c=$const., the mass function satisfies
$$
e^{\beta(v_c,r)}m(v_c,r)\geq e^{\beta(v_c,r_0)}m(v_c,r_0) \hspace{3mm} \forall r\geq r_0
$$
so that, if the mass function happens to be positive at any round sphere $(v_c,r_0)$, then it is positive for all round spheres $(v_c,r)$ with $r\geq r_0$. In particular, if the mass function is non-negative at $r=0$, then it is non-negative everywhere. Note also that, using (\ref{dec1}), (\ref{dec2}) implies that
$$
\frac{\partial m}{\partial r}\geq 0 \,\,\,\, \mbox{on} \,\,\,  \R .
$$
Similarly, from (\ref{dec1}---\ref{dec3}) we deduce
$$
\frac{\partial m}{\partial v}+\frac{1}{2}\left(1-\frac{2m}{r}\right)e^{\beta}\frac{\partial m}{\partial r}\geq0
$$
at the same region $\R$. These last two expressions can be physically reinterpreted if we note that they are equivalent to
\be
\vec \ell (m)\leq 0, \hspace{1cm} \vec k(m)\geq 0 \hspace{1cm} \mbox{on} \,\, \R
\label{dec4}
\ee
where the null vector fields $\vec\ell$ and $\vec k$ are given in (\ref{ks}).

In other words, the dominant energy condition implies that the mass function must be non-increasing (respectively, non-decreasing) along any future-pointing ingoing (resp.\ outgoing) radial null geodesic. Observe also that the mass function must be non-decreasing along any spacelike outward direction on $\R$ (see e.g. \cite{Hay2}), as follows from
$$
a\vec k (m) -b\vec\ell (m)\geq 0, \hspace{1cm} \forall a,b >0 \hspace{1cm} \mbox{on} \,\, \R .
$$
Yet another implication of the above conditions is that the hypersurfaces $m(v,r)=$ const. are non-spacelike everywhere on $\R$.

Observe also that, on AH, the dominant energy condition (\ref{dec3}) implies 
\be
\left.\frac{\partial m}{\partial v}\right|_{AH}\geq 0 \, .\label{dotmAH}
\ee

Only continuous piecewise differentiable mass functions will be considered, so that distributional singularities on the curvature tensor ---such as shells of matter or radiation---are avoided. We will restrict ourselves to mass functions bounded by a finite least upper bound $M>0$, so that $m(v,r)\leq M$ for all $v,r$, and \cite{D} there is a regular complete future null infinity $\scri^+$ for an asymptotically flat end. We shall also restrict ourselves to the physical case where the mass-energy starts flowing in from past infinity at a given advanced time, so that previous to that instant the spacetime has no mass-energy and is flat. The value of $v$ when the mass inflow starts will be chosen as $v=0$. Then, the mass function satisfies
\be
m(v,r)=0 \hspace{3mm} \forall v<0; \hspace{1cm} \forall v>0, \hspace{3mm} 
0< m(v,r)\leq M <\infty 
\label{massg}
\ee
We will not assume in general, however, that the energy-mass travels at the speed of light, so that the infalling mass can be composed of massive dust particles or more general matter. Therefore, the hypersurface $\sigma$ separating the flat portion and the rest of the spacetime can be timelike or null. The mass function can actually reach the value $M$ or not. In the former case, given that we are assuming that there is a regular future null infinity $\scri^+$ for an asymptotically flat end, there exists a value $v_1$ of $v$ such that $m(v,r)=M$ for all $v>v_1$. This implies that $m(v_1,r)=M$ for all $r$. Note that charged cases (such as those with an asymptotic, and static, Reissner-Nordstr\"om region) are included in the other case characterized by $m<M$ everywhere. 

\begin{figure}[!ht]
\includegraphics[width=13.5cm]{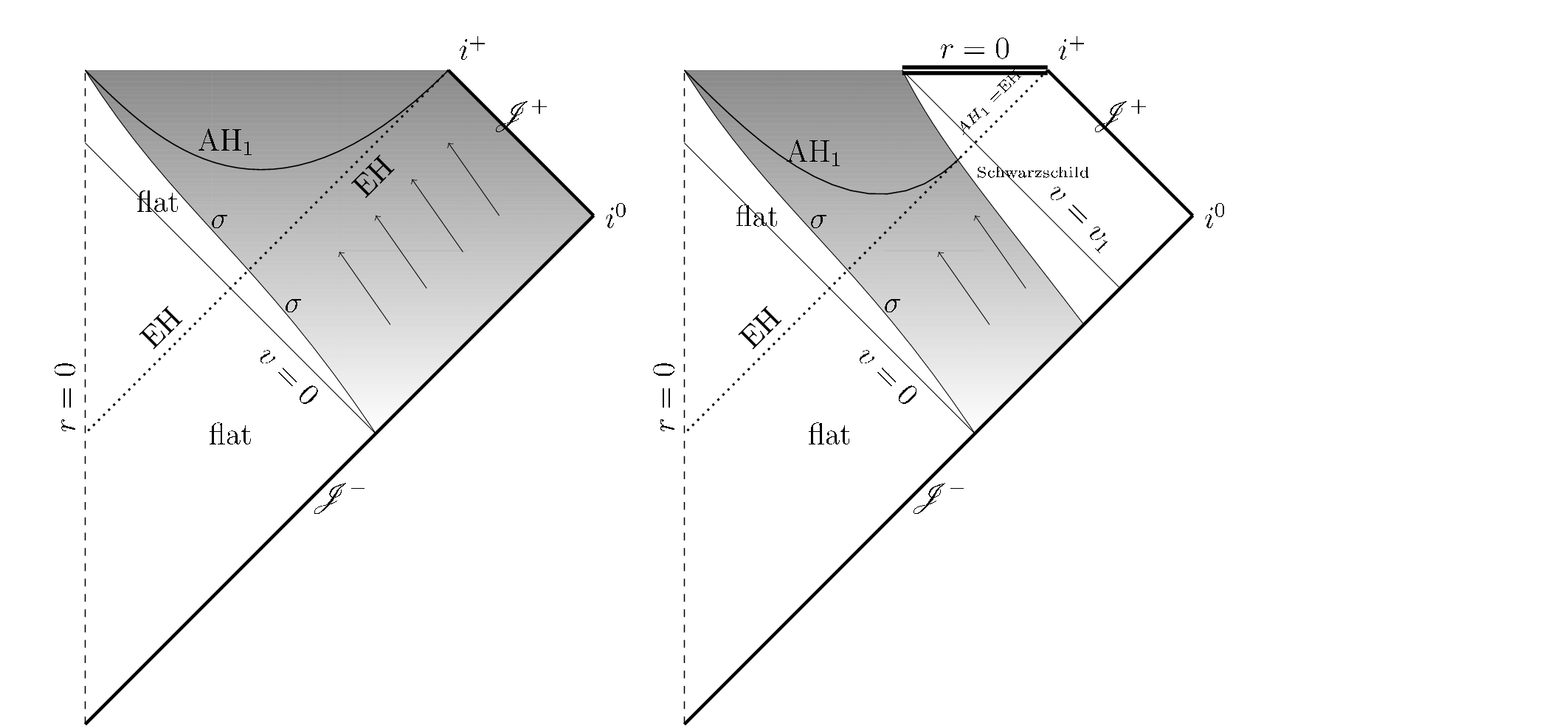}
\caption{\footnotesize{These are conformal diagrams of (\ref{gds2}) with (\ref{dotm}) and (\ref{massg}) when (i) $m(v,r)<M$ everywhere (left) and (ii) $m(v,r)=M$ in some open asymptotic region (right). As usual, null radial lines are drawn at 45$^o$ and the future direction is upwards. The discontinuous line marked as $r=0$ is the origin of coordinates. The spacetime is flat until energy starts falling to the centre in a spherical manner from past infinity at $v=0$. The hypersurface $\sigma$ separating the flat region from the rest is non-spacelike everywhere, so that material particles travel causally. (The particular case with a null $\sigma$ is depicted in figure \ref{fig:vaidya}). Thus, the shaded regions are non-flat spherically symmetric spacetimes with non-vanishing energy-momentum. In the figure on the right, the spacetime becomes Schwarzschild with mass $M$ at some other non-spacelike hypersurface (and therefore also for all $v>v_1$, where $v_1$ is the supremum of $v$ on that hypersurface). The connected component of the apparent 3-horizon $\AH_1$ approaches the event horizon EH either asymptotically (left) or at some finite value of $v\leq v_1$ and $r=2M$ (right). In the latter case, $\AH_1$ and EH merge and remain together for $v\geq v_1$ and $r=2M$. In both cases, EH starts developing in the flat region. In the two cases, $\AH_1$ is spacelike (ergo a dynamical or future outer trapping horizon) when approaching the EH, but this is not necessarily so in other regions. These possibilities are depicted in figure \ref{fig:General2}. Whether or not the initial collapsing shell $\sigma$ leads to the formation of a singularity when focusing at the centre $r=0$ depends on the particular properties of the mass function $m(v,r)$, see the main text. Thus, we have left open the future evolution of the spacetime for the shaded regions.}}
\label{fig:General1}
\end{figure}

\begin{figure}[!ht]
\includegraphics[width=16.5cm]{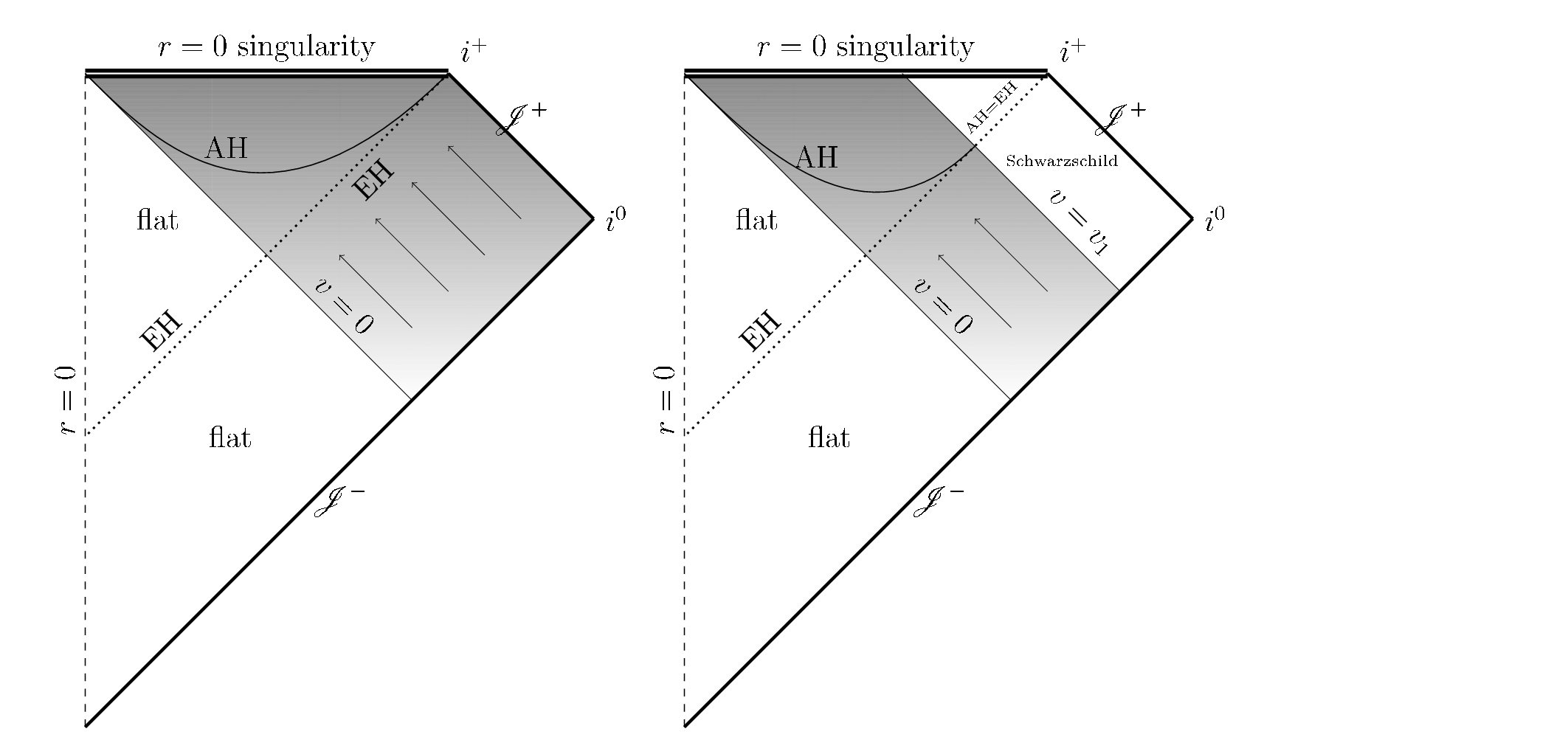}
\caption{\footnotesize{These are conformal diagrams similar to those of figure \ref{fig:General1} but with the energy flowing in along null incoming radial hypersurfaces, so that $\sigma$ is now the null hypersurface $v=0$. A particular case of this situation is given by the Vaidya imploding spacetime, analyzed thoroughly in the Appendix. As argued there, the generic situation now leads to the formation of a singularity in the future with $r=0$, as shown, and therefore there is a unique connected component of AH. The case on the left has $m(v,r)<M$ everywhere and that on the right has $m(v,r)=M$ for all $v>v_1$. In the picture on the left, the apparent 3-horizon AH approaches the event horizon EH asymptotically, while in that  on the right AH and EH merge together at $v=v_1$ and $r=2M$. In both cases, EH starts developing in the flat region. Note that the part of flat spacetime that lies inside the event horizon is the intersection of the interiors of two light cones; it is shown without conformal distortion.}}
\label{fig:vaidya}
\end{figure}

\begin{figure}[!ht]
\includegraphics[width=16.5cm]{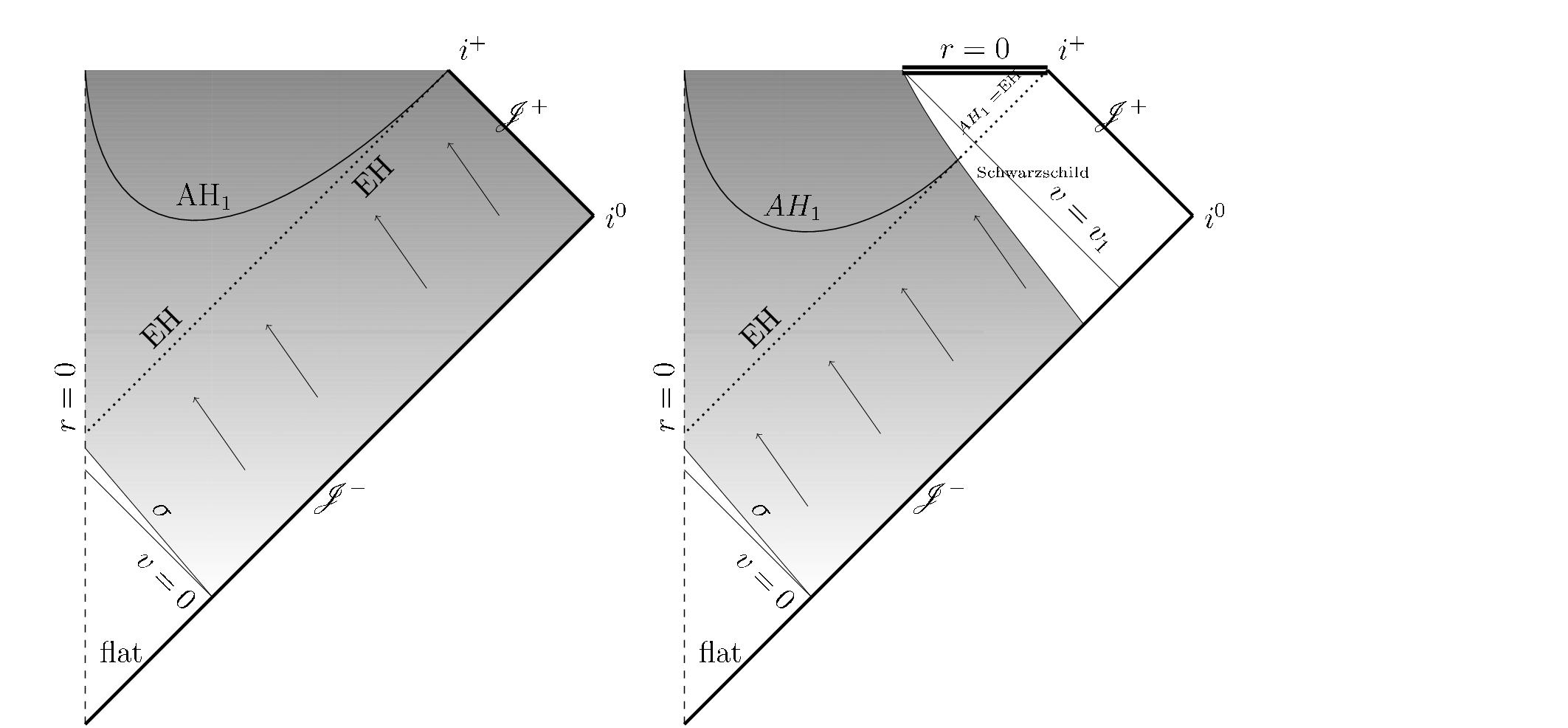}
\caption{\footnotesize{Once again, these are conformal diagrams of (\ref{gds2}) with (\ref{dotm}) and (\ref{massg}) when (i) $m(v,r)<M$ everywhere (left) and (ii) $m(v,r)=M$ in some open asymptotic region (right). All features are essentially the same as in figure \ref{fig:General1} except that EH does not intersect the flat region and that the connected component $\AH_1$ of the apparent 3-horizon may be timelike in some region (in the figures, close to the left upper corner). An explicit example of the latter behaviour is provided by the matching of a $p=\gamma \varrho$ Robertson-Walker metric with $\Lambda =0$ and $\gamma  < 1/3$ to the outwardly radiating Vaidya spacetime, see \cite{FST} (time reversal of figure 7 therein). See also \cite{BBGV}.}}
\label{fig:General2}
\end{figure}

Whether or not the spacetime becomes singular when the incoming matter reaches the centre depends on the particular properties of the falling matter and energy. The intersection of $\sigma$ with $r=0$ will {\em not} be a curvature singularity ---so that there will be a regular centre $r=0$ at a portion of the non-flat region--- if $m$ and $\beta$ satisfy there (e.g. \cite{FST2}):
\be
\frac{\partial \beta}{\partial r} (v,0)=0; \hspace{5mm} m(v,0)=\frac{\partial m}{\partial r} (v,0)=\frac{\partial^2 m}{\partial r^2} (v,0)=0\, .\label{regular}
\ee
In this case, some later singularities can develop. If (\ref{regular}) do not hold, then the singularity appears already at $\sigma(r\rightarrow 0)$. At this general level, and in any of the previous cases, one cannot know if the singularity will be spacelike, timelike, or null. Thus, we will not prejudge this, and leave the future evolution of the spacetime open, not showing it in some of the Penrose diagrams. These are depicted, for the essentially different possibilities of interest herein, in figures \ref{fig:General1}, \ref{fig:vaidya} and \ref{fig:General2}.

Summarizing, we consider spacetimes with line-element (\ref{gds2}) satisfying the dominant energy condition and subject to (\ref{massg}) so that there is an asymptotically flat end with regular $\scri^+$ and a non-degenerate EH, and such that $\AH_1$ is the connected component of $\AH$ associated to this EH. The actual position of EH depends on the particular properties of the mass function $m(v,r)$. Generically, $\AH_1$ separates the region $\R_1$, defined as the connected subset of $\rr$ which contains the flat region of the spacetime, from a region containing f-trapped round spheres. Under these assumptions, $\AH_1$ will eventually be spacelike (actually achronal) and asymptotic (probably merging) to the EH, see \cite{Wil}. (The recent counterexamples presented in \cite{Wil2} violate some of our assumptions.)
Thus, $\AH_1$ has a portion that is a spherically symmetric, regular, dynamical horizon. Nevertheless, in general $\AH_1$ can have timelike and null portions, see e.g. \cite{FST,S,BBGV}. This has been represented in the Penrose diagrams of figure \ref{fig:General2}.

Apart from the above, we will need a further assumption, given by
\be
\frac{\partial m}{\partial v}\geq 0 \hspace{5mm} \mbox{on}\hspace{5mm}  (\R_1\cup \AH_1)\cap J^+(EH) \label{dotm} 
\ee
To justify this assumption, and to understand its reasonability, let us make the following considerations.
The region $(\R_1\cup \AH_1)\cap J^+(EH)$ may contain a portion of the flat region, where $m(v,r)=0$ so that the assumption is trivial there, and it is bounded to the future by $\AH_1$, where (\ref{dotm})  holds due to (\ref{dotmAH}). The rest of its boundary is given by a portion of EH, whose null generators are given by $\vec k$. Observe also that $\vec k|_{AH_1}=\partial_v|_{AH_1}$. 

If there is part of the origin $r=0$ in  $(\R_1\cup \AH_1)\cap J^+(EH)$, then from (\ref{dec3}) we have $\partial m/\partial v|_{r=0}\geq 0$. And if there is a flat portion in $(\R_1\cup \AH_1)\cap J^+(EH)$, this is separated from the rest by the spherically symmetric hypersurface $\sigma$, where $m|_{\sigma}=0$ by continuity. Let us denote by 
$$
\vec w =\frac{\partial m}{\partial r}\, \partial_v -\frac{\partial m}{\partial v}\,  \partial_r
$$
the vector field tangent to the hypersurfaces $m(v,r)=$const. and orthogonal to the round spheres. Thus, we have
$$
\vec w(m)=0.
$$
Note that $\vec w$ is future-pointing on $\R$. 
The hypersurface $\sigma$ is imbedded in flat spacetime, so that the mass-energy is flowing in only if $\vec w(r)|_{\sigma}< 0$. 
But this implies that
$$\left.\frac{\partial m}{\partial v}\right|_{\sigma}>0 . $$

All in all, the dominant energy condition always ensures that there is a region of $(\R_1\cup \AH_1)\cap J^+(EH)$ where (\ref{dotm}) is automatically satisfied, and this includes its future boundary and its frontier with the flat region if non-empty. 
In summary, the assumption (\ref{dotm}) is equivalent to assuming that the mass function is non-decreasing to the future along any hypersurface $r=$const.\ on $(\R_1\cup \AH_1)\cap J^+(EH)$.

\section{The past barrier $\Sigma$}\label{sec:Kodama}
In this section, we are going to identify a past barrier for f-trapped surfaces in the spacetimes considered in the previous section. This barrier severely restricts the possible locations of marginally trapped tubes and dynamical horizons.

Consider the vector field
$$
\xiv=e^{-\beta} \partial_v
$$
which characterizes the spherically symmetric directions tangent to the hypersurfaces $r=$const. These hypersurfaces are timelike everywhere on $\rr$, and null at AH, while they are spacelike outside $\R$. $\xiv$ is hypersurface orthogonal, with the level function $\tau$ defined by
\be
\xi_{\mu}dx^\mu =-Fd\tau = dr-e^{\beta} \left(1-\frac{2m(v,r)}{r}\right) dv \label{gtau} .
\ee
The hypersurfaces $\tau =$const.\ are orthogonal to the hypersurfaces $r=$const.\ everywhere. Put in another way, the expansion of the round spheres along $\xiv$ vanishes, that is, the mean curvature vector defined in (\ref{Hspheres}) is such that
$$
\xi_\mu H^\mu_{sph}=0,
$$
hence $\xiv$ provides the invariantly defined direction in which the area of the round spheres remains constant.

We note that $\xiv$ is the Kodama vector field \cite{Ko}, which has been recently used in related investigations \cite{Racz,Tung,AV}. We have
$$
\xi_\mu\xi^\mu=-\left(1-\frac{2m(v,r)}{r}\right), \hspace{1cm} \ell_\mu\xi^\mu=-e^{-\beta}
$$
so that $\xiv$ is future-pointing timelike on the region $\rr$, future-pointing null at $\AH$, and spacelike outside $\R$. Therefore, $\tau$ can considered as a time function ---``the Kodama time" \cite{ABS,AV}--- in the whole region $\rr$.
Observe that
$$
\xiv =e^{-\beta}\vec k +\frac{e^{\beta}}{2}\left(1-\frac{2m(v,r)}{r}\right)\vec\ell
\ . $$

The deformation of the metric along $\xiv$ can be easily computed 
\be
(\lie_{\xiv} g)_{\mu\nu} =e^{\beta}\frac{2}{r}\frac{\partial m}{\partial v}\ell_\mu \ell_\nu -\frac{\partial\beta}{\partial r}\left(\delta_{\mu}^r \xi_{\nu}+\delta_\nu^r \xi_\mu \right) . \label{deforxi}
\ee
Thus, $\xiv$ is a Kerr-Schild vector field \cite{CHS} on any open region with $\partial\beta/\partial r=0$, an example is the Vaidya spacetime treated in the Appendix. $\xiv$ is Killing vector in the situations with $\partial m/\partial v=\partial\beta/\partial r=0$, such as the case of the Schwarzschild spacetime.

We can use the Kodama time to restrict the location of f-trapped surfaces.
\begin{theorem}\label{th:no-ming}
Assume that the spacetime (\ref{gds2}) satisfies (\ref{dotm}). Then, no f-trapped surface $S$ can have a local minimum of $\tau$ on $\rr$, nor they can have an open portion with $\tau=$ const. there.
\end{theorem}
\proof Let $q\in S\cap \R$ be a point where $S$ has a local minimum of $\tau$, or belonging to an open portion of $S\cap \{\tau=\tau_0\}$ for some constant $\tau_0$. Due to  Theorems \ref{th:no-min} and \ref{th:untrapped}, and since $\xiv$ is future-pointing on $\R$, it is enough to show that on any such point
$$\left.P^{\mu\nu}(\lie_{\xiv} g|_{S})_{\mu\nu} \right|_q\geq 0.$$ Projecting (\ref{deforxi}) onto $S$ we get
$$
(\lie_{\xiv} g|_{S})_{\mu\nu}\, e^{\mu}_Ae^{\nu}_B=\left.e^{\beta}\frac{2}{r}\frac{\partial m}{\partial v}\right|_S \bar\ell_A \bar\ell_B -\left.\frac{\partial\beta}{\partial r}\right|_S \left(\frac{\partial\bar{r}}{\partial \lambda_{A}}\bar\xi_{B}+\frac{\partial\bar{r}}{\partial \lambda_{B}}\bar\xi_A \right).
$$
In particular, given that $\bar\xi_A|_q=0$ we obtain
\be
\left.P^{\mu\nu}(\lie_{\xiv} g|_{S})_{\mu\nu} \right|_q=\left.e^{\beta}\frac{2}{r}\frac{\partial m}{\partial v} \bar\ell_A \bar\ell^A \right|_q \geq 0 \label{poscon}
\ee
whose non-negativity follows from (\ref{dotm}).\fin

By taking into account the third remark to Theorem \ref{th:no-min}, the same conclusion holds for weakly f-trapped surfaces unless the exceptional situation (\ref{nueva}) occurs.
\begin{coro}\label{cor:NOr>2m}
If the spacetime (\ref{gds2}) satisfies (\ref{dotm}), then no {\em closed} f-trapped surface can be contained in any connected component of $\R$. In particular, no {\em closed} f-trapped surface can be fully contained in the region $\R_{1}\cup\AH_{1}$.

The only weakly f-trapped surfaces contained in $\R$ are the marginally f-trapped surfaces foliating AH.
\end{coro}
Combining this corollary with the standard result \cite{HE,Wald} ---see \cite{Cl} for a rigourous derivation--- that no closed weakly f-trapped surface can penetrate outside the EH (actually, no {\em outer} f-trapped closed surface penetrates into this region \cite{HE,Wald}), we arrive at the following conclusion.
\begin{coro}\label{coro:NOD-}
Letting aside the marginally f-trapped surfaces in $\AH_1$, no closed weakly f-trapped surface can be fully contained in the region $(\R_{1}\cup\AH_{1})\cap J^+(EH)$, and thus they must penetrate outside $\R_{1}\cup\AH_{1}$.\fin
\end{coro}
Observe that, in the cases when $\AH_{1}$ is spacelike so that it is a dynamical horizon, and recalling that then EH is its past Cauchy horizon \cite{HE,Wald}, EH=$H^-(\AH_{1})$, Corollary \ref{coro:NOD-} can be rephrased as ``no closed weakly f-trapped surface can be fully contained in the past Cauchy development $D^-(\AH_{1})$'', which is in agreement with theorem 4.1 in \cite{AG}. 

Suppose that $r<2m(v,r)$ somewhere to one side of $\AH_{1}$. We are naturally led to the question of what is the extension of the connected f-trapped region $\mathscr{T}_{1}$ containing the f-trapped round spheres to that side of $\AH_{1}$. Equivalently, the question is what is the exact location of the connected component $\B_{1}$ of the boundary $\B$ which is to the past of $\AH_{1}$. At first, one is tempted to think that $\AH_{1}$ could actually be this boundary $\B_{1}$, so that all f-trapped closed surfaces remain outside of $\R_1\cup \AH_1$, but we already know that this is not the case in general, as follows from Theorem \ref{th:AHnotB}. Actually, f-trapped surfaces with spherical topology were explicitly exhibited in \cite{BS} for the self-similar Vaidya spacetime (see Appendix) such that they penetrate $\R_1$ and even extend to the flat region of the spacetime. The example from \cite{BS} is shown in figure \ref{fig:Improvement} below, other examples were constructed
  in \cite{ABS}. Thus, the connected f-trapped region $\mathscr{T}_1$ will enter into $\R_1\cap J^+(EH)$. One can then wonder if $\mathscr{T}_1$ will actually extend all the way down to EH. This was shown to be impossible in a particular Vaidya solution with a shell of null dust in \cite{BD}. We are going to prove in the following that this is a general property by identifying a past barrier to the connected f-trapped region $\mathscr{T}_1$.

Put by definition
$$
\Sigma\equiv \{\tau =\tau_\Sigma\} , \hspace{1cm} 
\tau_\Sigma\equiv \inf_{x\in AH_1} \tau |_x \, .
$$
Observe that $\tau_\Sigma$ is the least upper bound of $\tau$ on the event horizon EH. In other words, $\tau_\Sigma$ is either (i) the constant value of $\tau$ which defines the portion of the EH in the Schwarzschild region of the spacetime in the case when $m(v,r)=M$ for all $v>v_1$, or (ii) the common limit of $\tau$ on both AH and EH when $v\rightarrow \infty$ in the case that $m(v,r)<M$ everywhere. $\Sigma$ can be completely characterized as the {\em last} hypersurface orthogonal to $\xiv$ which is non-timelike everywhere. 

\begin{theorem}\label{th:aboveBg}
Assume that the spacetime (\ref{gds2}) satisfies the dominant energy condition, (\ref{massg}) and (\ref{dotm}). Then, no closed f-trapped surface can penetrate into $I^-(\Sigma)$ (i.e.\ the region with $\tau<\tau_\Sigma$.)
\end{theorem}

\begin{figure}[!ht]
\includegraphics[width=14.5cm]{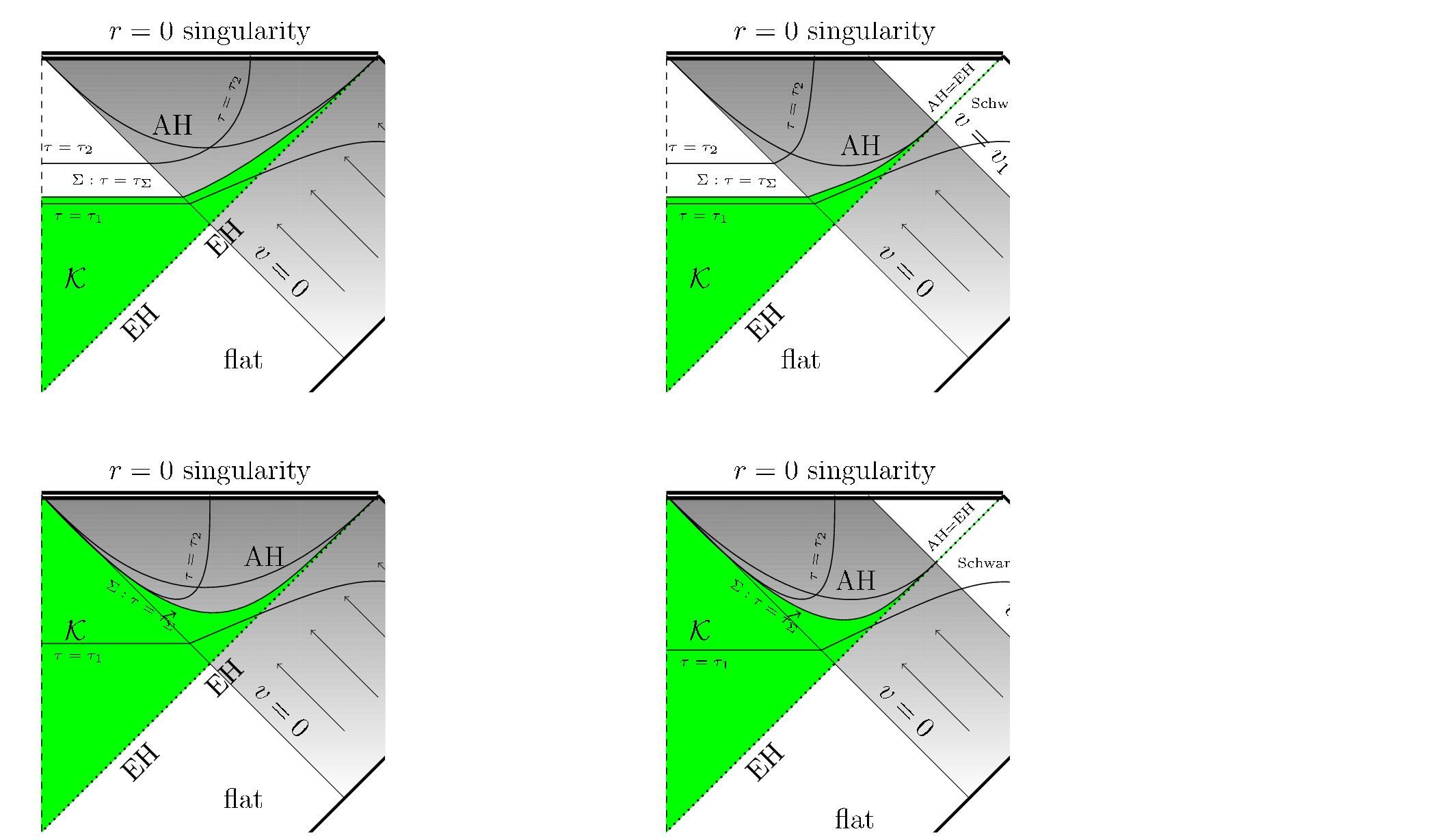}
\caption{\footnotesize{These are enlargements of relevant regions in figure \ref{fig:vaidya}, similar drawings could be performed for the cases of figures \ref{fig:General1} and \ref{fig:General2}.  On the left we picture the cases with $m(v,r)<M$ everywhere, and on the right with $m(v,r)=M$ for all $v>v_1$. The two pictures on top describe the cases when $\Sigma$ penetrates the flat region, for instance if  $8\dot m_{0}> (1-2m'_{0})^2$, see subsection \ref{subsec:SigmaLocation}, while the two bottom pictures depict the cases where $\Sigma$ never penetrates the flat region. Pertinent $\tau=$const. hypersurfaces are shown for all cases. Observe that these are spacelike everywhere (and approaching spacelike infinity $i^0$) if they have $\tau<\tau_{\Sigma}$ (example shown as $\tau=\tau_{1}$), while they are partly spacelike and partly timelike, becoming null at AH, if $\tau>\tau_{\Sigma}$ (example shown as $\tau =\tau_{2}$.) The hypersurface $\Sigma$ separating these two cases is non-timelike everywhere, and spacelike where it differs from EH: thus, it is spacelike everywhere for the left drawings, while it becomes null and identical with EH for all $v>v_1$ for the cases on the right. The closed set ${\cal K}$ used in the proof of Theorem \ref{th:aboveBg} limits to the future with $\Sigma$ and to the past with EH, and is shown in green in all cases.}}
\label{fig:SigmainVaidya}
\end{figure}

\proof Consider the closed set 
$${\cal K}\equiv \overline{D^-(\tau=\tau_\Sigma)}=\overline{J^-(\tau =\tau_\Sigma)\cap J^+(EH)}$$
This set is contained in the region $\R_{1}\subset \R$ where $\xiv$ is  future pointing. ${\cal K}$ is bounded to the future by $\tau=\tau_\Sigma$ and to the past by EH$\backslash \AH$ (see figure \ref{fig:SigmainVaidya}). Therefore, any compact surface $S$ such that $S\, \cap$ int${\cal K}\neq \emptyset$ will reach a minimum on ${\cal K}$. This minimum cannot be on $\tau=\tau_\Sigma$, because this is the maximum value of $\tau$ on ${\cal K}$. Thus it will have to be either a non-local minimum on EH$\backslash \AH$ or a local one attained on ${\cal K}\cap \R_{1}$. However these two possibilities forbid that $S$ be f-trapped, because the latter would contradict Theorem \ref{th:no-ming}, while the former would contradict the standard result \cite{HE,Wald,Cl,BD} that closed f-trapped surfaces never touch EH. \fin

Thus, the hypersurface $\Sigma$ is a limit, to the past, for f-trapped closed surfaces. In fact, they cannot even touch $\Sigma$.
\begin{theorem}\label{th:aboveB}
Under the assumptions of the previous theorem all  closed f-trapped surfaces must be contained in the region $I^+(\Sigma)$ (defined by $\tau >\tau_{\S}$) and penetrate outside $\R_{1}\cup \AH_{1}$.
\end{theorem}
\proof The last part of the theorem states that all f-trapped closed surfaces have points outside $\R$, that is points with $r<2m(v,r)$, but this is already known from Corollary \ref{coro:NOD-}. To prove the first part, observe that Theorem \ref{th:aboveBg} ensures that $\tau\geq \tau_\Sigma$ on any such f-trapped closed surface $S$. Thus, we only need to show that $\tau$ cannot reach the value $\tau_\Sigma$ on the surface, so that $S$ can never touch $\Sigma$. But  $S$ cannot touch the portion of $\Sigma$ which coincides with EH (if any), so this could only happen on the part of $\Sigma$ within $\R_{1}$, that is, with $r>2m(v,r)$. But if there were a point $x\in S\cap\R_{1}\cap\Sigma$, then Theorem \ref{th:no-min} would imply that $\tau |_S$ cannot have a local minimum on $x$, and that there cannot be any 1-dimensional line $L\subset S$ with $L\ni x$ such that $\tau |_L=\tau_\Sigma$; Theorem \ref{th:untrapped} would imply that no 2-dimensional piece of $S$ can be entirely contained in $\Sigma$. In summary, the existence of $x$ would lead to the existence of points on $S$ with $\tau<\tau_\Sigma$, in contradiction.\fin

\subsection{The location of the past barrier $\Sigma$}\label{subsec:SigmaLocation}
As we have shown, the hypersurface $\Sigma$ provides a strict limitation to the extension, towards the past, of the connected f-trapped region $\mathscr{T}_{1}$, and thereby to the location of its boundary $\B_{1}$. Therefore, it is important to know the exact location of $\Sigma$. This is the goal of this subsection.

Recall that $\Sigma$ is non-timelike everywhere, and actually spacelike on the entire portion where $\Sigma$ does not coincide with the EH. The location of $\Sigma$ depends, as is to be expected, on the particular properties of the mass function $m(v,r)$. However, one can deduce general properties of the hypersurfaces $\tau=$const. (One may also consult the Appendix where the specific case of the imploding Vaidya spacetime is treated in full, as then the equations can be explicitly integrated ---for appropriate choices of the mass function.) In particular, we are going to answer partially the question of whether or not $\Sigma$ can penetrate into the flat region of the spacetime. To that end, recall
the definition (\ref{gtau}) of $\tau$, so that the $\tau=$const. hypersurfaces are given by the solutions to the ODE 
\be
\frac{dv}{dr}=\frac{e^{-\beta}}{1-2m(v,r)/r}. \label{odeg}
\ee
Let $v_0\geq 0$ be the value of $v|_{\Sigma}$ at $r=0$. The ODE (\ref{odeg}) will not have any critical point at $(v=v_0^+,r=0)$ whenever 
\be
\lim_{r\rightarrow 0} \frac{m(v_0,r)}{r}=0\, .\label{nocrit}
\ee
In particular, this will be the case when there is no curvature singularity at $(v_0,0)$, due to (\ref{regular}).

On the other hand, if (\ref{nocrit}) does not hold, the ODE (\ref{odeg}) is equivalent to the autonomous system
$$
\frac{dv}{du}=r, \hspace{1cm} \frac{dr}{du}=e^{\beta}(r-2m(v,r))
$$
which has a critical point at $v=v_0, r=0$. Its linear stability is ruled by the eigenvalues $\lambda_{\pm}$ and eigenvectors $\vec{z}_{\pm}$ of the corresponding matrix
$$
\left(\begin{array}{ccc}
0 & 1\\
-2\dot m_0 e^{\beta_0} & e^{\beta_0}(1-2m'_0) \end{array}
\right), 
$$
where $\beta_0$, $\dot m_0$ and $m'_0$ are the limits of $\beta$, $\partial m/\partial v$ and $\partial m/\partial r$ when approaching $(v_0^+,0)$, respectively. These eigenvalues and eigenvectors are
$$
\lambda_{\pm}=\frac{e^{\beta_0}}{2}\left(1-2m'_0\pm \sqrt{(1-2m'_0)^2-8\dot m_0}\right), \hspace{1cm} \vec{z}_{\pm}=(1,\lambda_\pm).
$$
The character of the critical point is different depending on the sign of $8\dot m_0-(1-2m'_0)^2$. The different possibilities are
\begin{itemize}
\item If $8\dot m_0>(1-2m'_0)^2$, the critical point is a focus if $1-2m'_0\neq 0$ ---unstable or stable depending on whether $1-2m'_0$ is positive or negative. In this case no solution reaches $(v_0,0)$. The hypersurface $\Sigma$ always penetrates the flat region in this case. See the illustrative explicit solution for the Vaidya case in the Appendix. If on the other hand $2m'_0=1$, then the critical point is a centre at the linear level, and can become a focus, a node or remain as a centre depending on the properties of $m(v,r)$.
\item If $0< 8\dot m_0<(1-2m'_0)^2$, the critical point is a node, unstable or stable depending on whether $1-2m'_0$ is positive or negative. All possible solutions except one emerge from or approach $(v_0,0)$ with the same tangent direction, given by the eigenvector $(1,\lambda_-)$ in the unstable case or by $(1,\lambda_+)$ in the stable one. The exception for each case is given by one solution emerging from or approaching $(v_0,0)$ with the tangent direction of the other eigenvector. There exist three qualitatively different possibilities (with the same values of $\dot m_0$ and $m'_0$), depending on whether or not the special solutions $\tau=\tau_\pm$ ---corresponding to the eigenvalues $\lambda_\pm$ at $(v_0,0)$--- eventually meet $\AH_1$. This, in turn, depends on the specific properties of the mass function and on the total mass $M$. If at least one of the special solutions does {\em not} meet $\AH_1$, then the hypersurface $\Sigma$ cannot penetrate the flat region. On the other hand, if both special solutions meet $\AH_1$ then $\Sigma$ will have a portion in the flat region. Explicit illustrative cases are given in the Appendix for the Vaidya spacetime.
\item If $8\dot m_0=(1-2m'_0)^2\neq 0$, the critical point is a degenerate node (unstable or stable depending on whether $1-2m'_0$ is positive or negative) in the linear stability analysis, and it remains as such, or it  may become an unstable focus or node, depending on the specific properties of the mass function. Its properties are once more analogous to those of the Vaidya example in the Appendix.
\end{itemize}

\section{On the region $\mathscr{T}$ and its boundary $\B$}\label{sec:converse}
In this section we want to discuss the possible extension of the connected f-trapped region $\mathscr{T}_{1}$ associated to $\AH_1$, and the relation between its boundary $\B_{1}$ as defined in definition \ref{def:B} with marginally trapped tubes and closed weakly f-trapped surfaces. 

Having identified the past barrier $\Sigma$ for the connected f-trapped region $\mathscr{T}_{1}$, we can ask whether closed f-trapped surfaces can actually extend all the way down to $\Sigma$, in other words, if $\Sigma$ coincides with the connected component $\B_{1}$ of the boundary. This turns out not to be the case (Corollary \ref{cor:SigmanotB}.)

We already know that the region outside $\R$, such that $r<2m(v,r)$, belongs to $\mathscr{T}$. Now we collect some important properties of  how closed f-trapped surfaces can cross $\AH_{1}$ penetrating into $\R_{1}$.
\begin{theorem}\label{th:taum}
Assume that the spacetime (\ref{gds2}) satisfies the dominant energy condition, (\ref{massg}) and (\ref{dotm}). Any closed f-trapped surface $S$ crossing $\AH_1$ attains the minimum value $\tau_m$ of $\tau$ outside $\R_1\cup \AH_1$, and has
$$
\tau|_{S\cap \R_1} > \hat\tau_m > \tau_m >\tau_{\S}, \hspace{1cm} r|_{S\cap \R_1} < \hat r
$$
where $\hat\tau_m$ is the minimum value of $\tau|_S$ on $\AH_1$,
and $\hat r$ is the value of $r$ at the round sphere $\hat\varsigma \equiv \{\tau=\hat\tau_m\}\cap \AH_1$. (Let us note that $\hat\varsigma\in \AH_1\backslash$EH.) \fin
\end{theorem}
\proof
As $S$ is compact, it must attain a minimum of $\tau$ which is also a local minimum unless $\tau|_S=$const. This last possibility is not feasible for $S$ entering into $\R_{1}$ due to Theorem \ref{th:untrapped} and the non-negativity of $\left.P^{\mu\nu}(\lie_{\xiv} g|_{S})_{\mu\nu} \right|_S$ (if $\bar\xi_{A}=0$) as follows from (\ref{poscon}). For the former possibility, Theorem \ref{th:no-min} ensures that the local minimum cannot lie on $\R$, so that $\tau_m$ has to be attained outside $\R_1\cup \AH_1$. 

Pick up any value $\tau_2 > \tau_{m} > \tau_\Sigma$. As in the proof of Theorem \ref{th:aboveBg} consider the closed set 
$${\cal K}_2\equiv \overline{J^+(EH)\cap \{\tau\leq \tau_2\} \cap (R_{1}\cup \AH_{1})}.$$
${\cal K}_2$ is bounded to the future partly by $\tau=\tau_2$ and partly by $\AH |_{\tau\leq \tau_2}$, and to the past by EH$\backslash \AH_{1}$, see figure \ref{fig:SigmainVaidya}. 
Therefore, if $S\cap$ int${\cal K}_2\neq \emptyset$, $\tau |_{S\cap {\cal K}_2}$ will reach a minimum $\hat\tau_{m}$ on ${\cal K}_2$. As usual, this minimum cannot be on EH$\backslash \AH_{1}$ due to the standard result \cite{HE,Wald,Cl,BD} that $S$ can never touch EH; it cannot be on $\tau=\tau_2$ because this is the maximum value of $\tau$ on ${\cal K}_2$; and it cannot be on $\R_{1}\cap {\cal K}_2$ either, due to Theorem \ref{th:no-min}. Therefore, such a minimum has to be attained on $\AH_{1}$. Besides, there cannot be any point $x\in S\cap \R_{1}$ such that $\tau|_x =\hat\tau_m$, because this would contradict either Theorem \ref{th:no-min} or Theorem \ref{th:untrapped}. Thus, $\hat\tau_m< \tau|_{S\cap \R_1} $.

Consider now $\hat r$, and observe that it is the maximum value of $r$ on $S\cap \AH_{1}$, because $\tau|_{AH}$ is a monotonically decreasing function of $r$ on AH as follows from the definition (\ref{gtau}). To prove the result, recall that the hypersurfaces $\tau=$ const. and $r=$const. are orthogonal everywhere, the former are spacelike and the latter are timelike on $\rr$, and they both become null ---and tangent--- at $\AH$. Therefore, the hypersurface $\tau =\hat\tau_m$ is to the future of all hypersurfaces $r=r_c\geq \hat r$ everywhere on $\R_{1}$. Thus, if $S$ reached a value of $r\geq \hat r$ at a point $x\in S\cap \R_{1}$, then $x$ would be to the past of $\tau=\hat\tau_m$, that is, $\tau|_x <\hat\tau_m$ which is impossible.\fin

\begin{figure}[!ht]
\includegraphics[width=9.5cm]{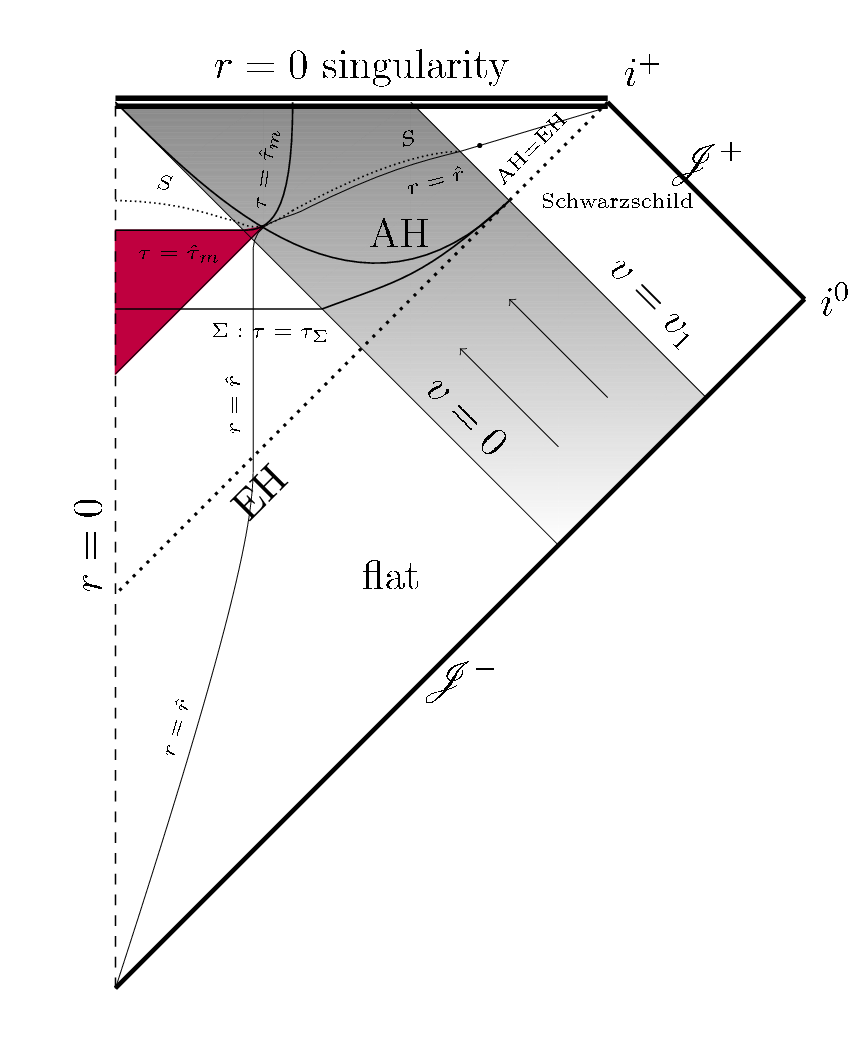}
\caption{\footnotesize{A Penrose diagram ---corresponding to the case on the right of figure \ref{fig:vaidya}, but any other choice would be similar--- showing a non-spherically symmetric closed f-trapped surface $S$ penetrating into the flat region of the spacetime. Such f-trapped surfaces were explicitly constructed in \cite{BS}. 
$S$ is shown as a dotted line emerging into the Schwarzschild part as a 
finite part of a line of constant $r$. The intersection of $S$ with AH occurs at $r=\hat r$ and $\tau =\hat\tau_{m}$, using the notation of Theorem \ref{th:taum}. The hypersurfaces $r=\hat r$ and $\tau =\hat\tau_{m}$ are explicitly shown. They meet at $AH$ where both of them become null and then change their causal character. Theorem 4.3 in \cite{AG} implies that $S$ cannot penetrate into the region $J^-(\AH|_{r>\hat r})$. However, Theorem \ref{th:taum} provides an improvement on that restriction, since $S$ cannot actually penetrate into the region with $\tau \leq \hat\tau_{m}$ below AH. Thus, the region shown in purple, allowed in principle by the former restriction, becomes a forbidden region for $S$. 
}}
\label{fig:Improvement}
\end{figure}

In the case that $\AH_{1}$ is spacelike ---a dynamical horizon---, Theorem 4.3 in \cite{AG} implies that no closed f-trapped $S$ can penetrate into the region $J^-(\AH|_{r>\hat r})$. Theorem \ref{th:taum} provides a stricter restriction, independently of the causal character of $\AH_{1}$, since $S$ cannot penetrate $J^-(\{\tau=\hat\tau_m\} \cap \R_{1})=\{\tau\leq \hat\tau_m\} \cap \R_{1}$. This is graphically explained in figure \ref{fig:Improvement}.

\begin{prop}\label{prop:tauBr=0}
Assume that the spacetime (\ref{gds2}) satisfies the dominant energy condition, (\ref{dotm}) and (\ref{massg}), and it has f-trapped round spheres to one side of $\AH_1$. Then, the connected component $\B_1$ cannot have a positive minimum value of $r$, and furthermore
$$
\tau_\B\equiv \inf_{x\in \B} \tau |_x = \inf_{x\in \B_1} \tau |_x=\tau_\Sigma \, .
$$
\end{prop}
\proof That $\tau_\B = \inf_{x\in \B_1} \tau |_x$ is obvious, as the spacetime is flat for $v<0$ so that there cannot be f-trapped surfaces penetrating the past of $\B_1$. To see that $\tau_\B=\tau_\Sigma$, we first note that $\tau_\B<\tau_\Sigma$ is impossible, as otherwise there would be f-trapped closed surfaces penetrating the region to the past of $\Sigma$ contradicting Theorem \ref{th:aboveBg}.

If $\tau_\B>\tau_\Sigma$, then $\B_1$ would be fully contained in the region $\tau \geq \tau_\B$. But this would mean, due to Properties \ref{pr:Bclosed} and \ref{pr:2sides}, that the region defined by $\tau <\tau_\B$ would be either part of $\mathscr{T}$, or completely external to it. However, this is again impossible because the part of this region with $\tau\in (\tau_\Sigma,\tau_\B)$ has f-trapped round spheres outside $\R_1\cup\AH_1$, and no closed f-trapped surface can penetrate its part with $\tau\leq \tau_\Sigma$.

The same reasoning serves to prove that
$$
\inf_{x\in \B_1} r |_x=0 \, .
$$
For, if this infimum were positive, say $r_a <2M$, it would follow that $\B_1$ would be fully contained in the region $r\geq r_a>0$. But there are f-trapped round spheres for {\em all} values of $r\in (0,2M)$ outside $\R_1\cup\AH_1$.\fin

\begin{prop}\label{prop:merge}
Under the same assumptions
$$\B_1\subset (\R_1\cup \AH_1) \cap J^+(\Sigma),$$
and $\B_1$ merges with, or approaches asymptotically, $\Sigma$, $\AH_1$ and EH in such a way that $\B_1\cap \AH_1=\emptyset$ at any portion of $\AH_1$ with $G_{\mu\nu}k^\mu k^\nu |_{AH_1}>0$. 

Furthermore, $\B_1$ cannot be non-spacelike everywhere.
\end{prop}
{\bf Remark:} In the case that $v_1$ is finite and the mass function is constant for $v>v_1$ the EH and $\AH_1$ coincide for all $v>v_1$, and so does $\B_1$. This portion of $\AH$ is an isolated horizon with $m=M$=constant, and $G_{\mu\nu}k^\mu k^\nu =0$ on that portion. However, there can be other portions of $\AH_1$ with $G_{\mu\nu}k^\mu k^\nu =0$ such that they are isolated horizons. This happens if $m=m(r)<M$ (possibly constant) for some interval $v\in (v_2,v_3)$ with $v_3<v_1$. Physically, this means that the inflow of matter and radiation stops between $v_2$ and $v_3$, and then it starts again. In principle, $\B_1$ may coincide with $\AH_1$ on these particular portions of $\AH_1$ of isolated-horizon type. 
\begin{figure}[!ht]
\includegraphics{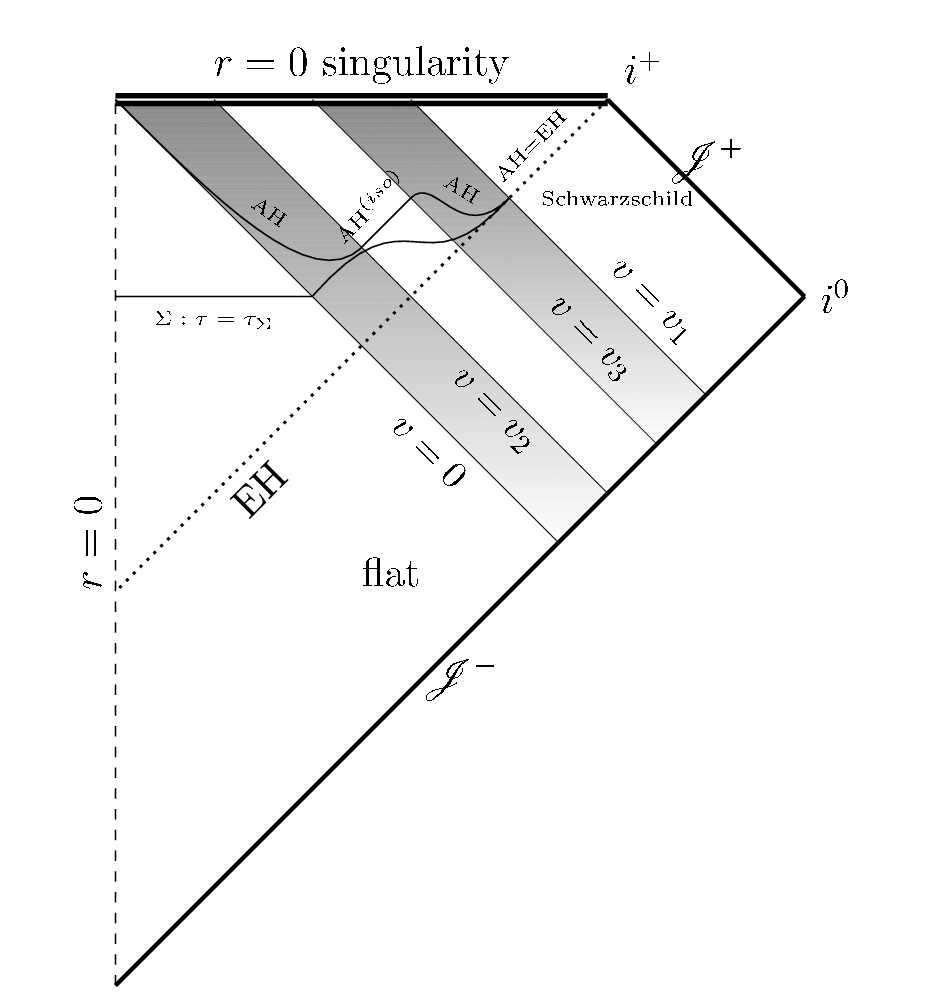}
\caption{\footnotesize{Conformal diagram for a particular example where the AH has a non-vanishing portion of isolated horizon type different from EH.  The energy flows in along null incoming radial hypersurfaces, starting at $v=0$ but suddenly stopping at $v=v_2$. There is no inflow of energy between $v_2$ and $v_3$,  and then the energy enters again from past null infinity until the radiation stops at $v=v_1$, and then for ever. A particular case of this situation is given by the Vaidya imploding spacetime (see Appendix) with $m(v)=$ constant for $v\in (v_2,v_3)$. In the picture, the AH is spacelike for $v\in (0,v_2)\cup (v_3,v_1)$, but it has two portions which are isolated horizons, one given by part of EH (for $v\geq v_1$) and the other one for $v\in (v_2,v_3)$, represented by $\AH^{(iso)}$. The past barrier $\S$ merges with AH and EH as usual. The boundary $\B$ will also merge with them at $r=2M$ and $v=v_1$, and it must be strictly below AH everywhere on $\AH\backslash \AH^{(iso)}$, but it is not guaranteed that it cannot touch, or even coincide with, $\AH^{(iso)}$.}}
\label{fig:AHiso}
\end{figure}
This is represented in figure \ref{fig:AHiso}. It will be useful to have a name for these portions, so that we set:
$$
\AH_1^{(iso)} \equiv \AH_1 \cap \AH^{(iso)} 
\, .
$$
Note that EH will belong to $\AH_1^{(iso)}$ if $v_1$ is finite and the mass function reaches the value $M$. Also, that $\tau$ is constant on $\AH_1^{(iso)}$, as $\AH_1^{(iso)}$ is defined by portions of $r=$constant hypersurfaces within $\AH_1$ which are null.
Observe that the condition $G_{\mu\nu}k^\mu k^\nu \neq 0$ ---required also in the perturbations of section \ref{sec:notAH}--- becomes $G_{\mu\nu}k^\mu k^\nu >0$ if the dominant energy condition holds. 

\proof There are f-trapped closed surfaces outside $\R_1\cup\AH_1$,
but due to Theorem \ref{th:aboveBg} there are none penetrating $J^-(\Sigma)$. Thus, $\B_1\subset  (\R_1\cup \AH_1)\cap J^+(\Sigma)$. However, $\B_1$ cannot meet  $\AH_1\backslash \AH_1^{(iso)}$ according to Theorem \ref{th:AHnotB}. As $\Sigma$ and $\AH_1$ merge together, or approach each other asymptotically, so does $\B_1$. Finally, consider the domain of dependence $D(\Sigma)$ of $\Sigma$. From the previuous observations, $\B_1\subset D^+(\Sigma)$. But $D(\Sigma)$ is globally hyperbolic with $\Sigma$ as a Cauchy hypersurface, therefore if $\B_1$ were non-spacelike everywhere it would have to cross $\Sigma$ \cite{HE,P5,S,Wald}, in contradiction with the fact that $\B_1\subset J^+(\Sigma)$.\fin

As in the case of Robertson-Walker spacetimes (Result \ref{res:noWTSinB}, section \ref{sec:RW}) we derive the following important result.
\begin{theorem}\label{th:noDH}
Under the same assumptions, $\B_1\backslash \AH_1^{(iso)}$ cannot be a marginally trapped tube, let alone a dynamical or trapping horizon. Furthermore, $\B_1\backslash \AH_1^{(iso)}$ does not contain any non-minimal closed weakly f-trapped surface.
\end{theorem}
\proof From the previous corollary $\B_1\backslash \AH_1^{(iso)}$ is contained in the region $\R_1$. But there are no closed weakly f-trapped surfaces completely contained in $\R_1$ due to Lemma \ref{lem:basic}.\fin

Thus, the only closed marginally f-trapped surfaces that can be contained in $\B_1$ are those which are actually on its part $\B_1\cap\AH_1^{(iso)}$, if any. In fact, this property could have been deduced more easily from Result \ref{res:AHunique}, as $\B_1$ is a spherically symmetric hypersurface. We have decided to include the alternative proof as it may probably be generalized to non-spherically-symmetric cases. 

Theorem \ref{th:noDH} implies that the notion of ``limit section" in \cite{Hay} (a spacelike 2-surface in $\B$ arising as the uniform limit of a sequence of trapped surfaces approaching $\B$, definition in p.6473) is generically non-existent or ill defined. Thus the assumptions of theorem 7 in \cite{Hay} are very rarely met.

\begin{prop}\label{prop:Bdecreasing}
Assume that the spacetime (\ref{gds2}) satisfies the dominant energy condition, (\ref{dotm}) and (\ref{massg}), 
and has f-trapped round spheres to one side of $\AH_1$. Then, 
$\tau$ is a non-increasing function of $r$ on any portion of the connected component $\B_1$ which is locally to the past of $\mathscr{T}_1$.
And it is actually strictly decreasing at least somewhere $\B_1\backslash \AH_1^{(iso)}$.
\end{prop}
{\bf Remark:} $\B_1$ is a connected component of the boundary $\B$ and, due to Property \ref{pr:2sides},  
$\B_1$ has two sides, one with and another without closed f-trapped surfaces. 
Thus, by ``locally to the past" we mean that the Kodama vector field points towards $\mathscr{T}_1$ at $\B_1\backslash \AH_1^{(iso)}$ ---at $\B_1\cap \AH_1^{(iso)}$ it is tangent.

\proof If $\B_1$ coincides partly with $\AH_1^{(iso)}$ the result is trivial there, as $\tau$ and $r$ are constant on $\AH_1^{(iso)}$.
Suppose then that $\tau|_{\B_1}$ were a non-decreasing function of $r$ around a round sphere 
$\varsigma\subset \B_1\backslash \AH_1^{(iso)}$ given by $\varsigma\equiv \{\tau =\tau_b,r=r_b\}$ for some constants 
$r_b>0$ and $\tau_b\geq\tau_\Sigma$, and assume that the f-trapped region $\mathscr{T}_1$ is to the future of $\varsigma$ (see next figure). 
\begin{center}
\begin{tikzpicture}
\foreach \x in {1,2,3,4}
\draw (\x,0) -- (\x,4);
\draw (0,1) -- (4.2,1);
\draw[very thick] (0.3,0.3) --  node[sloped,above,very near end] {$\B_1$} (4.1,4.1);
\filldraw (2,2) circle (2pt);
\draw (2.15,1.85) node {$\varsigma$};
\draw (0.7,2) -- (4.2,2);
\draw (4.7,1.8) node {$\tau=\tau_b$};
\draw (0.6,2.8) -- (4.1,2.8);
\draw (4.75,2.95) node {$\tau=\tau|_x$};
\draw (0.7,2.5) -- (4.2,2.5);
\draw (4.85,2.5) node {$\tau=\hat\tau_m$};
\draw (2,-0.2) node {$r=r_b$};
\draw (2.25,0.3) -- (2.25,4);
\draw (2.25,0.25) node {$r=\hat r$};
\filldraw (2,2.8) circle (1pt);
\draw (1.87,2.93) node {$x$};
\draw[dotted] (2,2.8) .. controls (0.5,2.6) and (1.5,2.6) .. (2.25,2.5);
\draw (1.35,2.7) node[anchor=east] {\tiny{$L$}};
\draw (2.25,2.5) circle (1pt);
\draw (2.4,2.63)  node{$\hat\varsigma$};
\draw (1.5,3.5) node {$\mathscr{T}_1$};
\end{tikzpicture}
\end{center}
Then, any point $x\in \R_1$ lying on the round spheres with $r=r_b$ and $\tau>\tau_b$ but near enough $\tau_b$ would belong to at 
least one f-trapped closed surface. Pick up any such $x$ and a f-trapped $S\ni x$, and let $\hat\tau_m$ and $\hat r$ be the minimum and 
the maximum values of $\tau|_S$ and $r|_S$ on $\AH_1$, respectively. From Theorem \ref{th:taum} we know that 
$\hat\tau_m<\tau|_{S\cap\R_1}$ and $\hat r > r|_{S\cap \R_1}$. In particular, $\hat\tau_m<\tau|_x$, $\hat r > r|_x =r_b$. 
From Theorem \ref{th:no-min} (see its third Remark) there should be a connected path $L\subset S\cap\R_1$ lying entirely on $S$ 
starting at $x$ and finishing on $\hat\varsigma \subset \AH_1$, $\hat\varsigma \equiv \{\tau=\hat\tau_m\}\cap \{r=\hat r\}$, 
such that $\tau|_L$ is non-increasing ---and strictly decreasing somewhere. Thus, $L$ would eventually cross all hypersurfaces 
$r=$const. with $r\in [r_b,\hat r]$, and each of the crossings would happen with a smaller value of $\tau$. 
Due to the results in Corollary \ref{cor:sym} and Proposition \ref{prop:tauBr=0}, 
and since $S$, and hence $L$, cannot intersect $\B_1$, this would mean that $\hat\tau_m$ and $\hat r$ should be such that $\tau_b<\hat\tau_m<\tau|_x$ and $r_b<\hat r <r|_{\B_1\cap\{\tau=\hat\tau_m\}}$. But this leads to a contradiction, because if such a result held for {\em all} $x\in \{r=r_b\}\cap \{\tau >\tau_b\}$, by taking an appropriate sequence $\{x_n\}$ of such $x$ approaching $\varsigma$, the sequence would have a limit on $\varsigma$, which would in turn produce a sequence of round spheres $\{\tau=\hat\tau_m(x_n),r=\hat r(x_n)\}$, all of them belonging to $\AH_1$, and converging to $\varsigma\in \B_1\backslash \AH_1^{(iso)}$. As $\AH_1$ is closed, $\varsigma$ would belong to $\AH_1$. But this is impossible as $(\B_1\backslash \AH_1^{(iso)})\cap \AH_1=\emptyset$ due to Proposition \ref{prop:merge}.\fin

\begin{coro}\label{cor:Anot0}
Under the same assumptions, $\B_1\backslash \AH_1^{(iso)}$ cannot be tangent to a $\tau=$const. hypersurface everywhere.
\end{coro}

In particular, $\B_1$ never touches $\Sigma$ before they merge together (and together with $\AH_1\cap$EH), that is to say, $(\B_1\cap\S |_{\R_1})=\emptyset$.
\begin{coro}\label{cor:SigmanotB}
The past barrier $\S\backslash EH$ is not part of the boundary $\B_1$.\fin
\end{coro}
\begin{coro}\label{cor:Bspacelike}
Under the same assumptions, $\B_1$ is spacelike close enough to its merging (or asymptotic approaching) to EH, $\Sigma$ and $\AH_1$.
\end{coro}
\proof We already know that $\B_1$ has to be spacelike somewhere. In the region of this corollary the f-trapped closed surfaces are to the future of $\B_1$, so that $\tau$ has to be a non-increasing function of $r$. The limitation to the past by $\Sigma$ and to the future by $\AH_1$ then implies the result.\fin

\begin{figure}[!ht]
\includegraphics{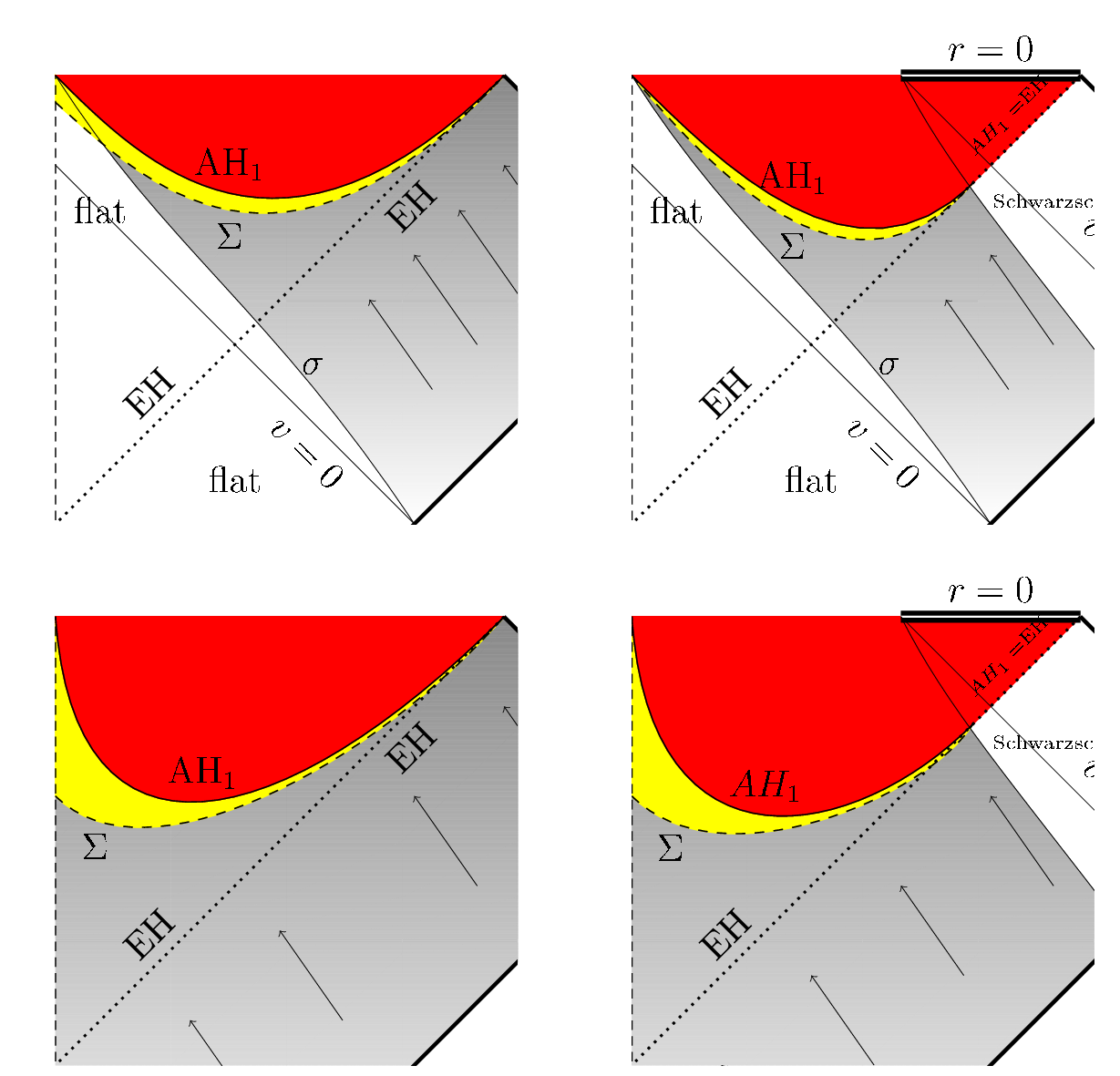}
\caption{\footnotesize{These are enlargements of some previous conformal diagrams showing the possible location of the boundary $\B_{1}$. The red regions are part of the f-trapped region $\mathscr{T}_{1}$, but this region actually extends further down and includes $\AH_{1}$, as follows from Theorem \ref{th:AHnotB}. On the other hand, Corollary \ref{cor:SigmanotB} informs us that $\B_{1}$ can never touch $\S$ outside EH. Thus, the yellow zones are the allowed regions for $\B_{1}$, keeping in mind that there is always a red zone below $\AH_{1}$, and $\B_{1}$ must be placed strictly above $\S$. In the left upper picture, $\Sigma$ and also $\B_{1}$ extend to the flat region of the spacetime.
}}
\label{fig:whereisB}
\end{figure}

The combinations of all results obtained hitherto can be schematically represented as in figure \ref{fig:whereisB}. 
We can see that the boundary $\B_1$ is highly non-local too, and it can have portions in flat regions of spacetime 
whose whole past is also flat. This is surely not a good candidate for the surface of a dynamical black hole. 
Nevertheless, it remains as an interesting puzzle to find the exact location and the defining properties of $\B$.
Some relevant results in this direction are collected in the remaining of this section.

Proposition \ref{prop:Bdecreasing} informs us that $\B_1\backslash \AH_1^{(iso)}$ has to bend down in the Kodama time $\tau$. This has
direct consequences on the extrinsic curvature of $\B_1$. Observe that, as follows from (\ref{deforxi}) and (\ref{dotm}),
we know that the level hypersurfaces $\tau=$const. have a non-negative semi-definite second fundamental form. 
Actually, they have two vanishing eigenvalues and the other one is proportional to $\partial m/\partial v$. 
Hence, one can prove that the boundary $\B_1$ must have a second fundamental form with a non-positive double eigenvalue.
\begin{prop}\label{prop:Knegative}
Assume that the spacetime (\ref{gds2}) satisfies the dominant energy condition, (\ref{dotm}) and (\ref{massg}). 
Then, any portion of the connected component $\B_1$ which is locally to the past of 
$\mathscr{T}_1\neq \emptyset$ has a
second fundamental form with a non-positive (and strictly negative whenever $\B_1$ is not tangent to a $\tau=$const.\ hypersurface) double eigenvalue at any point point where it is spacelike.
In particular, it cannot have a positive semi-definite second fundamental form there.
\end{prop}
\proof As $\B_1$ is part of the boundary, it has spherical symmetry. Take a portion $\hat \B_1$ of $\B_1$ where it is spacelike and 
to the past of $\mathscr{T}_1$, and include this portion in a local foliation by spherically symmetric spacelike hypersurfaces 
$t= $const. Locally, the future-pointing vector field orthogonal to the foliation can be given by
\bean
\vec\eta = e^{-\beta}\partial_v+A \partial_r \, , \hspace{2cm} 
\\
\eta_\mu dx^\mu =-F' dt = dr -e^\beta \left(1-\frac{2m}{r}-A\right)dv 
\eean
for some function $A(v,r)$. The level hypersurfaces are spacelike ($\vec\eta$ is
timelike) so that
\be
1-\frac{2m}{r}-2A>0 .\label{spacelike}
\ee
Observe that $\vec\eta = \vec\xi -A e^\beta \vec\ell$, and furthermore
$$
\eta_\mu dx^\mu =-F' dt =\frac{A}{1-2m/r} dr -F \left(1-\frac{A}{1-2m/r}\right)d\tau
$$
by using the Kodama time $\tau$ on $\R_1\supset \hat\B_1$. On $\hat\B_1$ we have that $t$ is constant, so that
$$
\left.\frac{d\tau}{dr}\right|_{\hat \B_1}=\left.\frac{1}{F}\,\frac{A}{1-2m/r-A}\right|_{\hat \B_1} \, .
$$
However, Proposition \ref{prop:Bdecreasing} implies that this is strictly negative, which requires necessarily (if $A\neq 0$)
$$
\left. \frac{1-2m/r}{A}\right|_{\hat \B_1}<1 .
$$
Given that $1-2m/r>0$ at $\hat\B_1\subset \R_1$, this together with (\ref{spacelike}) implies that (if $A\neq 0$)
$$
A<0 .
$$
We are now going to show that $A$ is essentially the double eigenvalue of the second fundamental form of $\hat\B_1$.
A straightforward calculation provides the deformation of the metric along $\vec\eta$
\bea
(\lie_{\etav} g)_{\mu\nu} =-2\frac{\partial \beta}{\partial r} \eta_\mu \eta_\nu +e^\beta  
\left[\frac{\partial \beta}{\partial r}\left(1-\frac{2m}{r}-A\right)-\frac{\partial A}{\partial r}\right]
\left(\ell_\mu \eta_\nu +\ell_\nu \eta_\mu \right)+\nonumber\\
\left[ e^{\beta}\frac{2}{r}\frac{\partial m}{\partial v} +2e^{\beta}\frac{\partial A}{\partial v}
+e^{2\beta}\frac{\partial A}{\partial r}(1-2m/r-A)-
A\frac{\partial\left[e^{2\beta}(1-2m/r)\right]}{\partial r}\right]\ell_\mu \ell_\nu \label{deforeta}\\
+2rA \, \, d\Omega_{\mu\nu} \hspace{2cm} \nonumber
\eea
where $d\Omega_{\mu\nu}$ represents the angular part of the metric, that is, the metric on the unit round sphere.
Of course, the projection to $\B_1$ of $\lie_{\etav} g$ is (proportional to) the second fundamental form $K_{\mu\nu}$ of $\B_1$ 
whenever it is spacelike, so that $K_{\mu\nu}|_{\hat \B_1}$ is proportional to the restriction to $\hat\B_1$ of the second and third
lines in (\ref{deforeta}). In particular, the double eigenvalue is proportional to $A$.
\fin

As a matter of fact, one can try to do better and try to set restrictions also on the third eigenvalue of $K_{\mu\nu}|_{\hat \B_1}$.
This is given by the expression on the second line of (\ref{deforeta}) (projected to $\hat\B_1$).
The idea is based on theorem \ref{th:no-min} and corollary \ref{cor:noposK}. There are closed f-trapped surfaces passing
through every point $x\in \mathscr{T}_1$ which is locally to the future of $\hat \B_1$. 
Choose a sufficiently smooth sequence of closed f-trapped surfaces approaching $\hat \B_1$. 
The smoothness is assumed such that one can take the limit and obtain a piece of a surface $\zeta$ touching $\hat \B_1$. 
The set obtained as limit of the surfaces may fail to be compact, 
e.g. the Robertson-Walker example mentioned after Result \ref{res:noWTSinB}; or to be spacelike everywhere, 
or even to be connected. However, all these problems will be irrelevant in what follows as long as $\zeta$ exists.
(This is the main problem when trying to turn this reasoning into a formal proof, as proving the existence of $\zeta$ encounters some technical
difficulties from a mathematical viewpoint).

Given that all the surfaces in the sequence are f-trapped, they all have $t > t_{\hat \B_1}$, where $t_{\hat \B_1}$ 
is the constant value of $t$ at $\hat \B_1$. Therefore, $\zeta$ is such that $t|_{\zeta}\geq t_{\hat \B_1}$, 
which implies that $\zeta$ has a local minimum of $t$ at the intersection $\zeta\cap \hat \B_1$, 
or that $\zeta$ has a two-dimensional portion within $\hat \B_1$. In particular, $\zeta$ is spacelike around 
that minimum because $\vec\eta$ is orthogonal to $\zeta$ there. The mean curvature vector of this spacelike portion 
of $\zeta$ (including $\zeta\cap \hat \B_1$) must be future-pointing or zero, given that $\zeta$ is limit of the sequence 
of f-trapped surfaces. Therefore, we have $\bar\nabla_A\bar\eta^A|_{\zeta\cap \hat \B_1}  \leq 0$ and also 
$\eta_\mu H^\mu |_{\zeta\cap \hat \B_1}\leq 0$.
Now, we can apply the same reasoning as in Theorem \ref{th:no-min}, for which the compactness is not necessary. 
Formula (\ref{main}) applied to the spacelike portion of $\zeta$ containing $\zeta\cap \hat \B_1$ gives 
$$
\bar\nabla_A\bar\eta^A|_{\zeta\cap \hat \B_1} +\eta_\mu H^\mu |_{\zeta\cap \hat \B_1}=\frac{1}{2} P^{\mu\nu}(\lie_{\etav} g)_{\mu\nu} |_{\zeta\cap \hat \B_1}
$$
where $P^{\mu\nu}$ is the orthogonal projector to $\zeta$. 
Thus, we deduce that
$$
P^{\mu\nu}(\lie_{\etav} g)_{\mu\nu} |_{\zeta\cap \hat \B_1} = P^{\mu\nu}K_{\mu\nu} |_{\zeta\cap \hat \B_1}\leq 0 \, .
$$
Observe that, provided this argument can be promoted into a rigourous proof, it may restrict the third eigenvalue severely, because it
has to hold for {\em all} projectors which are limits of the projectors to closed f-trapped surfaces close enough to $\hat\B_1$. 
If one could gain control on the variety of such projectors, then much more precise restrictions could be set on the boundary $\B$.

\section{The core of the trapped region and AH}\label{sec:new}
At this point we know that the EH is teleological, and also that closed f-trapped surfaces are clairvoyant: they are ``aware" of things that happen elsewhere, with spacelike separation. For instance, they can have portions in a flat region of spacetime whose whole past is also flat in clairvoyance of energy that crosses them
elsewhere to make their compactness and trapping feasible \cite{BS,ABS}, see figure \ref{fig:Improvement}. This non-local property of trapped surfaces is inherited by everything which is based on them, such as marginally trapped tubes including dynamical horizons.
In conjunction with the non-uniqueness of dynamical horizons, this poses a fundamental puzzle for the physics of black holes, a problem that has been recognized and discussed many times lately, see e.g. \cite{AK1,AG,B,BBGV,BF,E,Hay3,Kri,NJKS} and references therein. 

Four possible solutions have been put forward \cite{NJKS}. First, one can rely on the old and well defined event horizon. This encounters very serious problems because one needs to know the whole future evolution of the spacetime. The event horizon is unreasonably global \cite{Haji,AG}. Actually, the whole construction of trapping and dynamical horizons was developed to solve this problem and to have nice, local, definitions of the surface of a black hole \cite{AK1,B}. Second, one can treat all possible horizons on equal footing. The problem is how to associate unique physical properties to the corresponding black hole, because each dynamical horizon comes with its own set of magnitudes. And they do not agree. The third strategy consisted in finding the boundary $\B$ as defined in this paper. However, as we have shown, not only $\B$ will not be a marginally trapped tube in general, it also suffers from the non-local properties associated to f-trapped surfaces. For instance, 
we have seen that $\B$ can enter the flat regions of spacetime. Finally, the fourth approach consists in trying to define a preferred dynamical horizon. Hitherto, there has been no good definition for that.

In the following we are going to pursue a novel strategy. The idea is based on the simple question: what part of the spacetime is absolutely indispensable for the existence of the black hole? We already know that, in the cases considered so far with flat regions and matter imploding so that a black hole eventually forms, the flat region is certainly not essential for the existence of the black hole. What is? By answering this question we might actually get a bonus and provide a positive, constructive solution to the fourth strategy mentioned before: it may happen that a unique dynamical horizon is selected.

From Corollary \ref{coro:NOD-} and Theorem \ref{th:aboveB} it is clear that if the whole complement of $\R$ is removed from the spacetime, then no closed f-trapped surfaces remain. The question arises of whether or not proper subsets of that removed region suffice to achieve the same, that is, to short-circuit all closed f-trapped surfaces. To be precise, we give the following definition. 

\begin{defi}
A region $\mathscr{Z}_1\subset \varietat$ is called the {\em core} of a connected component $\mathscr{T}_1$ of the f-trapped region $\mathscr{T}$ 
if it is a minimal closed connected set that needs to be removed from the spacetime in order to get rid of all closed f-trapped surfaces in $\mathscr{T}_1$, and such that any point on the boundary $\partial\mathscr{Z}_1$ is connected to $\B_1=\partial \mathscr{T}_1$ in the closure of the remainder.
\end{defi}
Here, ``minimal" means that there is no other set $\mathscr{Z}'$ with the same properties and properly contained in $\mathscr{Z}_1$. The final condition states that the excised spacetime $(\varietat\backslash \mathscr{Z}_1,g)$, which no longer has a connected f-trapped region $\mathscr{T}_1$,  has the property that furthermore each point in the closure $(\varietat\backslash \mathscr{Z}_1)\cup \partial\mathscr{Z}_1$ can be joined to the original boundary $\B_1$ by a continuous curve fully contained in $(\varietat\backslash \mathscr{Z}_1)\cup \partial\mathscr{Z}_1$. (This curve may have zero length at points $x\in  \partial\mathscr{Z}_1\cap  \partial\mathscr{B}_1$). This is needed because one could identify a particular removable region to eliminate the f-trapped surfaces, excise it, but then put back a tiny but central isolated portion to make it smaller. However, this is not what one wants to cover with the definition.  

Obviously, $\mathscr{Z}_1\subset \mathscr{T}_1$, however, $\mathscr{Z}_1$ will be generally smaller than $\mathscr{T}_1$. As an example, take the Robertson-Walker spacetime of figure \ref{fig:RW}. There, the future trapped region $\mathscr{T}$ is the whole future of the recollapsing time, shown in red. However, one only needs to remove the triangle to the future of the AH in order to get rid of all f-trapped surfaces. (Note that the boundary of this region is an apparent 3-horizon AH). This example also proves that $\mathscr{Z}$ is not unique: one can choose any other  region $\mathscr{Z}$ equivalent to the chosen one by moving all its points by the group of symmetries on each homogeneous slice.

Actually this kind of non-uniqueness is rather trivial, and is due to the existence of a high degree of symmetry. Nevertheless, even in less symmetric cases the uniqueness of the cores $\mathscr{Z}$ cannot be assumed beforehand. We are actually going to show that it does not hold in general.

We can use the results found in section \ref{sec:notAH}, especially Theorem \ref{th:tiny}, to identify one core of the f-trapped region in spherically symmetric spacetimes. 
\begin{theorem}
Assume that $\AH^{iso}=\emptyset$ for simplicity. Then, the complement of $\rr$ (i.e. the region $\mathscr{Z}\equiv \{r\leq 2m(v,r)\}$) is the disjoint union of core f-trapped regions. Each of its connected components is the core of the corresponding connected components of $\mathscr{T}$.
\end{theorem}
{\bf Remark:} The cases with $\AH^{iso}\neq \emptyset$ are technically more involved, but one expects that the result will hold true too.

\proof First of all, it is clear that every closed f-trapped surface has points in $\mathscr{Z}\equiv \varietat \backslash \rr$, as follows from Corollary \ref{coro:NOD-} or Theorem \ref{th:aboveB}. Hence, if we remove $\mathscr{Z}$ from the spacetime all closed f-trapped surfaces disappear. To see that there is no proper subset of $\mathscr{Z}$ with this property, observe that its boundary is $\partial \mathscr{Z}=\AH$. Take then any closed connected proper subset $\mathscr{Z}'$ of $\mathscr{Z}$ such that all points of $\partial\mathscr{Z}'$ are connected to its nearest part of the boundary $\B$. This implies that there is a curve from every $x\in \partial\mathscr{Z}'$ to such a part of the boundary $\B$, and all these curves must therefore cross AH. 
In summary, $\mathscr{Z}\backslash\mathscr{Z}'$ always contains an open region around $\AH$ and outside $\R_0$. But then, theorem \ref{th:tiny} ensures that there are closed f-trapped surfaces fully contained in $\varietat\backslash \mathscr{Z}'$, so that $\mathscr{Z}'$ cannot be a core.\fin

As a bonus, we have obtained that the boundary of the identified core $\mathscr{Z}=\{r\leq 2m\}$ is formed by the marginally trapped tubes AH, in particular by their dynamical horizon portions. One can wonder if this property selects these marginally trapped tubes in spherically symmetric spacetimes. Observe, however, that according to the results in \cite{AG}, given any regular dynamical horizon $H$, there cannot be any closed weakly f-trapped surface fully contained in its past domain of dependence $D^-(H)$. Therefore, if we remove the appropriate future part of $H$ we also remove all possible closed f-trapped surfaces. The question now is whether or not these alternative would-be cores are actually optimal, or if they remove more than is needed from the spacetime to get rid of the f-trapped region. 
Independently of whether or not they are optimal, the result in \cite{AG} allows us to prove that there are non-spherically symmetric cores in spherically symmetric spacetimes. 

First we show that $\mathscr{Z}$ are the unique spherically symmetric cores.
\begin{prop}\label{prop:sphsymcore}
In spherically symmetric spacetimes, $\mathscr{Z}=\{r\leq 2m\}$ are the only cores of $\mathscr{T}$ which are
invariant under the action of the corresponding SO(3) group of isometries. Therefore, $\partial\mathscr{Z}=\AH$ are the only spherically symmetric boundaries of a core.
\end{prop}
\proof Suppose there were another spherically symmetric core $\mathscr{Z}'$. Obviously $\mathscr{Z}'$ could not be a proper subset of $\mathscr{Z}$, nor vice versa, because both are cores. Thus $\mathscr{Z}\backslash\mathscr{Z}'\neq \emptyset$ and this set would be spherically symmetric. However, every round sphere in $\mathscr{Z}$ is f-trapped, and therefore there would be f-trapped round spheres having no intersection with $\mathscr{Z}'$, contradicting the hypothesis that $\mathscr{Z}'$ was a core of the trapped region. \fin

\begin{prop}\label{prop:nonsphsymcore}
There exist non-spherically symmetric cores of the f-trapped region in spherically symmetric spacetimes.
\end{prop}
\proof Consider the case when the spacetime has only one connected component of AH, which is spacelike, such as for example the Vaidya spacetime in the Appendix. From Corollary \ref{cor:lars} or the general results in \cite{AG} we know that there are non-spherically symmetric dynamical horizons interweaving the AH ---see also the explicit constructions in \cite{BS,ABS,NJKS}. Take any of these, say $H$, so that $H$ lies partly to the future of AH (and partly to its past). From Theorem 4.1 in \cite{AG}, no weakly f-trapped surface can be fully contained in the past domain of dependence of $H$. Consider then the causal future $J^+(H)$ of $H$.  Removing $J^+(H)$ from the spacetime eliminates all  closed f-trapped surfaces. Nevertheless, it may happen that  $J^+(H)$ is not a core, because it is not minimal. In any case, there is a subset of $J^+(H)$ which is a core of the f-trapped region $\mathscr{T}$. This new core will never include those parts of the spacetime which are to the future of AH but to the past of $H$. Thus, this core of $\mathscr{T}$ is not $\mathscr{Z}$, and due to Proposition \ref{prop:sphsymcore}, it cannot be spherically symmetric.\fin

Still, the identified core $\mathscr{Z}=\{r\leq 2m\}$ may be unique in the sense that its boundary $\partial\mathscr{Z}=\AH$ is a marginally trapped tube. This would happen if, for instance in the example of the previous proof, any dynamical horizon $H$ other than AH is such that $J^+(H)$ is {\em not} a core of the f-trapped region, the core being a proper subset of $J^+(H)$. If this is the case, then AH would be selected as the unique dynamical horizon which is the boundary of a core of the f-trapped region $\mathscr{T}$. 

Whether or not this happens is a very interesting open question.

\section*{Acknowledgements}
We thank the Wenner-Gren Foundation for making this research possible. 
IB was supported by the Swedish Research Council. 
JMMS thanks the theoretical physics division at Fysikum in AlbaNova, Stockholms Universitet, for hospitality. 
He also acknowledges financial support from grants FIS2004-01626 (MICINN) and GIU06/37 (UPV/EHU). 
Finally we happily acknowledge the help we received from Robert Wald, Greg Galloway, an anonymous referee, and Jan \AA man. 

\section*{Appendix. The imploding Vaidya spacetime.}\label{sec:vaidya}
Consider the important case of  the Vaidya spacetime with incoming radiation. The line-element reads \cite{V,Exact}
\be
ds^2=-\left(1-\frac{2m(v)}{r}\right)dv^2+2dvdr+r^2d\Omega^2 \label{ds2}
\ee 
so that this is the particular case of (\ref{gds2}) with $\beta =0$ and a mass function independent of $r$, $m(v)\geq 0$.

The Einstein tensor of (\ref{ds2}) is of pure radiation type
$$
G_{\mu\nu}=\frac{2}{r^2}\frac{dm}{dv}\ell_\mu\ell_\nu, 
$$
and thus, if the Einstein field equations are assumed, the null convergence condition (which in this particular case implies the dominant energy condition) \cite{HE} requires that
\be
\frac{dm}{dv}\geq 0 \label{mdot}
\ee
so that the mass function cannot decrease as a function of $v$. Thus, in this case the condition (\ref{dotm}) is guaranteed.

There is only one connected component of AH defined by
$$
\mbox{AH:} \hspace{2mm} r-2m(v)=0
$$
and unique associated regions $\rr : r>2m(v)$ and $\R : r\geq 2m(v)$.
AH is a spacelike hypersurface for all $v$ such that $dm/dv>0$, and it is null where $dm/dv=0$. Therefore, AH is a dynamical horizon \cite{AK1} on the region where it is spacelike, and an isolated horizon \cite{AK1} on any open region where $m(v)=$const. 

We adopt all the assumptions as in the general case so that the mass function satisfies
\be
m(v)=0 \hspace{3mm} \forall v<0; \hspace{1cm} m(v)\leq M <\infty \hspace{3mm} \forall v>0
\label{mass}
\ee
together with (\ref{mdot}).

The spherically symmetric collapse of null radiation may lead to the formation of a naked singularity \cite{Pap,K}. To avoid this possibility one has to assume that \cite{K}
$$
\lim_{v\rightarrow 0^+}\frac{m(v)}{v}> \frac{1}{16}\, .
$$
A hidden, non-naked, curvature singularity is always present at $r=0,\, v>0$. This is a spacelike future singularity.

Under all the above conditions, the Penrose diagrams for the imploding Vaidya spacetime are depicted in figure \ref{fig:vaidya}, the first of them for the case with $m(v)<M$ everywhere, the second with $m(v)=M$ from $v=v_1$ on.

The Kodama vector field $\xiv=\partial_v$ defined in subsection \ref{sec:Kodama} is actually a proper Kerr-Schild vector field \cite{CHS} (KSVF from now on) of type (\ref{KSVF}) relative to the null direction $\vec\ell$ for the Vaidya spacetime. 
It is immediate to get
$$
(\lie_{\xiv}g)_{\mu\nu}=2\frac{dm}{dv}\ell_\mu\ell_\nu \, , \hspace{1cm} (\lie_{\xiv}\ell )_\mu=0
$$
so that the function $h$ in (\ref{KSVF}) is $h=dm/dv\geq 0$. This is one of the requirements of Theorem \ref{th:no-min}. The other requirements are satisfied as in the general case, so that $\xiv$ is future pointing on the region $\R=\rr\cup\AH$, and the level function $\tau$ is defined by
\be
\xi_{\mu}dx^\mu =-Fd\tau = dr-\left(1-\frac{2m(v)}{r}\right) dv . \label{tau}
\ee
Besides, $\xiv$ is timelike on $\rr$ and null at the AH: $r=2m(v)$. 

Notice that the KSVF $\xiv =\partial_v$ coincides with the standard static Killing vector on the regions with $m(v)=$ const.\ and in particular with a static Killing vector in the flat region $v<0$ where $m(v)=0$. Thus, all portions of AH that are isolated horizons are actually Killing horizons, including the portion of the EH with $v>v_1$ in the case that $m(v)=M$ for all $v>v_1$.

\subsection*{The level hypersurfaces $\tau=$ const.\ and the past barrier $\Sigma$ in the Vaidya spacetime.}

Now, we analyze the shape and position of the level hypersurfaces $\tau=$const. for the KSVF $\xiv =\partial_v$ in the Vaidya spacetime (\ref{ds2}) subject to (\ref{mdot},\ref{mass}). Observe that these hypersurfaces are characterized by being spherically symmetric and orthogonal to the hypersurfaces $r=$const. 

\begin{figure}[!ht]
\includegraphics[width=10cm]{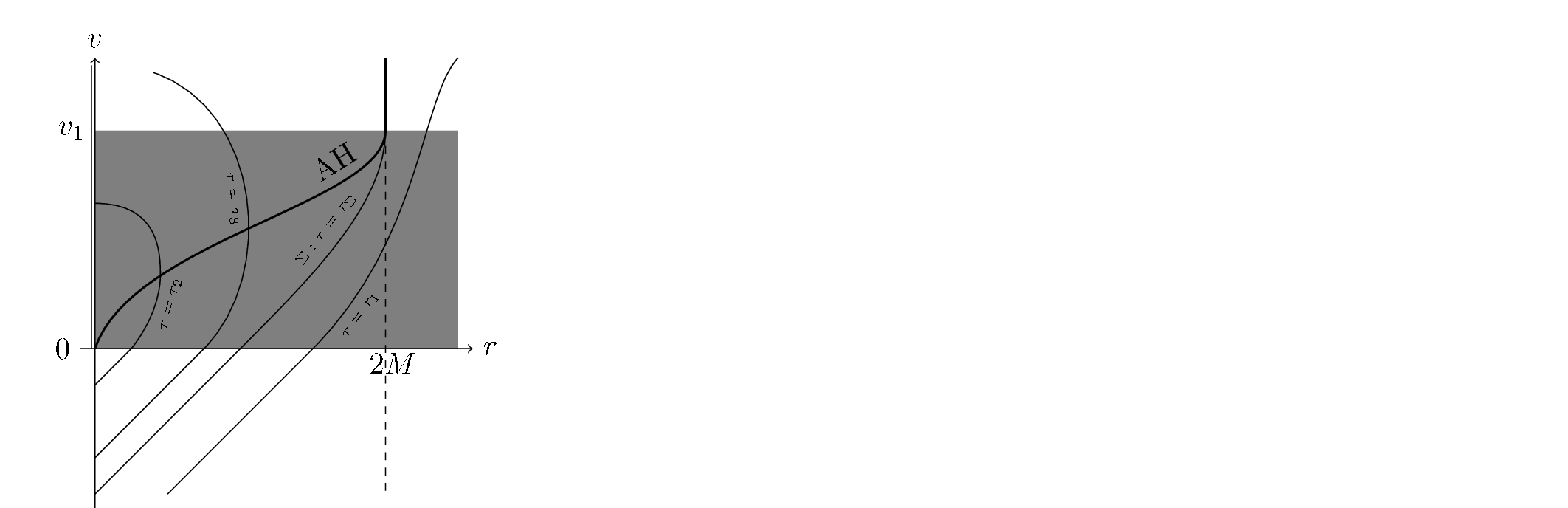}
\caption{\footnotesize{The phase plane $\{v,r\}$ around the critical point $(0,0)$ for the equation (\ref{ode}) in the spacetime (\ref{ds2}) subject to (\ref{mdot}) and (\ref{mass}). Only the case with $m(v)=M$ for all $v\geq v_1$ is represented. It should be noticed that horizontal lines ($v=$const.) are null in the spacetime, while vertical lines ($r=$const.) are timelike below AH, null at AH, and spacelike above it. Several $\tau=$const. hypersurfaces are represented. They are spacelike while they remain below AH, null when they meet AH, if they do, and timelike above AH. Here, we have represented the case with $m_0>1/8$ (see main text), so that the origin is a {\em focus}, which implies that all $\tau=$const. hypersurfaces, and in particular $\Sigma$, never reach the origin.}}
\label{fig:Focus}
\end{figure}

The definition of $\tau$ is (\ref{tau}), so that the sought hypersurfaces are defined by the solution to the differential equation
\be
\frac{dv}{dr}=\frac{1}{1-2m(v)/r}. \label{ode}
\ee
This is equivalent to the autonomous system
$$
\frac{dv}{du}=r, \hspace{1cm} \frac{dr}{du}=r-2m(v)
$$
so that its orbits on the phase plane $\{v,r\}$ provide the required hypersurfaces. Given that $m(0)=0$, the ODE (\ref{ode}) has a critical point at $v=0, r=0$. Its linear stability is ruled by the eigenvalues $\lambda_{\pm}$ and eigenvectors $\vec{z}_{\pm}$ of the corresponding matrix
$$
\left(\begin{array}{ccc}
0 & 1\\
-2m_0 & 1 \end{array}
\right), \hspace{1cm} m_0\equiv \lim_{v\rightarrow 0^+}\frac{m(v)}{v}
$$
which are given by
$$
\lambda_{\pm}=\frac{1}{2}\left(1\pm \sqrt{1-8m_0}\right), \hspace{1cm} \vec{z}_{\pm}=(1,\lambda_\pm).
$$
The character of the critical point is different depending on whether $m_0<1/8$ or not. If $m_0>1/8$, the critical point is an unstable focus, so that no solution actually reaches $(0,0)$. The schematic phase plane is represented in figure \ref{fig:Focus}. As we can check, the hypersurface $\Sigma$ always penetrates the flat region in this case.

If $m_0<1/8$ the critical point is an unstable node, with $\lambda_\pm >0$ real and positive. In fact, one can see that
$$
\lambda_+\in\left(\frac{1}{2},1\right], \hspace{5mm} \lambda_-\in \left[0,\frac{1}{2}\right), \hspace{1cm} \lambda_+ +\lambda_- =1\, .
$$
In this case, all possible solutions except one emerge from $(0,0)$ with the same tangent direction given by the eigenvector $(1,\lambda_-)$. The exception is given by one solution emerging from $(0,0)$ with the tangent direction of the other eigenvector $(1,\lambda_+)$. Recall that $m_0\geq 1/16$ is necessary to avoid (locally) naked singularities \cite{K}, and thus the allowed intervals for $\lambda_\pm$ can be further restricted if such singularities are to be avoided. There are three qualitatively different possibilities in this case (with the same value of $m_0$), depending on whether or not the special solutions $\tau=\tau_\pm$ ---corresponding to the eigenvalues $\lambda_\pm$ at $(0,0)$--- eventually meet the AH. 
\begin{figure}[!ht]
\includegraphics[width=10cm]{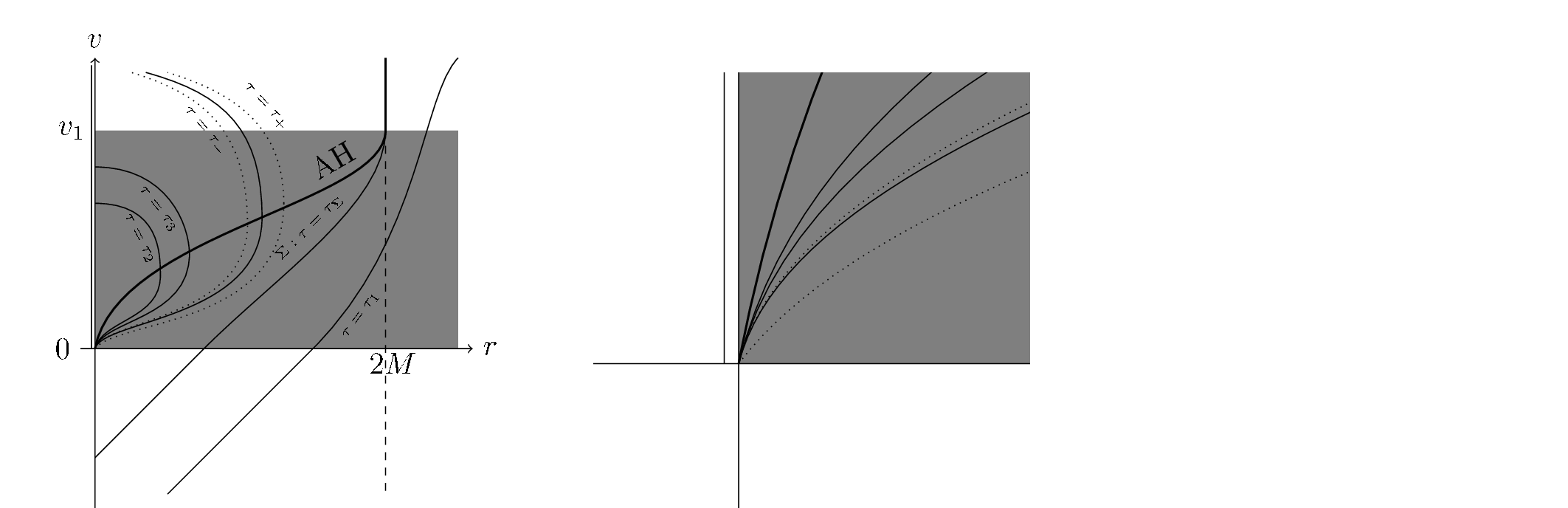}
\caption{\footnotesize{Same phase plane as in figure \ref{fig:Focus} but with $m_0<1/8$. The origin is  now a {\em node}, with two special solutions given by $\tau=\tau_{\pm}$ and represented by dotted lines. All solutions but $\tau=\tau_+$ emerge from the origin with the same slope given by that of the solution $\tau=\tau_-$. This is represented in the right picture, which is a magnification of a small  neighborhood of (0,0). If the mass function $m(v)$ is such that both solutions $\tau=\tau_{\pm}$ cross the AH, as depicted, then $\Sigma$ will never reach the origin. The other possibilities are given in figure \ref{fig:Node2}.}}
\label{fig:Node1}
\end{figure}
This, in turn, will depend on the specific properties of the mass function $m(v)$ for $v\neq 0$ and on the total mass $M$. Loosely speaking, if there is a period when the mass function has a large derivative, then the chances are that the special solutions will meet the AH. If at least one of the special solutions does {\em not} meet the AH, then the hypersurface $\Sigma$ cannot penetrate the flat region. On the other hand, if both special solutions meet AH then $\Sigma$ will have a portion in the flat region. These possibilities are schematically represented in the figures \ref{fig:Node1}-\ref{fig:Node2}.
\begin{figure}[!ht]
\includegraphics[width=10cm]{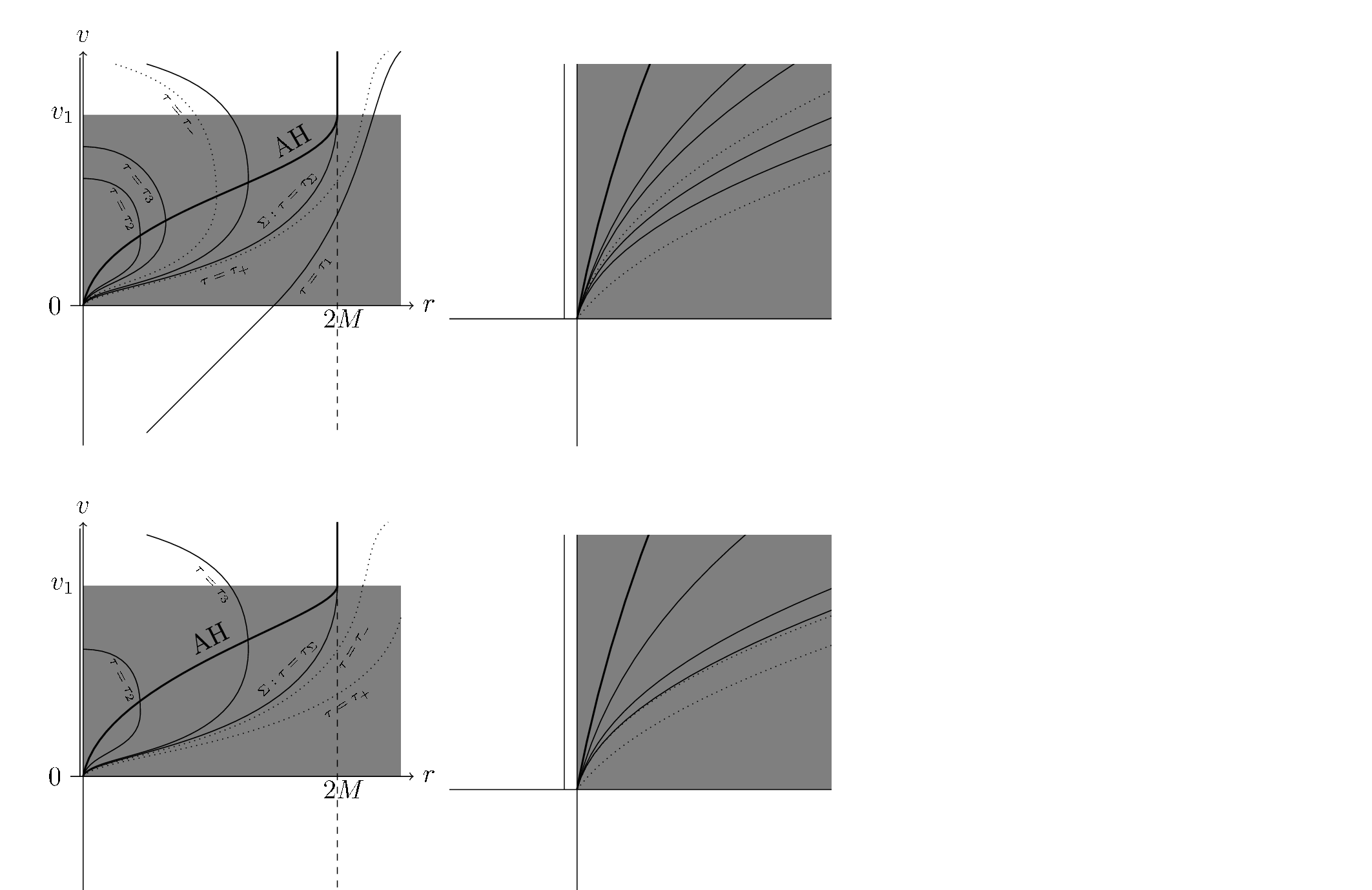}
\caption{\footnotesize{Same case $m_0<1/8$ as in figure \ref{fig:Node1} so that the origin is a {\em node}, with their corresponding magnifications on the right. If the mass function $m(v)$ is such that at least one of the solutions $\tau=\tau_{\pm}$ crosses the AH, then $\Sigma$ will reach the origin tangent to $\tau=\tau_-$. This implies that $\Sigma$ will never penetrate into the flat part of the spacetime, which is the half-plane $v<0$. The upper figure represents the case when only one of the two solutions crosses AH, while the bottom does the same when both $\tau=\tau_{\pm}$ are spacelike everywhere never crossing AH.}}
\label{fig:Node2}
\end{figure}

The limit case with $m_0=1/8$ has the character of a degenerate unstable node (only one universal direction through which all solutions emerge from the critical point) in the linear stability analysis, and it remains as such, or it  may become an unstable focus or node, depending on the specific properties of the function $m(v)$ around $v=0$. The schematic structure of the phase portraits for this case are thus analogous to those already shown, with the small difference that the solutions $\tau=\tau_+$ and $\tau=\tau_-$ coincide when $(0,0)$ is a degenerate node for the full, non-linear, system.

The corresponding Penrose diagrams, including the most relevant $\tau=$const. hypersurfaces, are presented in the next figure \ref{fig:vaidya2}.

\begin{figure}[!ht]
\includegraphics[width=16.5cm]{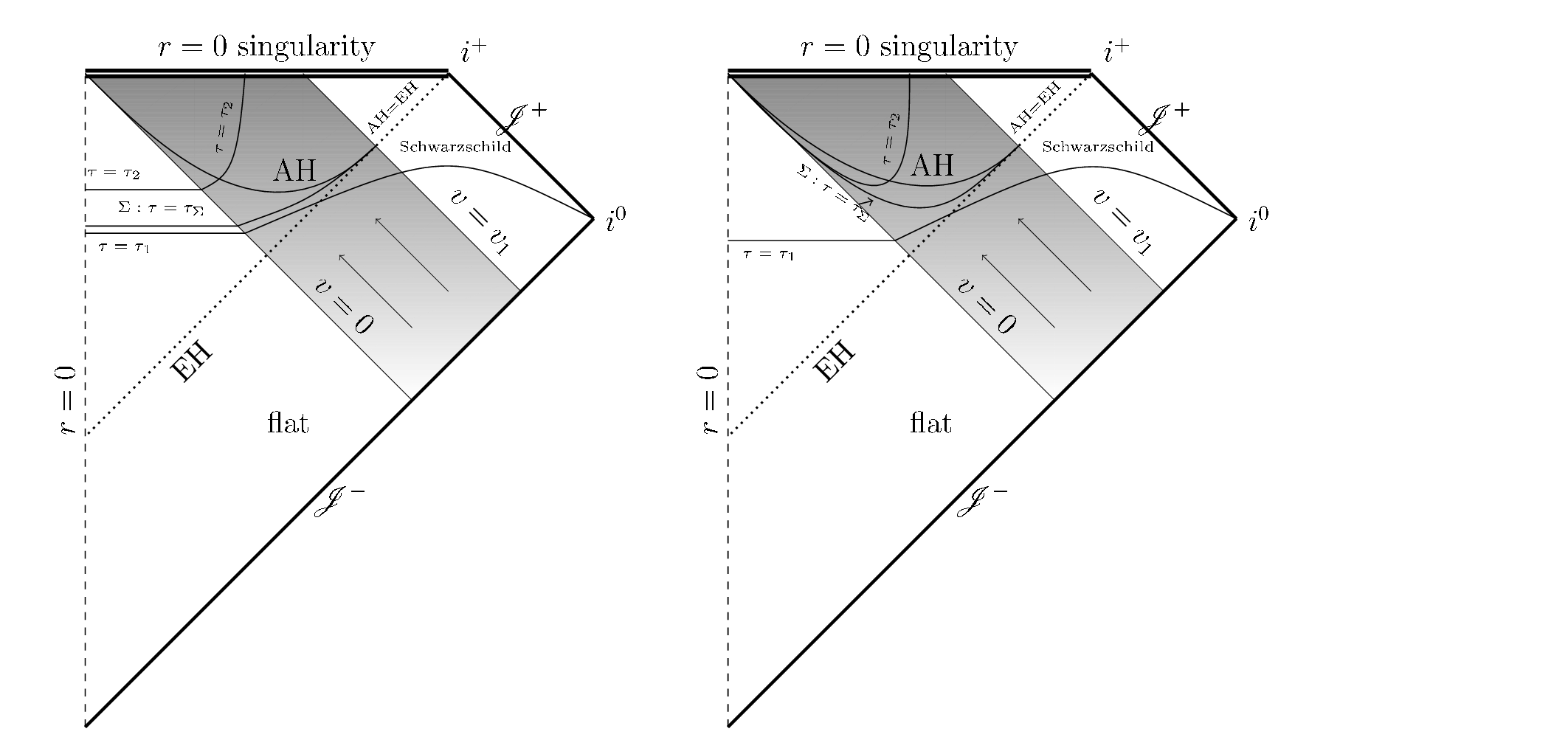}
\caption{\footnotesize{These are conformal diagrams of spacetime (\ref{ds2}) with (\ref{mdot}) and (\ref{mass}) when $m(v)=M$ for all $v>v_1$. The notation is as in figure \ref{fig:vaidya}. Here, on the left picture we have represented the case in which the hypersurface $\Sigma$ has a portion in the flat region of the spacetime (e.g. if $m_0>1/8$, but not only). On the right picture, the other possibility is represented, when $\Sigma$ never touches the flat region. These two pictures are the global Penrose diagrams corresponding to the pictures on the right of figure \ref{fig:SigmainVaidya}.}}
\label{fig:vaidya2}
\end{figure}

To illustrate the above, we present a particular example where the solutions of (\ref{ode}) can be given explicitly in full. This is given by the self-similar Vaidya spacetime, with a linear mass function
$$
m(v)=\left\{\begin{array}{lc}
M & v> v_1=M/\mu \\
\mu v & 0\leq v\leq v_1\\
0 & v<0
\end{array} \right. 
$$
which admits the following homothetic Killing vector field for all $v<v_1$
$$
\vec\zeta =v\partial_v+r\partial_r, \hspace{1cm} (\lie_{\vec\zeta}g)_{\mu\nu}=2g_{\mu\nu} \, .
$$
Observe that $m_0=\mu$ in this case, so that $\lambda_\pm=(1\pm\sqrt{1-8\mu})/2$. The solutions to the ODE (\ref{ode}) provide the level hypersurfaces for $\xiv$. They are given by $\tau =v -r$ for all $v< 0$. For $v>0$ we have:
\begin{itemize}
\item $\mu >1/8$. The critical point is an unstable focus and the solutions are
$$
\tau (v,r)=-\sqrt{r^2-vr+2\mu v^2}\exp\left\{\frac{1}{\sqrt{8\mu -1}}\arctan \left(\frac{1-4\mu v/r}{\sqrt{8\mu -1}} \right)  \right\}
$$
\item $\mu =1/8$. The critical point is a degenerate unstable node whose particular special solution is simply
$$
\tau =\tau_+=\tau_- \Longleftrightarrow v=2r
$$
and the rest of solutions, all of them emanating from the origin tangent to the special solution $v=2r$, are
$$
\tau (v,r)=(v-2r)\exp\left\{\frac{2r}{v-2r} \right\}
$$
\item $\mu <1/8$. The critical point is an unstable node with the special solutions
$$
\tau =\tau_+ \Longleftrightarrow 2\mu v=\lambda_+ r , \hspace{15mm}
\tau =\tau_- \Longleftrightarrow 2\mu v=\lambda_- r
$$
the first of them being the exceptional one. The rest of solutions, all of them emanating from the origin tangent to the second special solution $2\mu v=\lambda_- r$, are
$$
\tau (v,r)= (2\mu v -\lambda_- r)^{\frac{\lambda_+}{\sqrt{1-8\mu}}} (2\mu v -\lambda_+ r)^{-\frac{\lambda_-}{\sqrt{1-8\mu}}}
$$
\end{itemize}
In all three cases, the hypersurface $\Sigma$ is given by $\tau=\tau_\Sigma$, where $\tau_\Sigma \equiv \tau(M/\mu ,2M)$. Observe that $\Sigma$ does not enter the flat region if $\mu \leq 1/8$, and therefore no closed f-trapped surface can penetrate the flat region  in these cases \cite{BS}.

We note as a final remark that the AH in this particular Vaidya spacetime is an intrinsically flat hypersurface, and that the trace of its second fundamental form (its expansion) is proportional to $1-8\mu$. Therefore, the AH is non-expanding exactly for the limit case with $\mu =1/8$. In the cases where the past barrier $\S$ enters the flat portion of the spacetime, the AH is contracting.

\end{document}